\documentclass[twocolumn,nofootinbib,prc,amsmath,amssymb,10pt,superscriptaddress,floatfix,a4paper,fleqn]{revtex4-1}

\usepackage{amssymb}
\usepackage{amsfonts}
\usepackage{amsmath}
\usepackage{makeidx}
\usepackage[bookmarksnumbered,colorlinks,bookmarks,breaklinks=true,linkbordercolor={1 1 1},linktocpage,citecolor=blue]{hyperref}
\usepackage{graphicx,color}
\usepackage{subfigure}
\usepackage{multirow}
\usepackage{dcolumn}
\usepackage{bm}
\usepackage{fancyhdr}

\def\utwo{$^{232}$U}
\def\uthree{$^{233}$U}
\def\ufour{$^{234}$U}
\def\ufive{$^{235}$U}
\def\usix{$^{236}$U}
\def\useven{$^{237}$U}
\def\ueight{$^{238}$U}
\def\uall{$^{233-238}$U~}

\def\xs{cross section}
\def\xss{cross sections}

\def\r{Ref.~}
\def\csin{Ref.~\cite{Sin:2017}}

\def\ce{\cite{EMPIRE}}
\def\ceem{\cite{EMPIRE,EMPIRE-man}}

\newcommand\T{\rule{0pt}{2.6ex}}       
\newcommand\B{\rule[-1.2ex]{0pt}{0pt}} 
\newcommand{\eqnarrow}{\setlength{\mathindent}{3mm}}
\newcommand{\eqnormal}{\setlength{\mathindent}{10mm}}

\def\ql{``}
\def\qr{''\hspace{0.5mm}}
\def\qrs{''}
\def\etal{et al.~}
\def\etals{et al.}

\begin{document}
\title{Modeling photon--induced reactions on $^{233-238}$U actinide targets}

\author{M. Sin}
\affiliation{University of Bucharest, Faculty of Physics, Bucharest-Magurele, RO-077125, Romania}
\email{mihaela.sin@g.unibuc.ro}

\author{R.\thinspace Capote}
\affiliation{NAPC--Nuclear Data Section, International Atomic Energy Agency, A-1400 Vienna, Austria}
\email[Corresponding author: ]{r.capotenoy@iaea.org}

\author{M.\thinspace W.\thinspace Herman}
\affiliation{Los Alamos National Laboratory, Los Alamos, NM 87544, USA}

\author{A.\thinspace Trkov}
\affiliation{NAPC--Nuclear Data Section, International Atomic Energy Agency, A-1400 Vienna, Austria}

\author{B.\thinspace V.\thinspace Carlson}
\affiliation{Instituto Tecnol\'ogico de Aeron\'autica, Brazil}

\begin{abstract}
Comprehensive calculations of cross sections of photon induced reactions on \uall~targets for incident photon energies from 3 up to 30~MeV are undertaken with the statistical model code EMPIRE-3.2 Malta.  Results are compared with the experimental data from EXFOR and with the current evaluations. The differences and the similarities between the models and parameters used in calculations of photon- and neutron-induced reactions on the same nuclei are discussed with focus on fission. 
The role of the extended optical model for fission in improving the description of the measured data and in determining  consistent sets of barrier parameters is pointed out.



\end{abstract}

\maketitle
\section{Introduction}\label{Intro}

The fission model and model parameters represent one of the largest sources of uncertainty when performing model reaction calculations for actinide targets.

To address this issue, the International Agency for Atomic Energy is coordinating an ongoing research project
to deliver comprehensive sets of fission parameters corresponding to recommended well-documented models~\cite{CRP-RIPL4}. Newly proposed parameterized models are expected to enhance the use of modelling in evaluation practice and to meet target uncertainties for applications.

As pointed out in \csin, ``a consistent and reliable  set of fission  parameters  as  model independent as
possible is the one which provides simultaneously a reasonable  description  of  multiple  fission  chances  induced
by neutrons, photons, protons or direct transfer reactions
leading to the same fissionable compound nucleus''. Such consistent sets of fission  parameters have been obtained for the Uranium isotopic chain from the simultaneous description of the experimental neutron-induced reactions \xss~and Neutron Standard cross sections~\cite{Carlson:2009,Carlson:2018} by model calculations in Refs.~\cite{Sin:2017,Capote:2014a,Capote:2014b,Capote:2014c}. The evaluations based on those calculations for \ufive~and \ueight~ have been produced within the NEA CIELO project~\cite{CIELO-U,CIELO:2018}, and have been adopted by the ENDF/B-VIII.0 library~\cite{ENDFVIII}.

A new step in obtaining ``consistent and reliable''  sets of fission  parameters for the Uranium isotopes is presented in this work by testing the compatibility of the fission parameters deduced from the fit of the neutron induced fission \xss~in \csin~with the input parameters specific to photon-induced reaction modelling. For this purpose,  photon-induced reaction \xs~calculations for \uall have been performed with the statistical model code EMPIRE--3.2 Malta \ceem~in the incident energy range 3--30 MeV.

Photon-induced reactions are important for a large range of applications and can provide useful information for the data evaluation of reactions induced by other particles. In the last decade, new photon sources became available and others are under construction (e.g., the Extreme Light Infrastructure - Nuclear Physics (ELI-NP) \cite{Filipescu:2015}). In response to the growing needs for photonuclear data, IAEA-NDS initiated a  research project on Photonuclear Data and Photon Strength Functions, with the primary task to create a new
IAEA Photonuclear Data Library~\cite{photo-CRP}. Several evaluations based on photo-reaction calculations performed with the EMPIRE code have been already included in this library~\cite{Kawano:2020}.

The calculated photo-reaction cross sections for \uall~which are in agreement with the experimental data, as well as the models (e.g., the extended optical model for fission) and the parameters (e.g., for the Giant Dipole Resonances and for the fission barriers) reported in this paper integrate into this context of scientific interest.

\section{Reaction models and parameters}
\label{models}
The models and parameters implemented in the EMPIRE--3.2 code \ce~and used for the present photo-reaction calculations on Uranium isotopes are briefly outlined,  mentioning the differences and the similarities with the models and parameters used in \csin~for the calculations of neutron-induced reactions on the same target nuclei. Initial values of the model parameters are automatically retrieved in the EMPIRE code from the Reference Input Parameter Library (RIPL-3) \cite{ripl3}.

\subsection{Incident channel}
The photo-nuclear excitation process is described by two mechanisms: the excitation of the isovector Giant Dipole Resonances (GDR) which dominates at low energies, below about 30 MeV, and the photo-absorption on a neutron-proton pair (a quasi-deuteron, QD) which dominates at higher energies.
\\
The total gamma cross section is calculated in the EMPIRE code  as the sum of two components \ceem
\begin{equation}
\sigma_{\gamma t}(E_{\gamma})=\sigma_{GDR}(E_{\gamma})+\sigma_{QD}(E_{\gamma}).
\label{sigma_tot}
\end{equation}
The QD component $\sigma_{QD}(E_{\gamma})$ has a small contribution in the studied energy range and is not discussed in this paper.
The GDR component, $\sigma_{GDR}(E_{\gamma})$, is calculated in terms of the photo-excitation (upward) strength function $\overrightarrow{f}$
\begin{equation}
\sigma_{GDR}(E_{\gamma})=3(\pi\hbar c)^2\cdot E_{\gamma}\cdot\overrightarrow{f}(E_{\gamma}).
\label{sigma_GDR}
\end{equation}
In RIPL-3  there are several Lorentzian-type closed-form expressions for the dipole radiative (downward)
$\overleftarrow{f}$ and excitation (upward)  $\overrightarrow{f}$ strength functions, as well as microscopic
Hartree-Fock-Bogoliubov plus quasi-particle random phase
approximation model  predictions for these quantities. Theoretical details about these formulations can be found in \cite{ripl3}. All of them are implemented in the EMPIRE code.

The phenomenological expression of the excitation strength function for the cold and deformed nuclei (in units of MeV$^{-3}$) is the sum of two Lorentzian shapes
\eqnarrow
\begin{equation}
 \overrightarrow{f}(E_{\gamma})=c\sum_{i=1}^2\sigma_{ri}\Gamma_{ri}
\frac{E_{\gamma}\Gamma_{i}(E_{\gamma})}      {(E_{\gamma}^{2}-E_{ri}^{2})^{2}+(E_{\gamma}\Gamma_{i}(E_{\gamma}))^{2}},
\label{MLO}
\end{equation}
\eqnormal
where $c=8.674\cdot 10^{-8}$, and $\sigma_{ri}$, $E_{ri}$ and $\Gamma_{ri}$ are the GDR peak cross
section (in mb), energy and width (in MeV), respectively. The different closed-form expressions treat differently   $\Gamma_{i}(E_{\gamma})$, a quantity which takes into account the collective state damping.
After testing all of them, in the EMPIRE code was selected as default option the Modified Lorentzian 1 (MLO1) strength function  that involves the use of the Landau-Vlasov equation with a collision term. More on the calculation of $\Gamma_{i}(E_{\gamma})$ in MLO1 approach can be found in Refs.~\cite{ripl3,Plujko:2007,Plujko:2011,Plujko:2018}.

If the excitation of the GDRs is considered the only (or the dominant) excitation mechanism, one assumes that only electric dipole transitions or only photons with zero orbital momentum
are involved. Because of the conservation laws which act as selection rules,
the photo-excited compound nucleus  is populated only in states with spins and parities
$J=J_0\pm 1,{\pi=-\pi_0}$, where $J_0,{\pi_0}$ are the spin and parity of the target  in the ground state. In reality this is true only at low energies, because at higher energies, due to the gamma cascade, the compound nucleus can have different spins and parities.  In the reactions induced by fast neutrons, which may have higher orbital momenta, such a strict selectivity in spin and parity does not appear.

\subsection{Exit channels}
According to the Bohr hypothesis \cite{Bohr:1936}, the compound nucleus should have decay probabilities independent of its formation. In photon and neutron induced reactions the compound nucleus is not populated in the same states, therefore the decay probabilities in the two cases are not expected to be the same, but it is expected to be described by the same models and parameters. This assumption was tested by using, for the present photo-reaction calculations, the same models and parameters used in \csin~to describe the outgoing channels in neutron induced reactions.

The main outgoing channels up to 30 MeV incident energy are gamma decay ($\gamma,\gamma$), neutron emission ($\gamma,n$), ($\gamma,2n$), ($\gamma,3n$) and fission ($\gamma,f$).
The charged particle emission ($p,\alpha, d, t, ^3$He) become comparable with the neutron emission around 30 MeV, but have a small contribution below 20 MeV. Not being relevant for the aim of this paper, they
are not further discussed, but are considered in calculations as competing channels.
 The photon, neutron and charged particles emission have a preequilibrium and a compound nucleus component, while fission is a compound nucleus process.

Preequilibrium emission was described by the one-exciton  model with  gamma,  nucleon  and  cluster emissions implemented in the EMPIRE  module  PCROSS \cite{PCROSS}.
The Hauser-Feshbach model \cite{hf52}  with full gamma cascade and exact angular momentum and parity coupling was employed for the compound nucleus reaction calculations. It should be noted that width-fluctuation corrections do not play an important role for photon-induced reactions~\cite{HRTW75}, nor the effects of deformation studied in Ref.~\cite{Kawano:2016}. The particle transmission coefficients have been calculated with the same optical potentials as in \r\cite{Sin:2017}. For the gamma transmission coefficients was used the MLO1 radiative strength function with GDR parameters obtained in this paper.
The discrete levels for the compound nucleus and the residual nuclei were retrieved from RIPL-3.
The level densities, both at the  equilibrium deformation and at the saddle points, have been described with the Enhanced Generalized Superfluid Model (EGSM)  and the same parameters as in \r\cite{Sin:2017}.
\begin{figure*}[!t]
\vspace{-3mm}
\subfigure{\includegraphics[trim={0cm 11cm 14cm 0cm},clip,width=0.98\columnwidth]{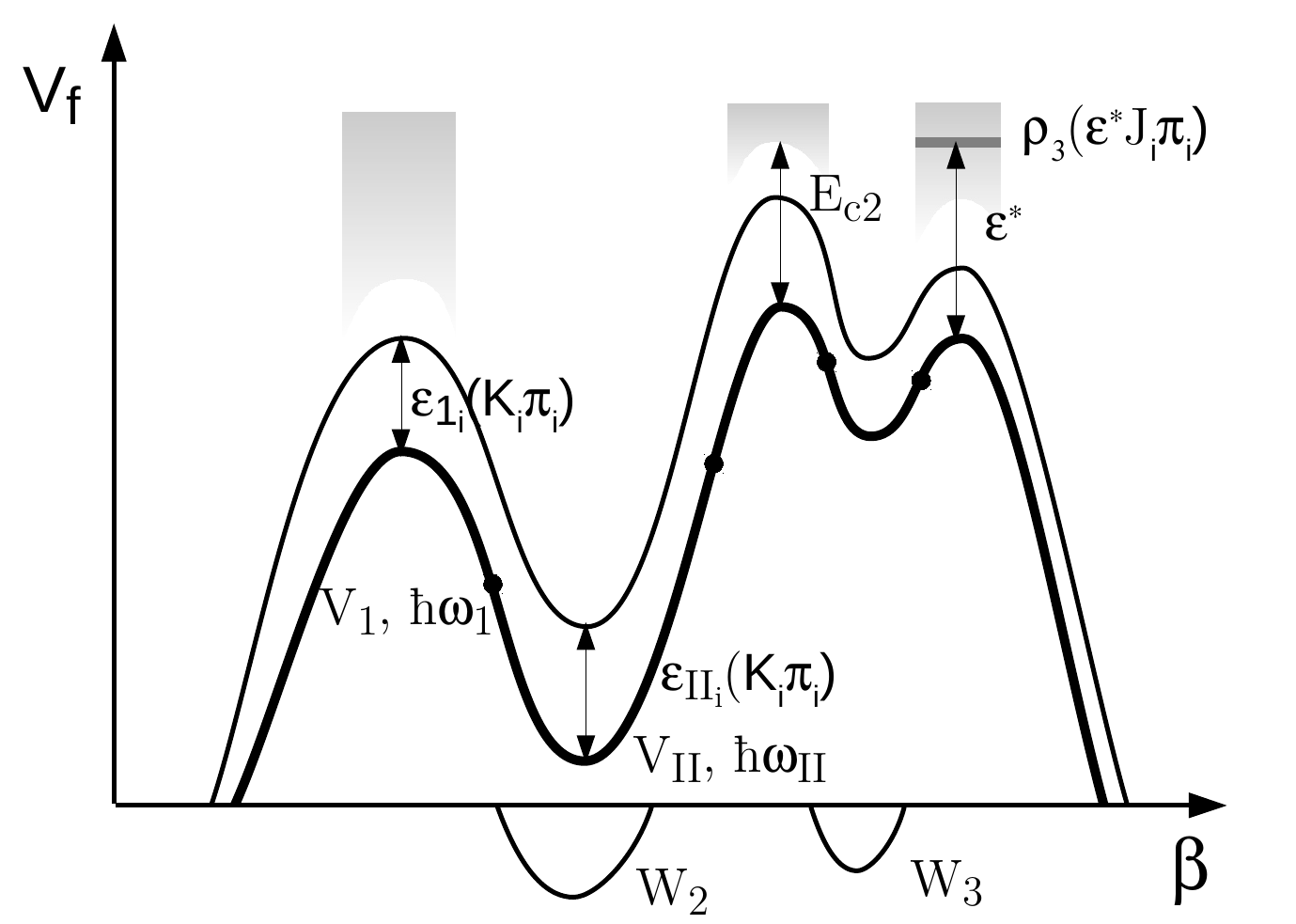}}\hfill
\subfigure{\includegraphics[trim={0cm 11cm 14cm 0cm},clip,width=0.98\columnwidth]{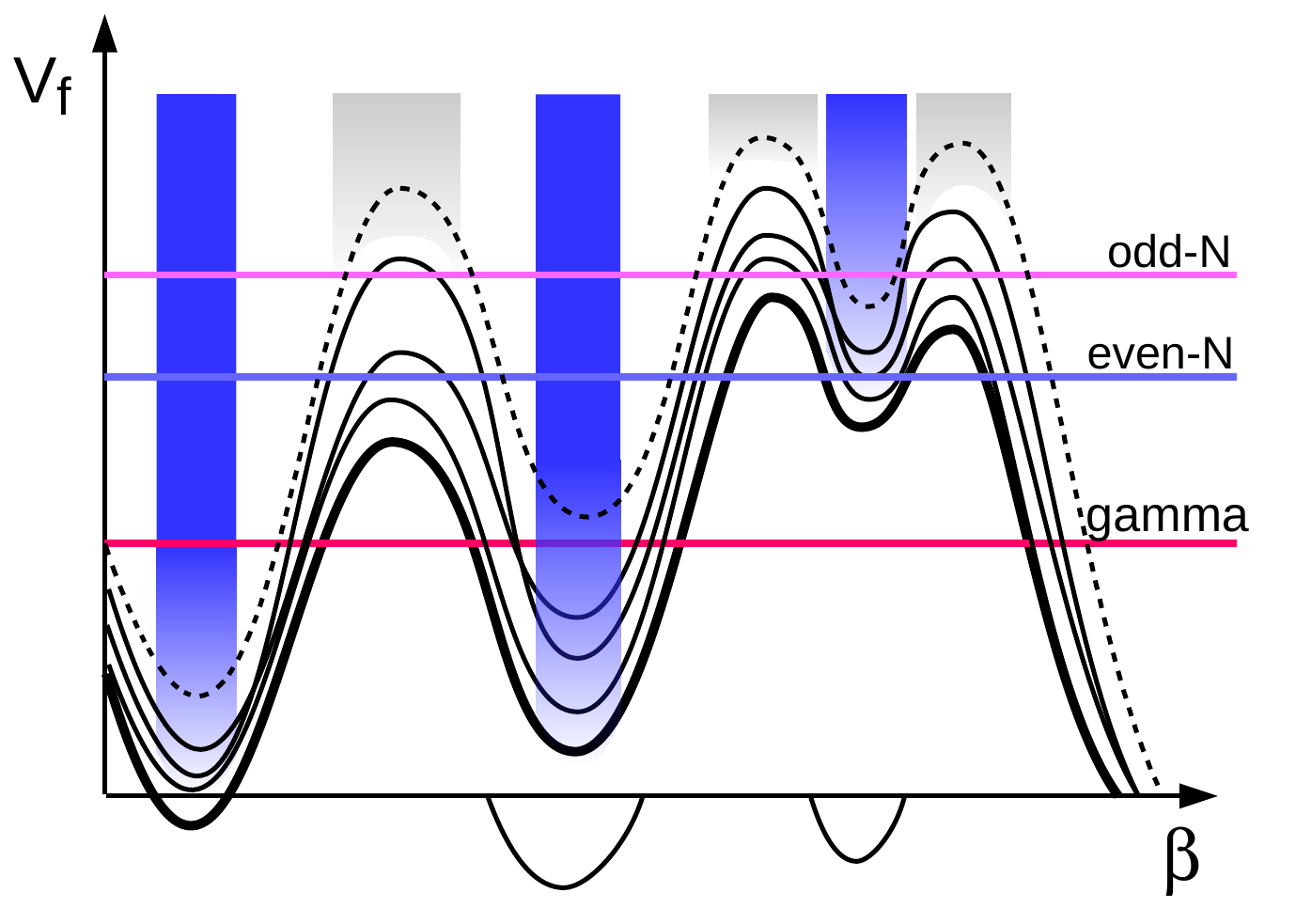}}
\vspace{-4mm}
\caption{Triple-humped fission barriers of light actinides (explanations provided in text).}
\label{fisbar-triple}
\vspace{-3mm}
\end{figure*}
The fission coefficients have been calculated with the extended optical model for fission (OMF).
In OMF the main fission mode is associated to the nuclear vibrational motion, so that the key role is played by: (i) the coupling between the vibrational states with similar excitation energies, the same spin projection on the symmetry axis and parity, which are located in different wells of the fission barrier, and (ii) by the coupling between the vibrational states in each well and other degrees of freedom which increases with increasing the excitation energy above the bottom of the well. The first type of coupling is responsible for the direct resonant transmission across the multi-humped barrier at the excitation energies  of the
vibrational states in the wells. The second type of coupling which dissipates or damps the vibrational strength of the states is interpreted as an absorption out of the fission mode, and it is simulated  by adding to the real part of the deformation potential imaginary term(s) in the region of the well(s). This second type of coupling is responsible for the indirect transmission mechanism representing transmission through the outer hump(s) after absorption in the well(s).

In the EMPIRE code it is implemented a very compact and elegant formulation of OMF which describes transmission through barriers with any number of humps and absorption in any number of wells \cite{Sin:2017, Sin:2016, Sin:2008}.

In the present work one has considered, as in \csin, triple-humped barriers  for $^{231-237}$U  ($^{231,232}$U are responsible for the second and third fission chances in $^{233}$U($\gamma,f$) reaction) and a double-humped barrier for $^{238}$U. The model implemented in EMPIRE was applied before for the neutron induced fission of light actinides with triple-humped barriers as $^{232}$Th, $^{231,233}$Pa \cite{Sin:2006} and $^{232-237}$U \cite{Sin:2017}. It is worth remembering that the evaluations performed based on the model calculations for $^{232}$Th, $^{231,233}$Pa, and $^{235}$U were adopted by the ENDF/B-VIII.0 library \cite{ENDFVIII}, the evaluation for $^{235}$U being included also in the CIELO library \cite{CIELO-U}.

The OMF formalism and the parameterization of the fission barriers are fully described in Ref.~\cite{Sin:2016} and rehashed in \csin, therefore only selected features of interest for the present work are reviewed in this contribution.

In Fig.~\ref{fisbar-triple} a typical triple-humped barrier for light actinides considered to have the inner hump wider and lower than the outer humps is  sketched. The second well which accommodates the super-deformed (class II) states is much deeper than the third well which accommodates the hyper-deformed (class III) states (the class I states are the normal states situated in the minimum corresponding to the equilibrium deformation not included in the barrier description). The bold black curve represents the fundamental barrier, with thinner black lines representing the barriers associated to the fission paths of the nucleus in different discrete excited states, and the continuum spectra of the transition states represented in gray shadows. The imaginary potentials (W$_2$,W$_3$) are introduced in the wells region to simulate the damping of the vibrational strength of the class II and III states. As exemplified in the left panel of Fig.~\ref{fisbar-triple}, for barriers with parabolic representation the fission input parameters are: (i) the heights/depths and widths of the humps/wells of the fundamental barrier, (ii) the sets $\varepsilon_i(K\pi)$ representing the excitation energy of discrete levels with respect to the fundamental barrier at saddle points and in wells, the spin projection along the symmetry axis, and the parity, for each discrete barrier, (iii) the parameters defining the transition states densities for the continuum above the humps, and (iv) the strengths of the imaginary potentials. Note that the barrier continuity condition requires the same number of discrete levels at all humps and in all wells.

In the right panel of Fig.~\ref{fisbar-triple} the gradients in blue suggest the vibrational strength's degree of damping for the class I, II and III states.  The horizontal lines indicate the excitation energy of the compound nucleus (CN) formed in three situations: after absorption of photons with incident energy of about 3 MeV (red line), after absorption of neutrons with incident energy of about 10 keV by an even-N fertile target (blue line) and by an odd-N fissile target (magenta line). The incident energies selected for this illustration are the lowest ones considered in the present work for photons and in \csin~for neutrons. This picture reveals several aspects important for the fission modeling below the excitation energy of approximately 6-7 MeV: (i) at 3 MeV the class I vibrational state (in the minimum corresponding to the equilibrium deformation) are already completely damped, therefore this first well is not included in the parameterization of the deformation potential, (ii) the different shades of blue (degree of damping) in the second and third well at the three excitation energies taken as example explain the different behavior of the photon and neutron induced fission cross sections at low energies, as discussed in the next Section,  (iii) the dashed line barrier indicates that the class III vibrational states associated to barriers with maxima in the continuum still can be only partially damped, (iv) the full damping approximation implemented in most of the statistical reaction codes obviously can not be used in this energy range.
\begin{figure*}[!t]
\vspace{-3mm}
\subfigure{\includegraphics[trim={0cm 11.cm 14cm 0cm},clip,width=0.98\columnwidth]{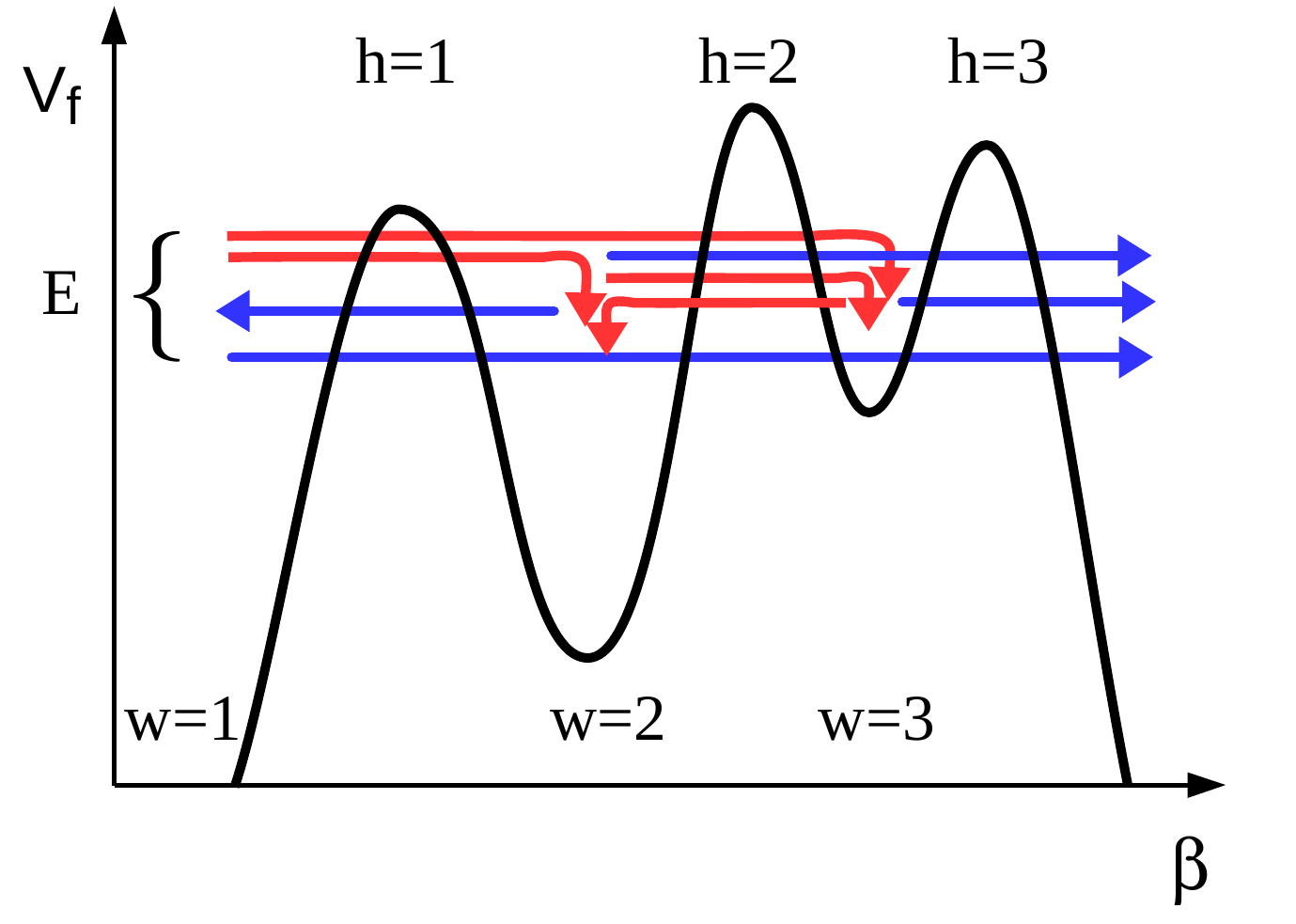}}\hfill
\subfigure{\includegraphics[trim={0cm 11.cm 14cm 0cm},clip,width=0.98\columnwidth]{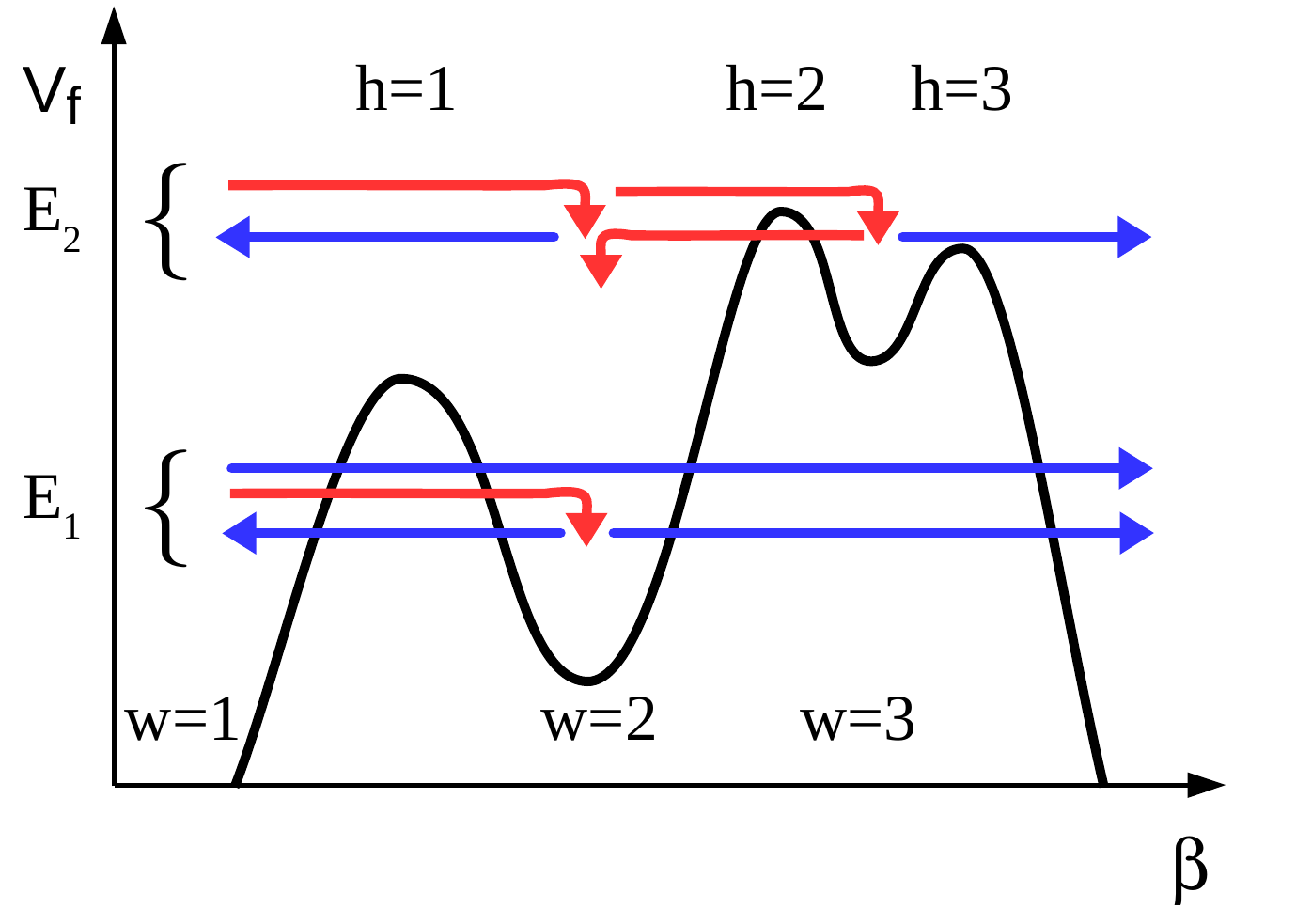}}
\vspace{-4mm}
\caption{Transmission mechanisms through triple-humped fission barriers (explanations provided in text).}
\label{fisbar-triple-m}
\vspace{-3mm}
\end{figure*}

Expressions relevant for photon-induced fission on nuclei with triple-humped barriers have been extracted from the extended OMF formalism and will be discussed below.
In the left panel of Fig.~\ref{fisbar-triple-m} are represented the transmission mechanisms through a triple-humped fission barrier at the excitation energy $E$: the blue arrows represent forward and backward direct transmission through one or more humps, and the bent red arrows describe the absorption in the wells. The total fission coefficient is the sum of the direct transmission through the entire barrier ($T_d^{(1,3)}$) and two indirect terms representing re-emission in the fission channel after absorption in the second ($T_i^{(2)}$) and third well ($T_i^{(3)}$)
\begin{equation}
T_f=T_d^{(1,3)}+R[T_i^{(2)}+T_i^{(3)}].
\label{tf-triple}
\end{equation}
The normalization factor $R$, defined by Eq.~\eqref{r}, takes into account the infinite sequence of shape transitions  of the nucleus between wells ensuring the flux conservation. The indirect transmission coefficients have the expressions
\begin{equation}
T_i^{(2)}=T_a^{(1,2)}
           \left[\frac{T_{d}^{(2,3)}}{\sum T(2)} +
                 \frac{T_{a}^{(2,3)}}{\sum T(2)}\cdot
                 \frac{T_3}{\sum T(3)}\right],
\end{equation}
\begin{equation}
T_i^{(3)}=T_a^{(1,3)}
           \left[\frac{T_3}{\sum T(3)} +
                 \frac{T_{a}^{(3,2)}}{\sum T(3)}\cdot
                 \frac{T_{d}^{(2,3)}}{\sum T(2)}\right],
\end{equation}
where: $T_d^{(h,h')}$ represent the direct transmission coefficients through the humps $h\div h'$,  $T_a^{(w,w')}$ represent the absorption coefficients from well $w$ into well $w'$, and $T_h$ stands for the transmission through the hump $h$ ($T_h=T_d^{(h,h)}$).
These transmission coefficients are calculated using the recursive procedure presented in \r\cite{Sin:2008}, having as starting point the expressions for a double-humped barrier proposed by Bhandari in \r\cite{Bhandari:79}. These expressions are derived in the
first order JWKB approximation \cite{Froman:65,FD70}, in terms of momentum integrals for the
humps and the wells of the real barrier and for the imaginary potential(s).
The denominators in the above equations represent the sum of the transmission coefficients for the competing channels specific to the second and third wells
\begin{eqnarray}
  \sum T(2) &=& T_1+T_d^{(2,3)}+T_a^{(2,3)}+T_{\gamma}^{(2)} \\ \nonumber
  \sum T(3) &=& T_d^{(2,1)}+T_3+T_a^{(3,2)}+T_{\gamma}^{(3)}.
\label{denom3}
\end{eqnarray}
The gamma-decay in the second and third wells has been considered in an approximate way, but the contribution of the isomeric (delayed) fission has been ignored. From a super-deformed state a shape isomer would decay mainly by gamma-emission, while the fission of a hyper-deformed shape isomer would occur at energies higher than the third well ($\approx$ 5 MeV) where the delayed fission contribution would be negligible. However, this subject needs further studies.\\
The normalization factor $R$ reads:
\begin{equation}
R=\left[1-\frac{T_a^{(2,3)}}{\sum T(2)}\frac{T_a^{(3,2)}}{\sum T(3)}\right]^{-1}.
\label{r}
\end{equation}

The right panel of Fig.~\ref{fisbar-triple-m} presents two particular situations: transmission at an excitation energy $E_1$ lower than the third well, and  transmission at an excitation energy $E_2$ at which full-damping limit of the vibrational class II and III states is reached.
\\
Below the third well the absorption coefficients $T_a^{(1,3)}$, $T_a^{(2,3)}$ are zero and
Eq.~\eqref{tf-triple} becomes an expression typical for the fission coefficient of a double-humped barrier
\begin{equation}
T_f=T_d^{(1,23)}+T_a^{(1,2)}
           \frac{T_{d}^{(23)}}{T_1+T_d^{(23)}+T_{\gamma}^{(2)}}.
\label{tf-triple-under}
\end{equation}
In the full-damping limit (which is equivalent to full flux absorption) corresponding to the excitation energy $E_2$, the direct
transmissions through more than one hump disappear (consequently  $T_a^{(1,3)}\rightarrow 0$), and the transmission across each hump is fully absorbed in the next well
\begin{eqnarray}
T_d^{(1,2)}\rightarrow 0, \quad T_d^{(1,3)}\rightarrow 0, \quad T_d^{(3,2)}\rightarrow 0
\qquad\\\nonumber
T_a^{(1,2)}\rightarrow T_1,\quad T_a^{(2,3)}\rightarrow T_2,\quad T_a^{(3,2)}\rightarrow T_2.
\end{eqnarray}
As expected, the Eq.~\eqref{tf-triple} takes the classical form
\begin{equation}
{T_f}=\frac{T_1T_2T_3}{T_1T_2+T_1T_3+T_2T_3}.
\label{tf-classic3}
\end{equation}

Another important aspect for the description of the fission cross section at excitation energies lower than 5--6 MeV is the treatment of the fission channels in the lower part of the continuum spectrum. The Eqs.~\eqref{tf-triple}--\eqref{tf-classic3} refer to the transmission coefficients for a single barrier. However, the spin and parity ($J\pi$) dependent fission coefficients (which we call effective fission coefficients) enter the Hauser-Feshbach formula for the compound nucleus cross sections \cite{hf52}. Those effective coefficients represent the transmission through all the barriers associated to the discrete and the continuous spectrum of the transition states with the same $J\pi$.
Therefore, the single-hump transmission and the direct and absorption coefficients are the sum of two contributions corresponding to the discrete and to the continuous part of the transition  state spectrum. The calculation of these effective fission coefficients is presented in detail in Refs.~\cite{Sin:2017, Sin:2016}. In this paper only simplified expressions for the continuum contribution are reproduced in which the explicit dependence on energy, spin and parity is omitted.
\\
The continuum contribution to the transmission coefficient across the hump $h$ is calculated as
\begin{eqnarray}
\label{tjt}
T_{h,cont}=\int_{E_{ch}}^{\infty}
             \frac{\rho_h(\varepsilon^*)d\varepsilon^*}
{{1+\exp\left[-\frac{2\pi}{\hbar\omega_{h}}(E-V_{h}-\varepsilon^*)\right]}},
\label{th-cont}
\end{eqnarray}
where $V_h,~\hbar\omega_{h}$ are the parameters of the hump $h$ of the fundamental barrier,
$\rho_h$ is the transition states density function, $E$ is the excitation energy and $E_{ch}$ is the energy where continuum starts with respect to the top of hump $h$ (see Fig.~\ref{fisbar-triple}).
\\
For a triple-humped barrier with a deep second well and a shallow third well the continuum contribution to different direct and absorption coefficients is different. The super-deformed (class II) vibrational states are completely
damped at the energies where the transmission across the barriers in continuum
becomes significant, so there is no direct transmission via these states, only full absorption in the second well (obviously there is no direct transmission across an entire barrier in continuum)
\begin{eqnarray}
T_{d,cont}^{(1,2)}=0, \quad T_{d,cont}^{(3,2)}=0, \quad T_{d,cont}^{(1,3)}=0 \nonumber \\
T_{a,cont}^{(1,2)}=T_{1,cont},\quad T_{a,cont}^{(3,2)}=T_{2,cont}\quad.\nonumber
\label{tcont_2}
\end{eqnarray}
On the other hand, the hyper-deformed vibrational states might not
be fully damped at the excitation energies where the transmission through the barriers
in continuum becomes important (see the dashed line barrier in the right panel of Fig.~\ref{fisbar-triple}). As explained in Refs.~\cite{Sin:2017, Sin:2016}, the treatment of partial damping for discrete barriers which cannot be applied  for those in continuum is replaced by a surrogate for the optical model for fission \cite{Sin:2008a,Goriely:2009}.
In this approach, the degree of damping is simulated by using a linear combination of a direct transmission coefficient through the outer humps corresponding to the zero-damping limit, and an indirect transmission coefficient corresponding to the full damping of the class III vibrational states. The continuum contributions to the direct and absorption coefficients involving the third well are
\begin{eqnarray}
T_{dir,cont}^{(2,3)}&=&(1-p_3)T_{d_{(0)},{cont}}^{(2,3)},\\ \nonumber
T_{abs,cont}^{(2,3)}&=&p_3T_{a_{(f)},cont}^{(2,3)}.
\label{tcont_3}
\end{eqnarray}
The expression for direct transmission coefficient corresponding to the zero-damping limit
$T_{d_{(0)},{cont}}^{(2,3)}$ is provided in Refs.~\cite{Sin:2017, Sin:2016}, and
$T_{a_{(f)},cont}^{(2,3)}\rightarrow T_{2,cont}$.
The definition of the energy dependent weight $p_3$ given in Refs.~\cite{Sin:2017, Sin:2016}
was changed to become valid at excitation energies lower than the third well and became
\begin{equation}
p_3 = \frac{E^2}{V_{d}^2{\rm exp}[-(E-V_{d})/b_3]},
\label{p2}
\end{equation}
where $V_d$ is the excitation energy where the full-damping limit is supposed to be reached and $b_3$ is a parameter which controls the energy dependence of the weight.
\\
Considering the partial damping of the fission channels in the lower part of continuum represents the extension of the optical model for fission.

The impact of the optical model for fission and of its extended version on the photo-fission cross sections of odd- and even-N Uranium isotopes is shown in Fig.~\ref{u34-OMF} and commented in Section~\ref{fis-res}.

\section{Results and discussions}\label{res}
The results of our calculations performed with the EMPIRE code for
 the photo-absorption, ($\gamma,n$), ($\gamma,2n$) and
($\gamma,f$) cross sections are compared to the available experimental data
from the EXFOR library \cite{EXFOR} and to the evaluated data from
JENDL/PD-2016 (JENDL-PD)~\cite{JENDL-PD} and IAEA-Photonuclear Data Library 1999 (IAEA-PD)~\cite{IAEA-PD}.
\\
To explain some of the similarities and differences between EMPIRE calculations and evaluations, it is worth mentioning that
JENDL-PD relies mainly on model calculations performed with the CCONE code \cite{CCONE} (which in many respects is close to EMPIRE code \cite {Capote:2017}), while IAEA-PD is mostly based on least-squares fit of the experimental data.

\subsection{Photo-absorption cross sections}

\begin{figure*}[!t]
\subfigure{\includegraphics[trim={1.4cm 2.2cm 3.0cm 4.54cm},clip,width=0.5\textwidth]{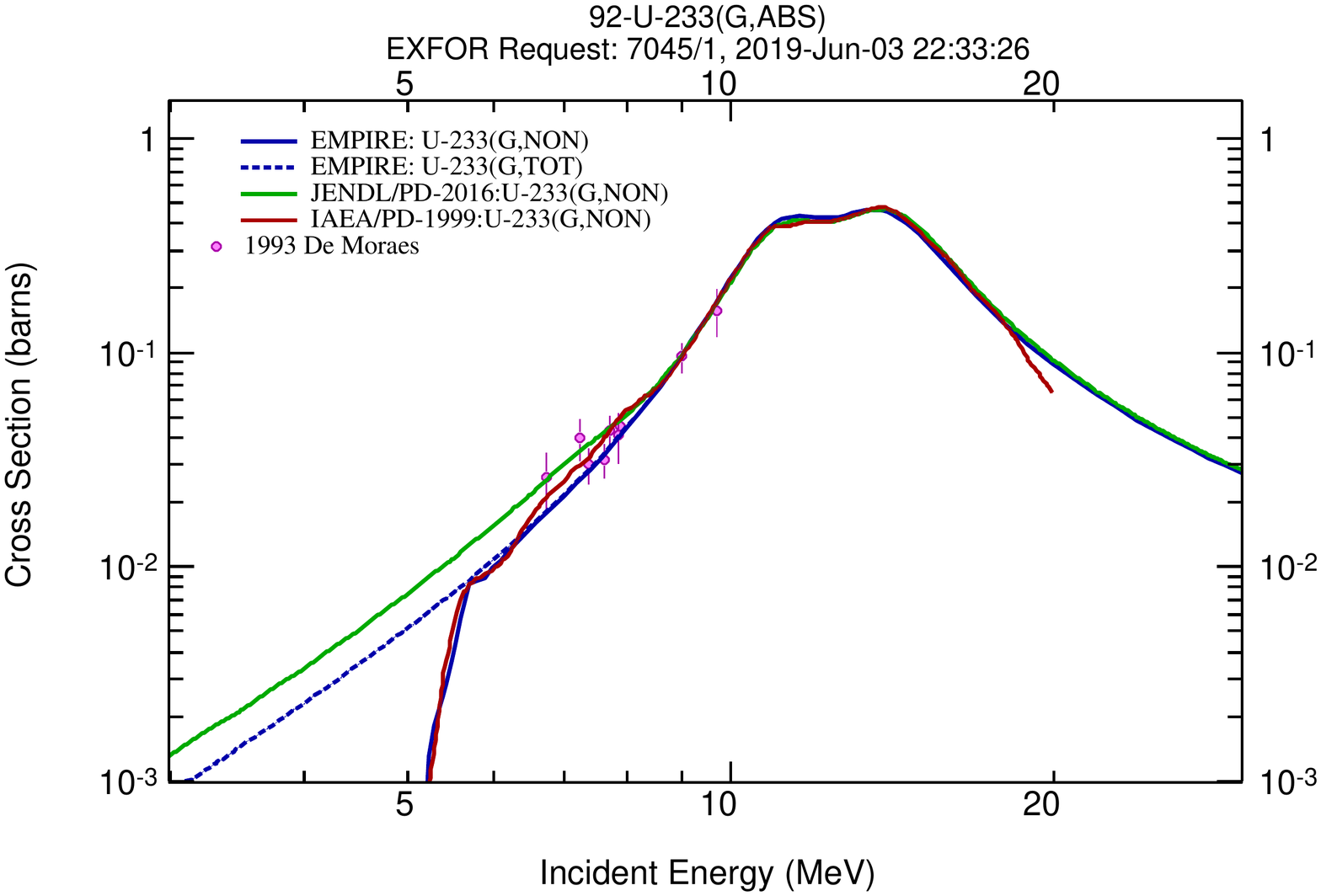}}
\put (-60, 140){\makebox{{\boldmath{$^{233}\rm{U(}\gamma,\rm{abs)}$}}}}\hfill
\subfigure{\includegraphics[trim={1.4cm 2.2cm 3.0cm 4.54cm},clip,width=0.5\textwidth]{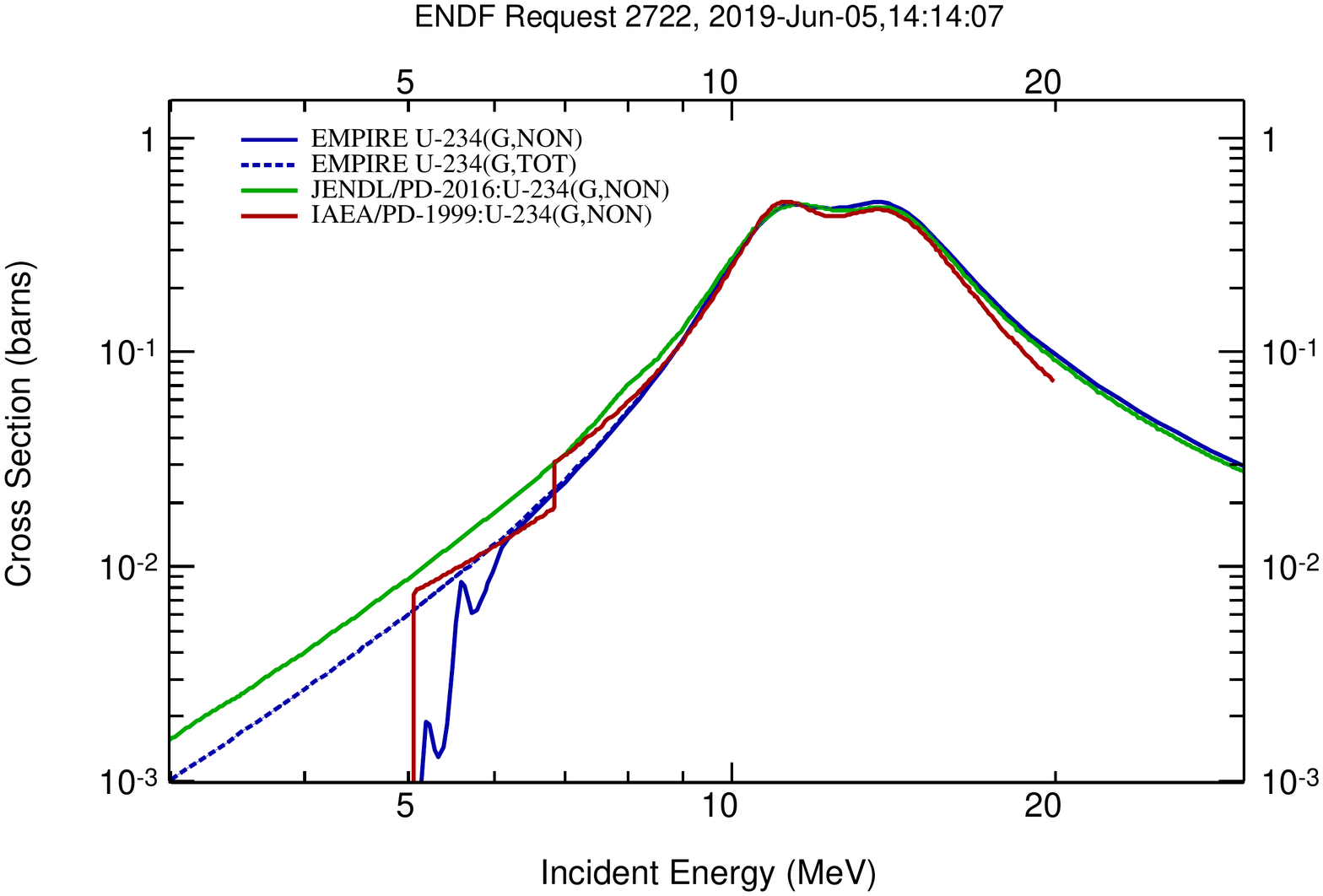}}
\put (-60, 140){\makebox{{\boldmath{$^{234}\rm{U(}\gamma,\rm{abs)}$}}}}\hfill
\vspace{-4mm}
\\
\subfigure{\includegraphics[trim={1.4cm 2.2cm 3.0cm 4.54cm},clip,width=0.5\textwidth]{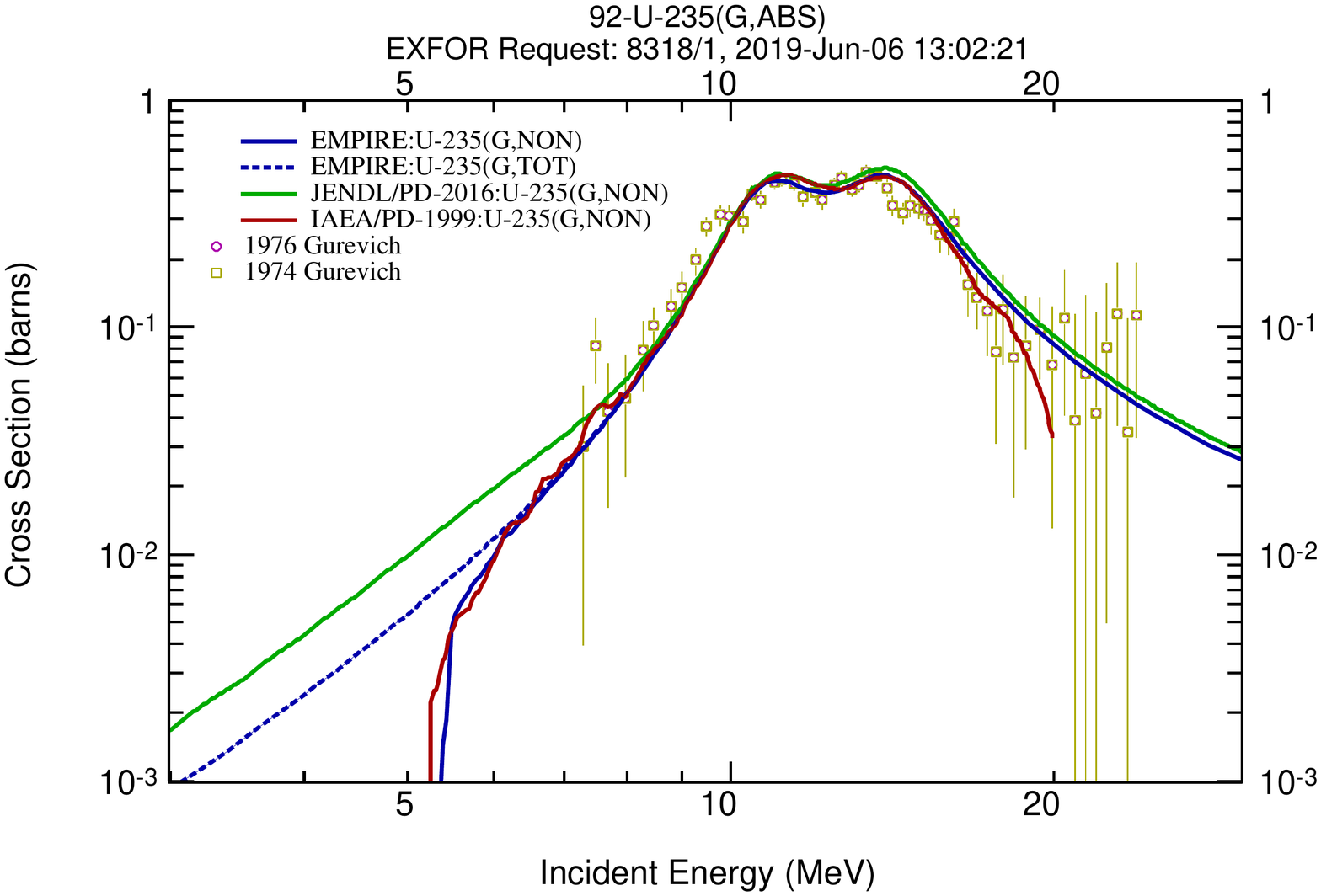}}
\put (-60, 140){\makebox{{\boldmath{$^{235}\rm{U(}\gamma,\rm{abs)}$}}}}\hfill
\subfigure{\includegraphics[trim={1.4cm 2.2cm 3.0cm 4.54cm},clip,width=0.5\textwidth]{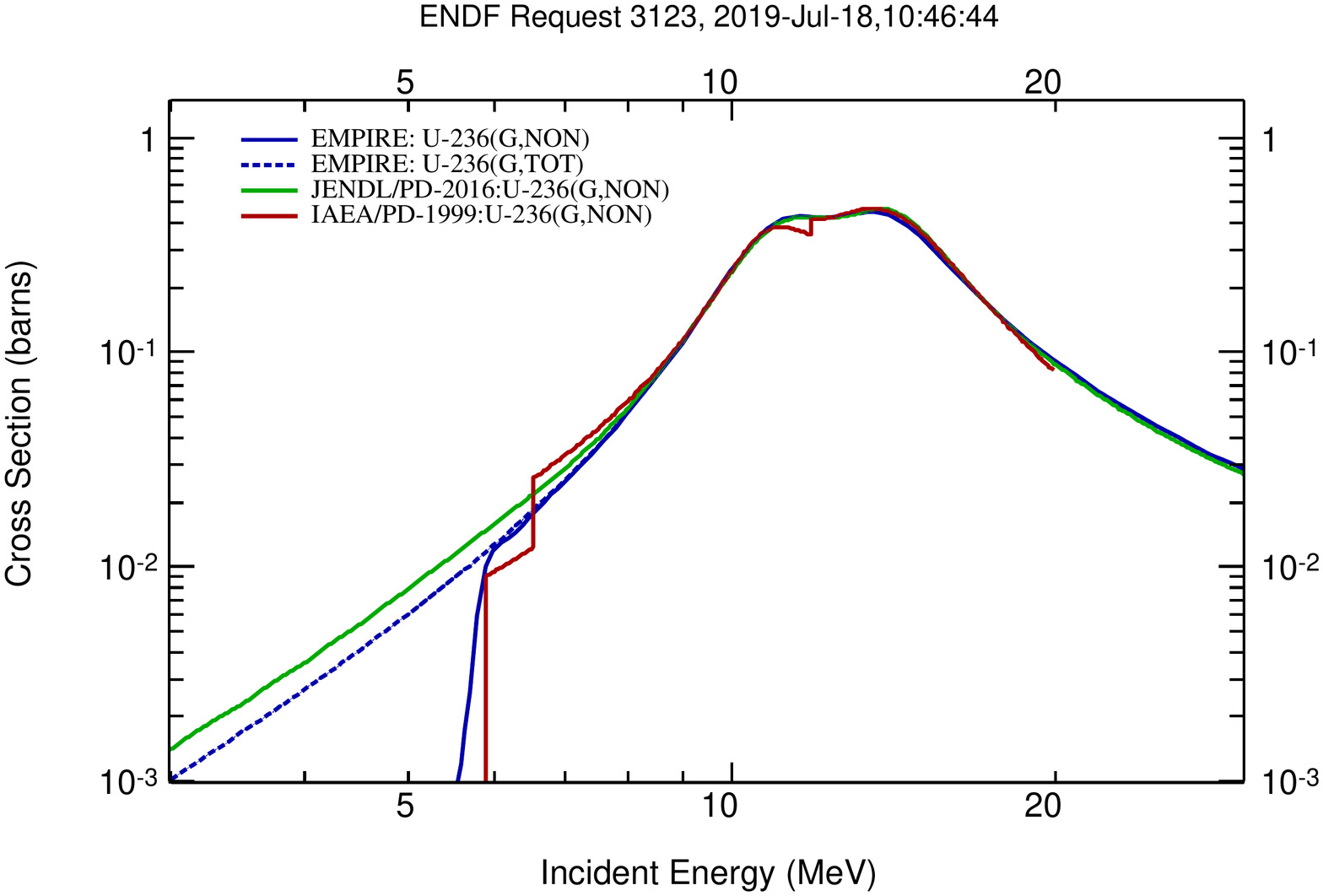}}
\put (-60, 140){\makebox{{\boldmath{$^{236}\rm{U(}\gamma,\rm{abs)}$}}}}\hfill
\vspace{-4mm}
\\
\subfigure{\includegraphics[trim={1.4cm 2.2cm 3.0cm 4.54cm},clip,width=0.5\textwidth]{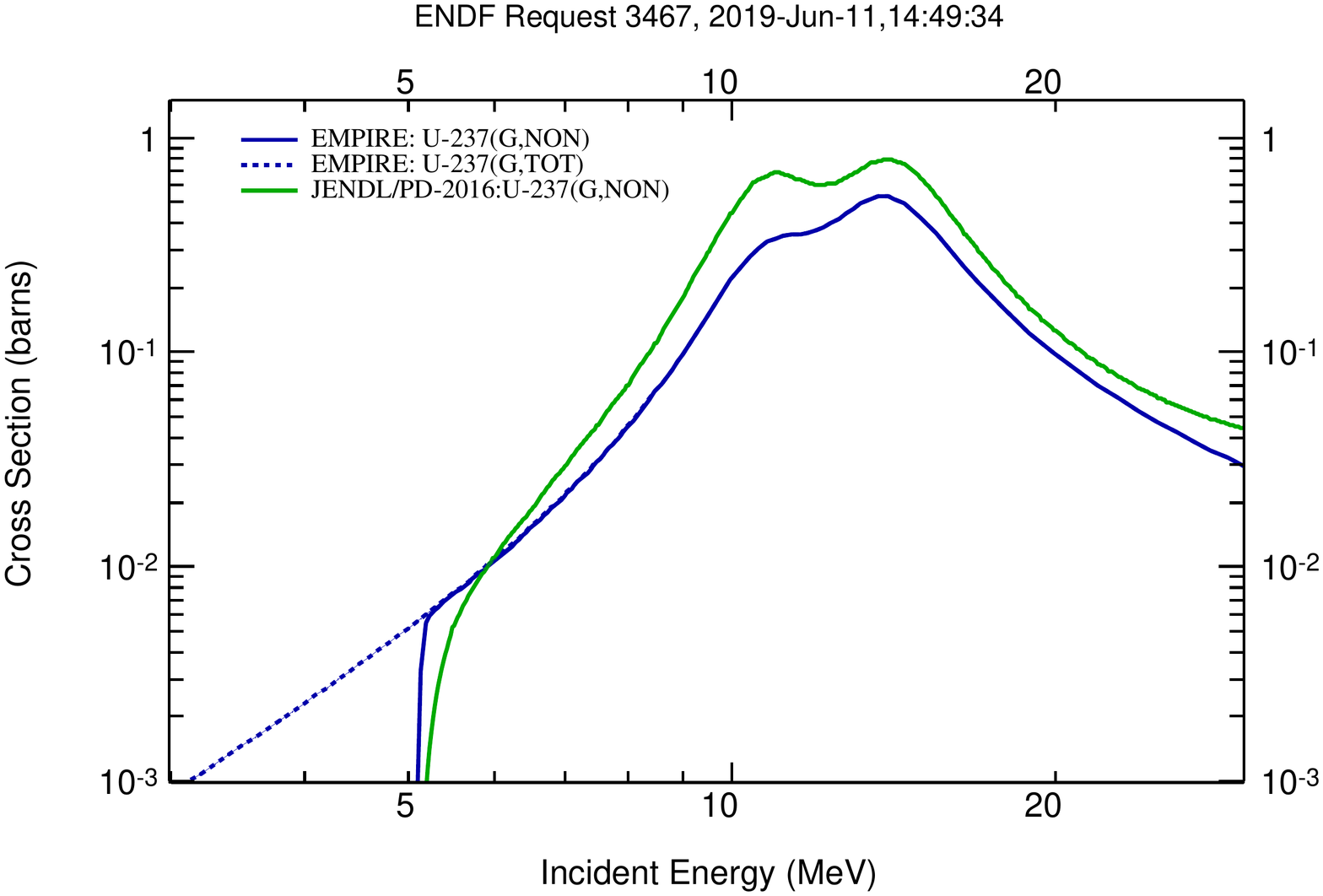}}
\put (-60, 140){\makebox{{\boldmath{$^{237}\rm{U(}\gamma,\rm{abs)}$}}}}\hfill
\subfigure{\includegraphics[trim={1.4cm 2.2cm 3.0cm 4.54cm},clip,width=0.5\textwidth]{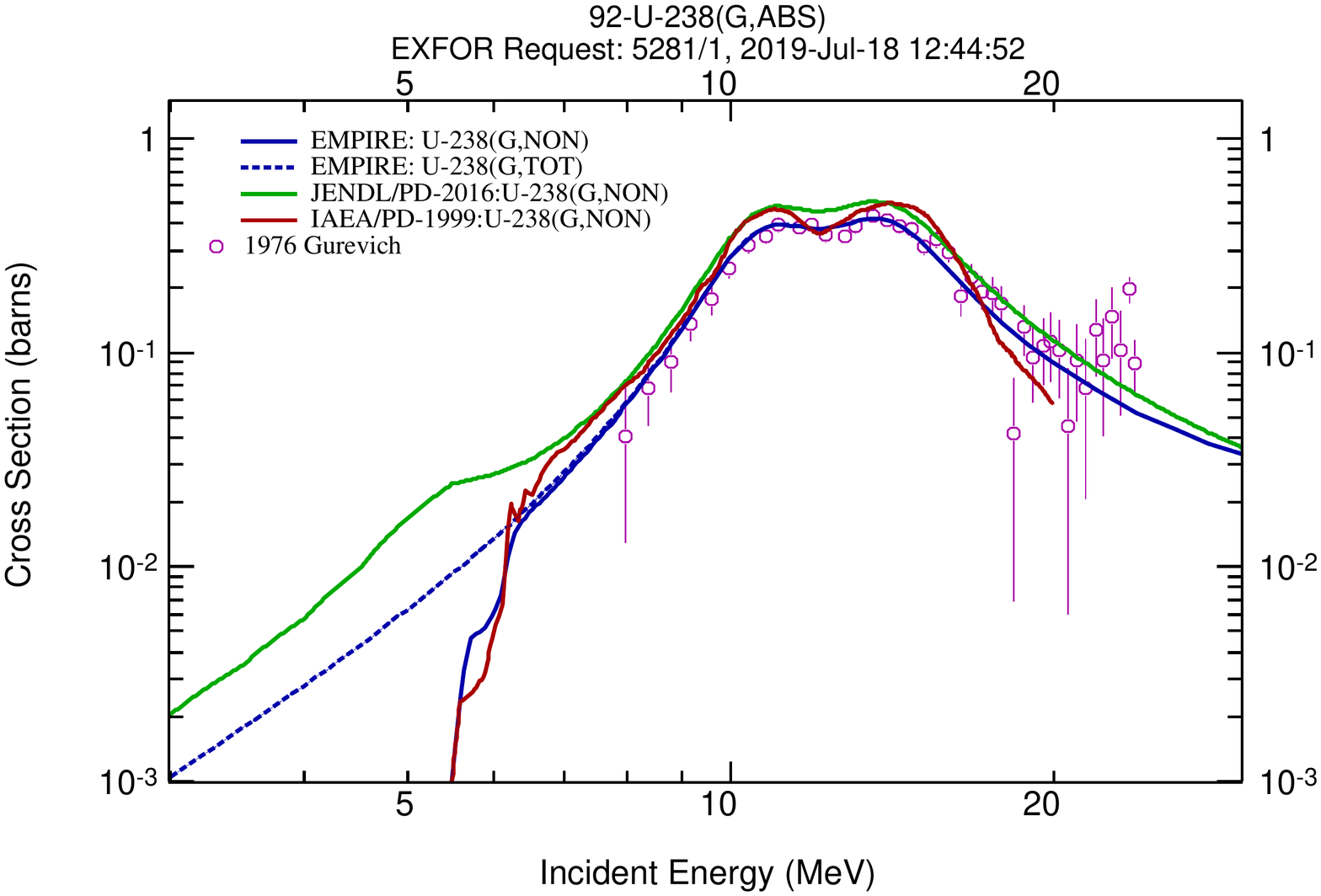}}
\put (-60, 140){\makebox{{\boldmath{$^{238}\rm{U(}\gamma,\rm{abs)}$}}}}\hfill
\vspace{-4mm}
\caption{JENDL-PD total \xs~(green line), IAEA-PD nonelastic \xs~(red line), calculated total (dashed blue line) and nonelastic (blue line) cross sections for $^{233-238}$U  compared to experimental data from EXFOR~\cite{Moraes:93,Gurevich:86}.}
\label{u_all-ABS}
\end{figure*}

\begin{figure*}[!t]
\subfigure{\includegraphics[trim={1.4cm 2.2cm 3.0cm 4.54cm},clip,width=0.5\textwidth]{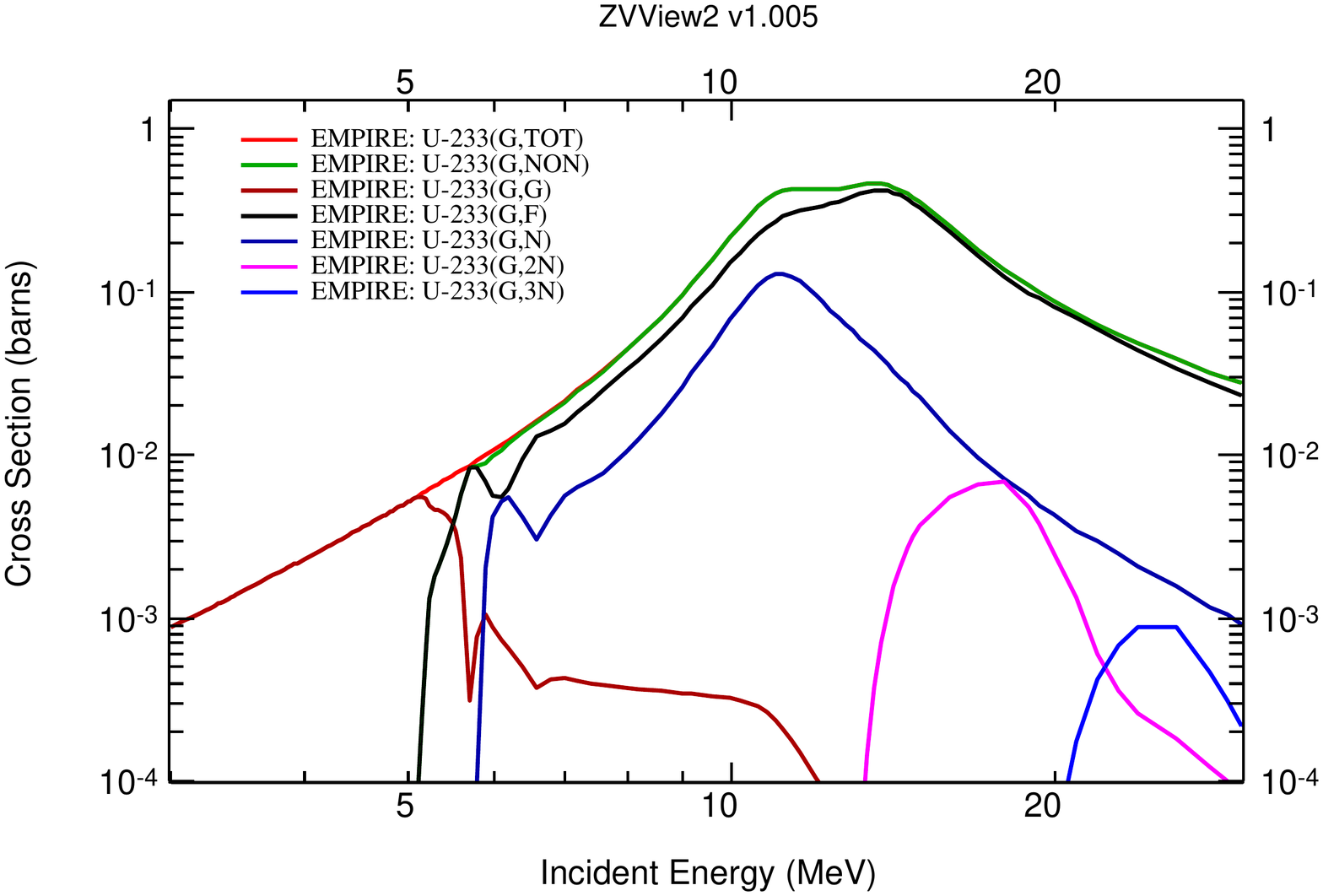}}
\put (-60, 140){\makebox{{\boldmath{$\gamma+^{233}\rm{U}$}}}}\hfill
\subfigure{\includegraphics[trim={1.4cm 2.2cm 3.0cm 4.54cm},clip,width=0.5\textwidth]{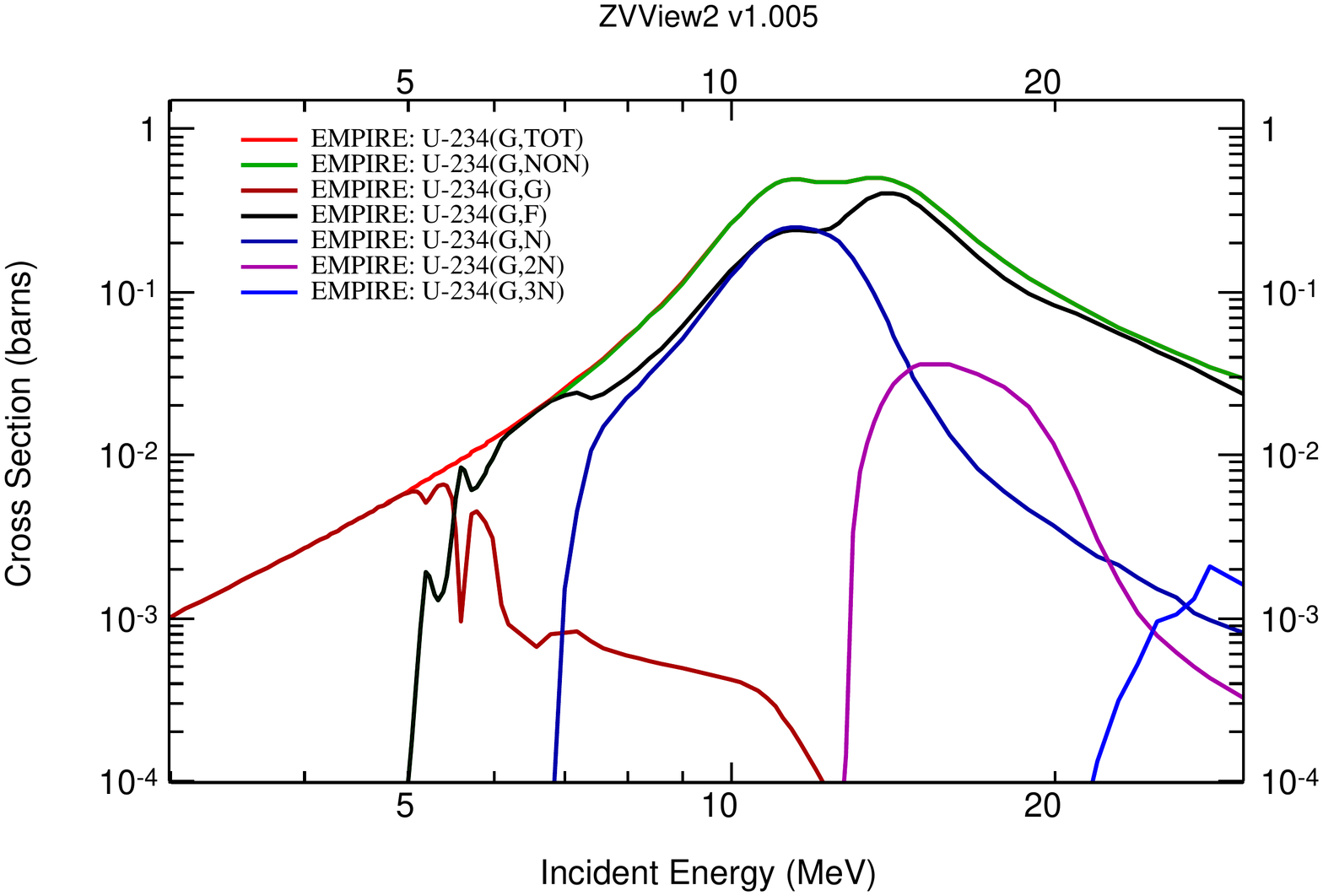}}
\put (-60, 140){\makebox{{\boldmath{$\gamma+^{234}\rm{U}$}}}}\hfill
\vspace{-4mm}
\\
\subfigure{\includegraphics[trim={1.4cm 2.2cm 3.0cm 4.54cm},clip,width=0.5\textwidth]{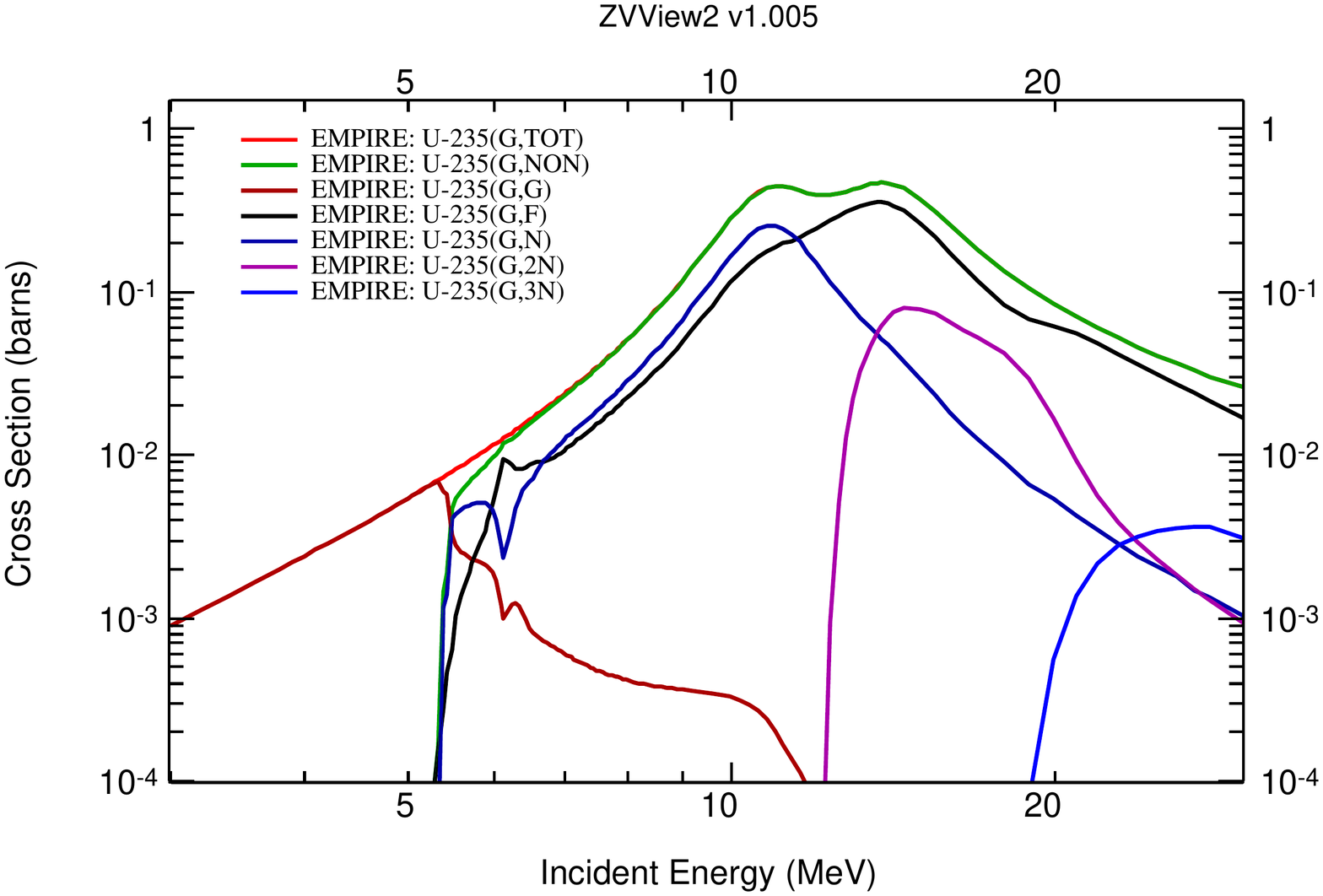}}
\put (-60, 140){\makebox{{\boldmath{$\gamma+^{235}\rm{U}$}}}}\hfill
\subfigure{\includegraphics[trim={1.4cm 2.2cm 3.0cm 4.54cm},clip,width=0.5\textwidth]{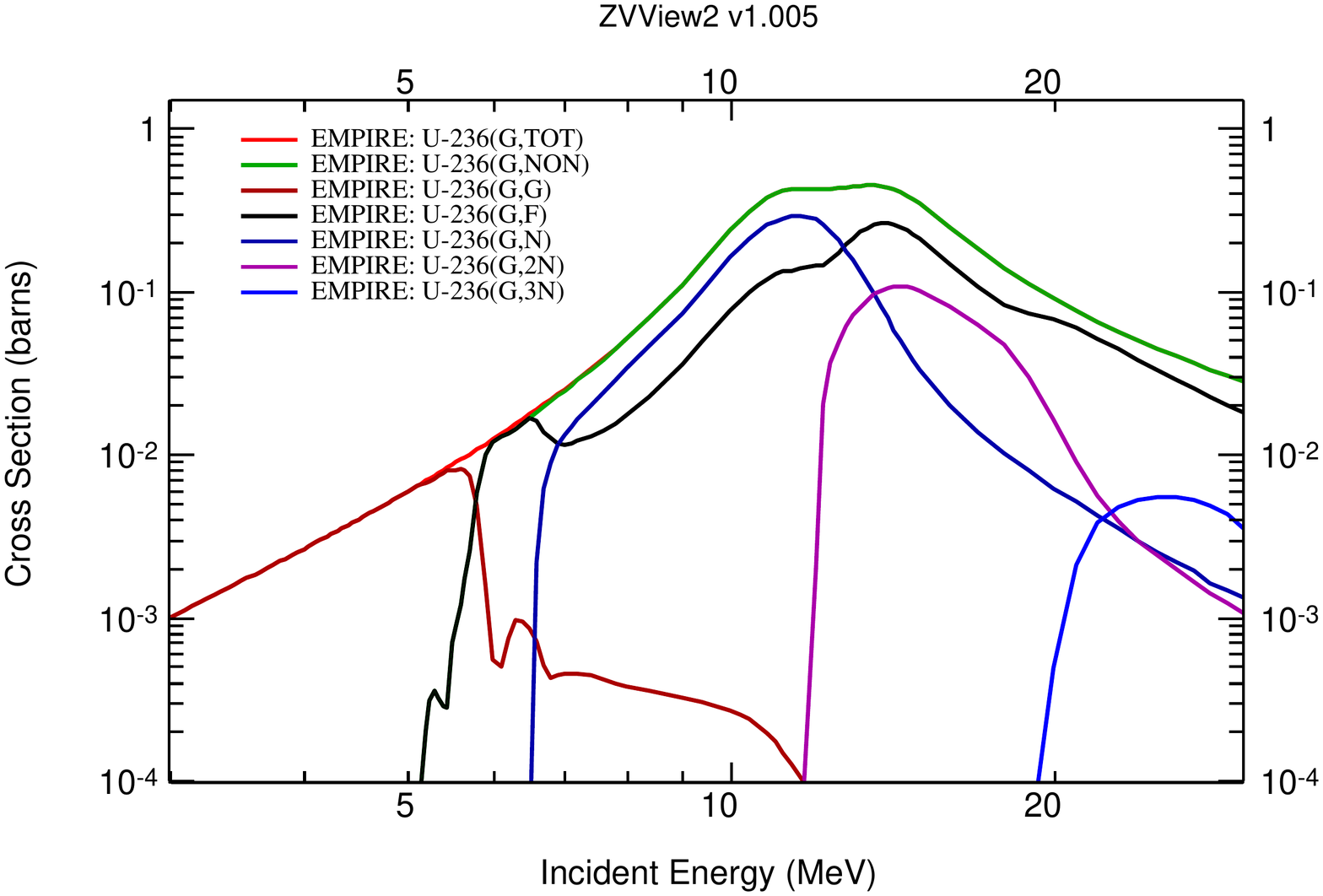}}
\put (-60, 140){\makebox{{\boldmath{$\gamma+^{236}\rm{U}$}}}}\hfill
\vspace{-4mm}
\\
\subfigure{\includegraphics[trim={1.4cm 2.2cm 3.0cm 4.54cm},clip,width=0.5\textwidth]{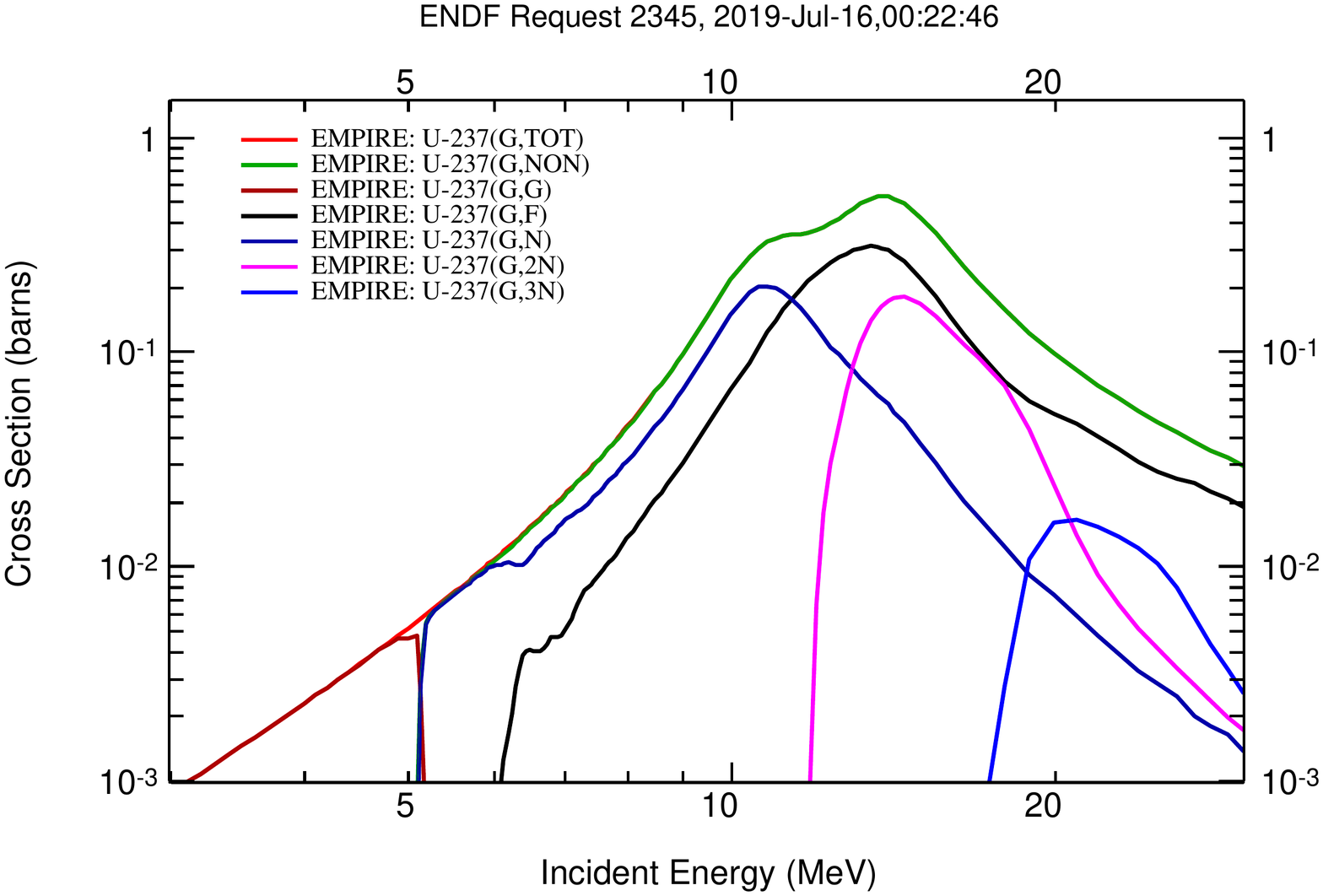}}
\put (-60, 140){\makebox{{\boldmath{$\gamma+^{237}\rm{U}$}}}}\hfill
\subfigure{\includegraphics[trim={1.4cm 2.2cm 3.0cm 4.54cm},clip,width=0.5\textwidth]{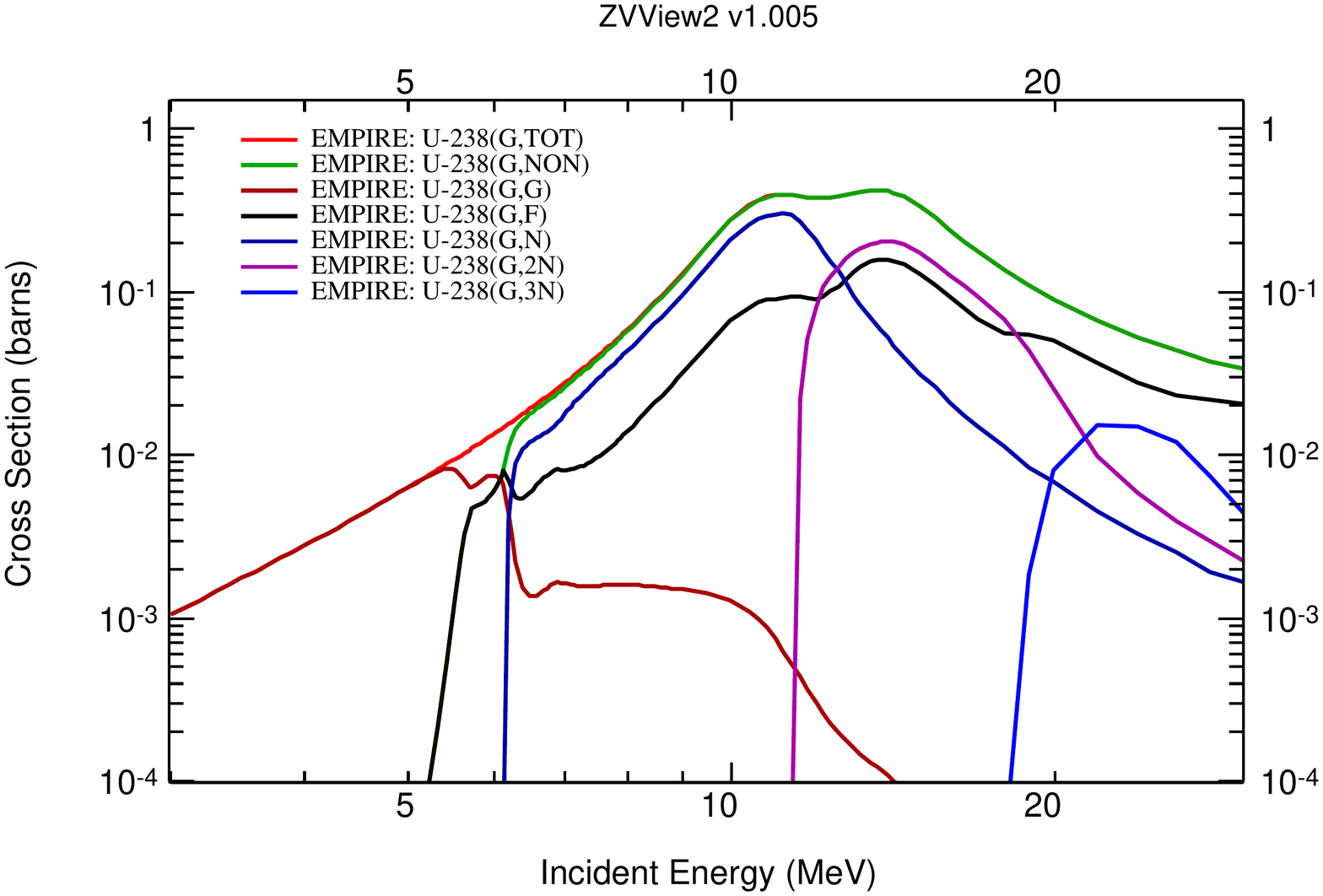}}
\put (-60, 140){\makebox{{\boldmath{$\gamma+^{238}\rm{U}$}}}}\hfill
\vspace{-4mm}
\caption{Photo-reaction cross sections for $^{233-238}$U calculated with EMPIRE code: total (light-red), photo-absorption (green), fission (black), ($\gamma,\gamma$) (red), ($\gamma$,n) (blue), ($\gamma$,2n) (magenta), ($\gamma$,3n) (light blue).}
\label{u_all-ALL}
\end{figure*}

The photo-absorption cross section is defined as the difference between the total gamma cross section Eq.~\eqref{sigma_tot} and the elastic scattered gamma $\sigma_{\gamma\gamma}$ cross section.
\begin{equation}
\sigma_{abs}(E_{\gamma})=\sigma_{\gamma t}(E_{\gamma})-\sigma_{\gamma\gamma}(E_{\gamma}).
\label{sigma_abs}
\end{equation}
Generally, the reaction evaluated data libraries include the photo-absorption as nonelastic cross section and do not include the total gamma cross section \cite{ENDF-6}.
In IAEA-PD this formatting rule is applied, but in JENDL-PD it is assumed that the nonelastic is equal to the total gamma cross section (except for \useven). Therefore, in Fig.~\ref{u_all-ABS} both total gamma and photo-absorption calculated cross sections are presented for comparison. From the Fig.~\ref{u_all-ALL} one can notice that the total gamma cross section (light-red line) is visible only in a small energy range around the fission threshold, being overwritten at lower energies by the gamma emission cross section and by the nonelastic cross section at higher energies.

For photon-induced reactions, the total gamma and photo-absorption cross section calculation requires GDR parameters, see Eqs.\eqref{sigma_tot}--\eqref{MLO}, the same way the optical potentials are needed for the total cross section calculation in the particle-induced reactions.
For particle-induced reactions, in particular for neutrons, the global or regional optical potentials are usually quite reliable, providing accurate total cross sections, as well as a proper partition between direct elastic and nonelastic cross sections.
For photon-induced reactions on actinides there are not enough experimental information for a reliable parameterization of the GDR parameters \cite{Plujko:2011,Plujko:2018}, and the microscopic strength functions are useful to set trends but the normalization is typically more uncertain. Therefore, if in the neutron induced reactions the nonelastic cross section provided by optical model calculations has to be distributed in the outgoing channels, for those photon induced reactions where no experimental photoabsorption data are available, the order is somehow reversed: the GDR parameters are adjusted to produce a nonelastic cross section equal to the sum of the cross sections for the open channels which fit the corresponding experimental data.
The GDR parameters used in this paper and presented in  Table \ref{t-gdr} have been obtained by adjusting the RIPL-3 ``experimental'' parameters (derived as explained above) to improve the description of the experimental data.

\begin{table}[h]
\begin{center}
\caption{GDR parameters for \uall used in the present work.}
\begin{tabular}{c|c|c|c|c|c|c}
\hline\hline
\T
        & $E_{\gamma_1}$ & $\Gamma_{\gamma_1}$ & $\sigma_{\gamma_1}$ & $E_{\gamma_2}$ & $\Gamma_{\gamma_2}$ & $\sigma_{\gamma_2}$\\
CN      & (MeV)   & (MeV) & (mb) & (MeV)   & (MeV) & (mb)\\
\hline
\T
\uthree & 11.3 & 3.0 & 320.0 & 14.0 & 4.3 & 360\\
\ufour  & 11.2 & 3.2 & 370.0 & 14.1 & 4.3 & 380\\
\ufive  & 11.0 & 2.8 & 370.0 & 14.1 & 4.0 & 380\\
\usix   & 11.3 & 3.4 & 320.0 & 14.0 & 4.5 & 320\\
\useven & 11.2 & 3.2 & 360.0 & 14.0 & 4.3 & 380\\
\ueight & 11.0 & 3.3 & 316.0 & 14.1 & 4.4 & 320\\
\hline\hline
\end{tabular}
\end{center}
\label{t-gdr}
\end{table}

Fig.~\ref{u_all-ABS} shows that our calculations agree well with the experimental data, and that there are not significant differences between our calculations and the two evaluations around the GDR energies, except for \useven. The behavior at low energies (underestimation of the JENDL-PD total gamma cross section and the agreement with the IAEA-PD nonelastic cross section) is related mainly to the description of the fission cross section, as can be seen in  Fig.~\ref{u_all-FISS}. The lower values of the IAEA-PD nonelastic cross section at higher energies are reflected in the corresponding values of the fission and ($\gamma,2n$) cross sections.

\useven~is different from the other isotopes for several reasons: (i) no experimental information is available, (ii) there is no IAEA-PD evaluation,  (iii) JENDL-PD evaluation is based on other codes than the CCONE used for the rest of the isotopes, and  provides the photo-absorption cross section as nonelastic.
Fig.~\ref{u_all-ABS} shows  a big difference (\~40\%) between EMPIRE
and JENDL-PD absorption cross sections, confirming that without experimental constraint, the model predictions can be very discrepant.

Small differences between JENDL-PD and IAEA-PD evaluations on one side,  EMPIRE calculation and
Gurevich experimental data \cite{Gurevich:86} on the other hand, appear for \ueight~also.
The main reason is that the evaluations and our calculations describe different sets of experimental data for the ($\gamma,f$), ($\gamma,n$), and ($\gamma,2n$) processes, as discussed in the next Sections.

\subsection{Neutron emission cross sections}
\begin{figure*}[!tbh]
\subfigure{\includegraphics[trim={1.4cm 2.2cm 3.0cm 4.54cm},clip,width=0.5\textwidth]{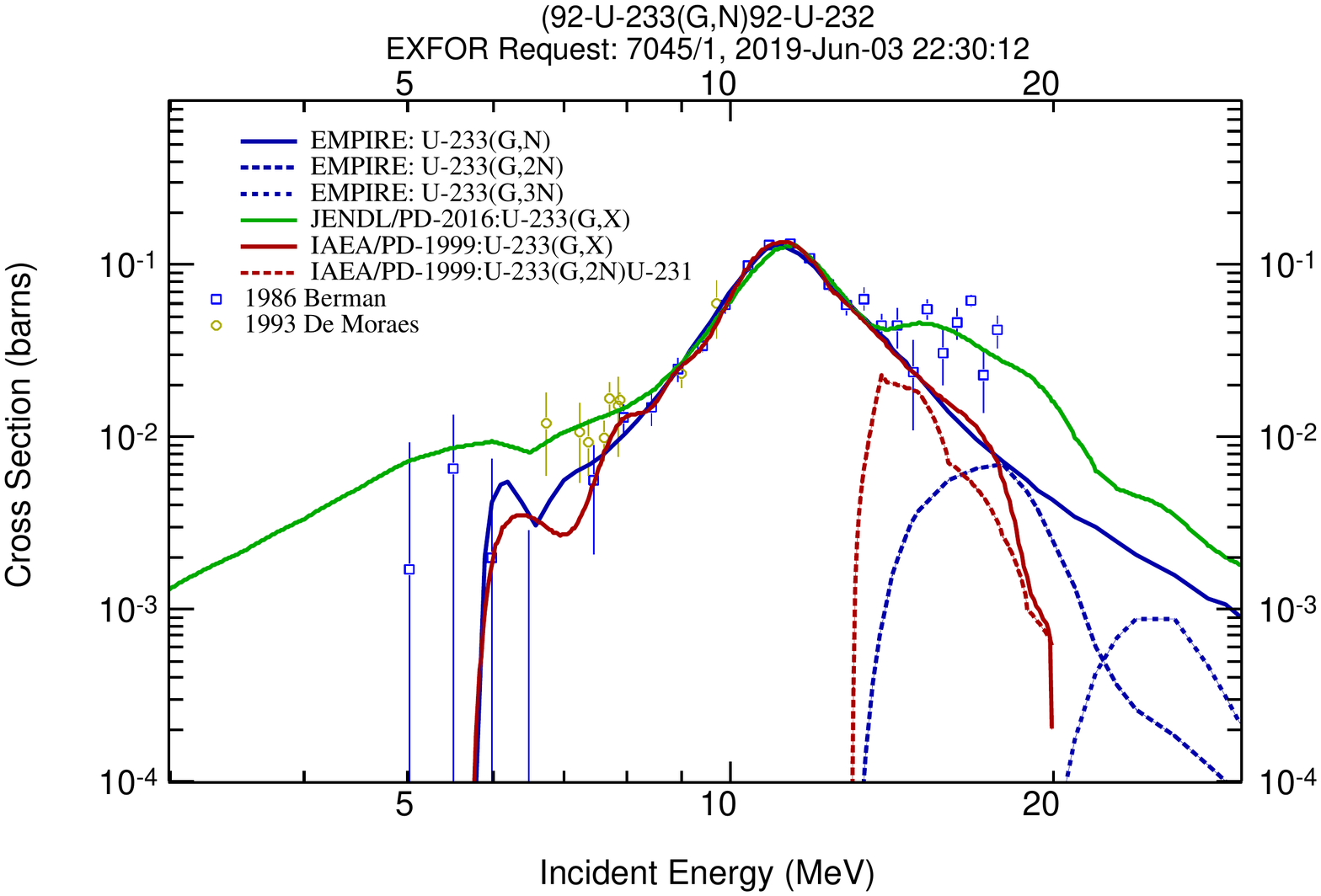}}
\put (-60, 140){\makebox{{\boldmath{$^{233}\rm{U(}\gamma,\rm{xn)}$}}}}\hfill
\subfigure{\includegraphics[trim={1.4cm 2.2cm 3.0cm 4.54cm},clip,width=0.5\textwidth]{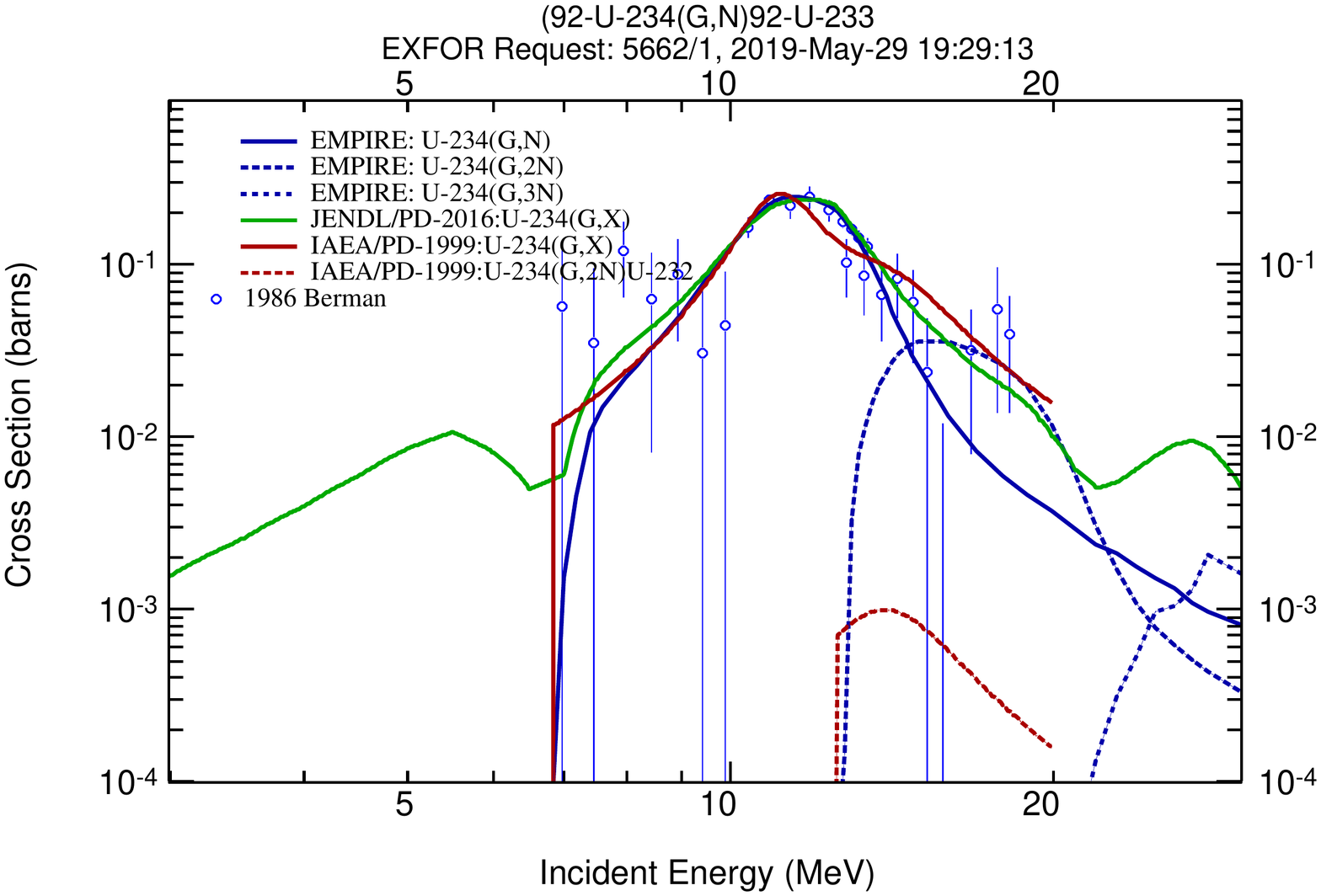}}
\put (-60, 140){\makebox{{\boldmath{$^{234}\rm{U(}\gamma,\rm{xn)}$}}}}\hfill
\vspace{-4mm}
\\
\subfigure{\includegraphics[trim={1.4cm 2.2cm 3.0cm 4.54cm},clip,width=0.5\textwidth]{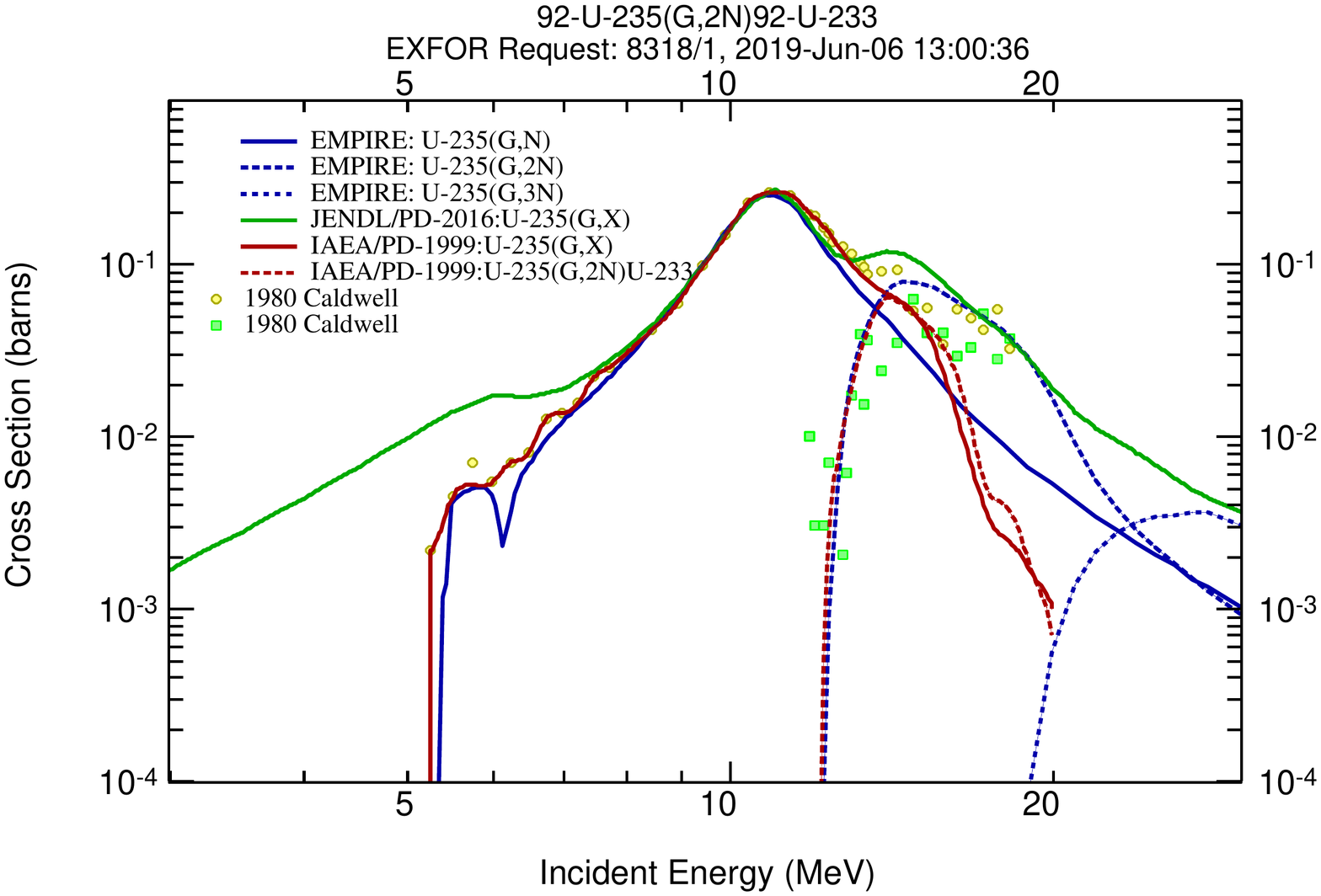}}
\put (-60, 140){\makebox{{\boldmath{$^{235}\rm{U(}\gamma,\rm{xn)}$}}}}\hfill
\subfigure{\includegraphics[trim={1.4cm 2.2cm 3.0cm 4.54cm},clip,width=0.5\textwidth]{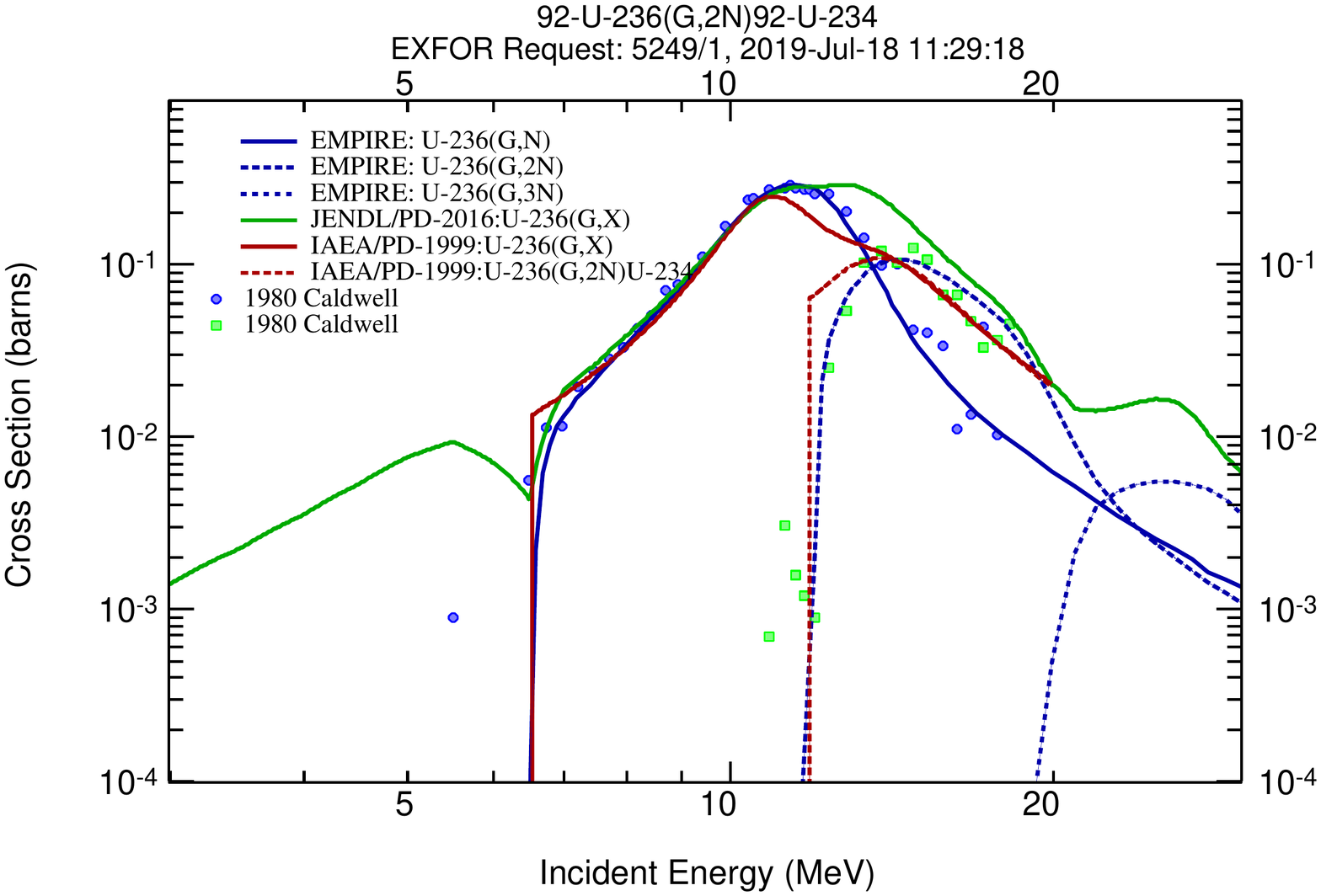}}
\put (-60, 140){\makebox{{\boldmath{$^{236}\rm{U(}\gamma,\rm{xn)}$}}}}\hfill
\vspace{-4mm}
\\
\subfigure{\includegraphics[trim={1.4cm 2.2cm 3.0cm 4.54cm},clip,width=0.5\textwidth]{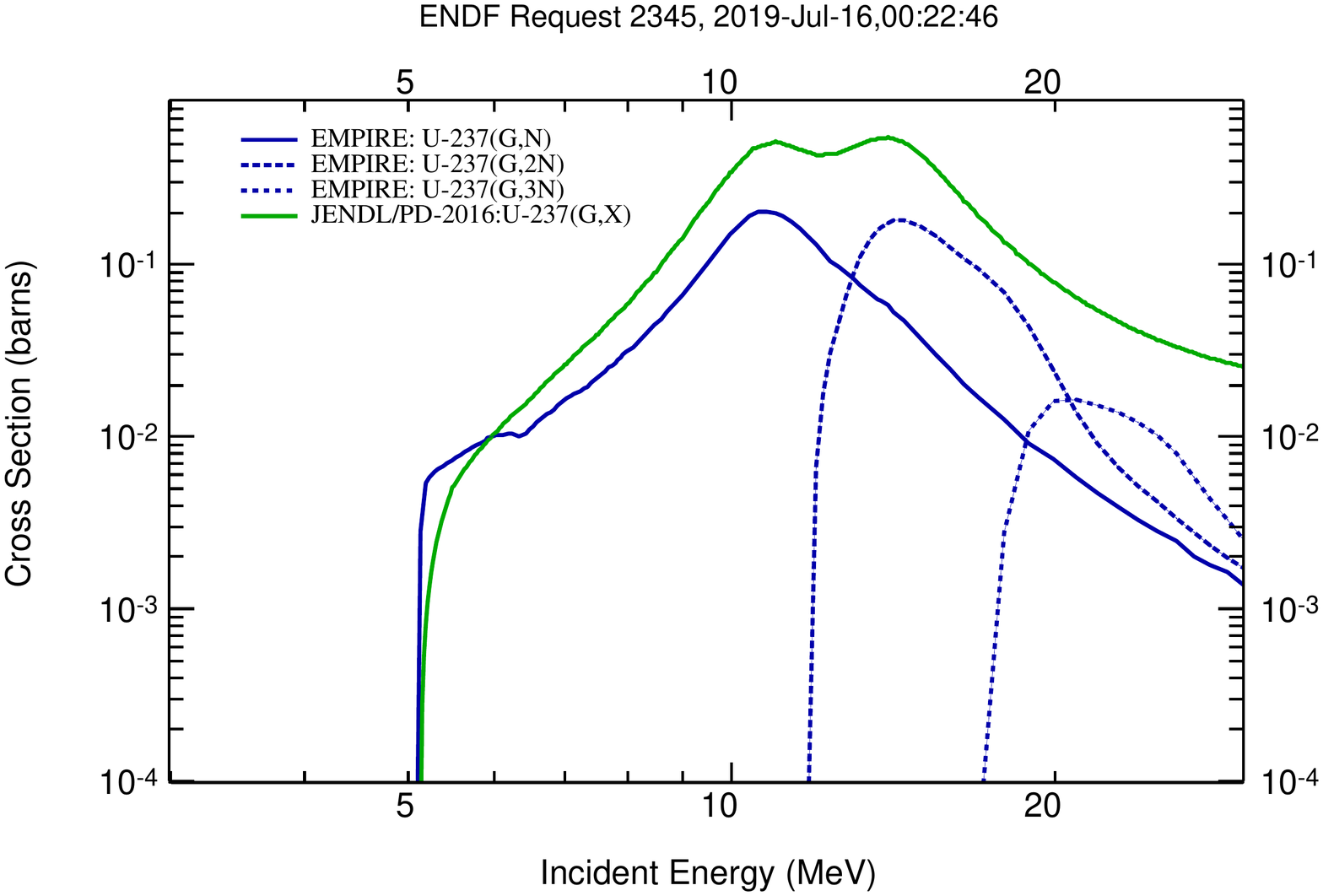}}
\put (-60, 140){\makebox{{\boldmath{$^{237}\rm{U(}\gamma,\rm{xn)}$}}}}\hfill
\subfigure{\includegraphics[trim={1.4cm 2.2cm 3.0cm 4.54cm},clip,width=0.5\textwidth]{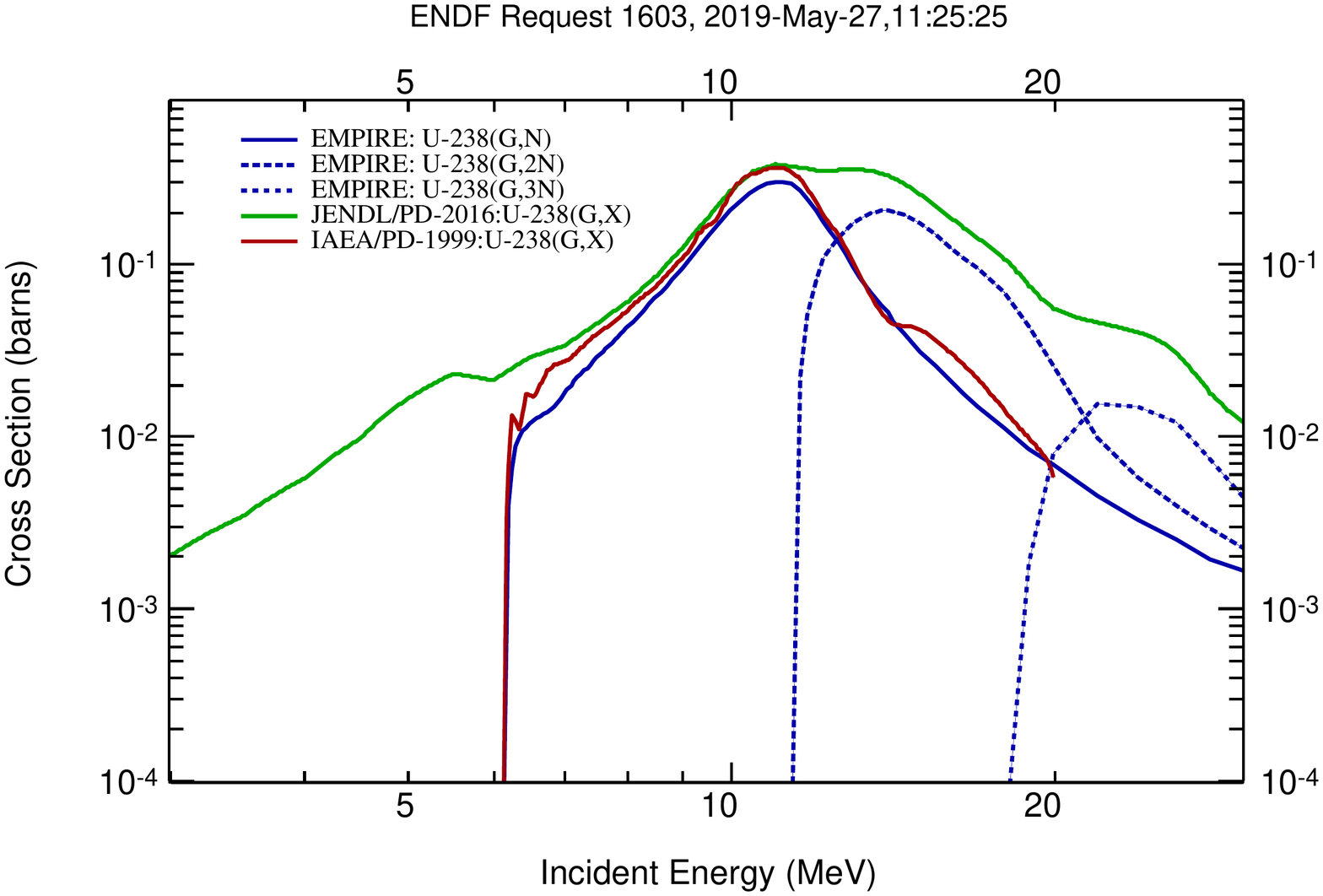}}
\put (-60, 140){\makebox{{\boldmath{$^{238}\rm{U(}\gamma,\rm{xn)}$}}}}\hfill
\vspace{-4mm}
\caption{Calculated neutron emission cross sections (($\gamma$,n) blue solid line, ($\gamma$,2n) blue dashed line, ($\gamma$,3n) blue dotted line)  for $^{233-238}$U compared to evaluated $(G,X)$ cross sections (JENDL-PD green line, IAEA-PD red line) and the  experimental data from EXFOR~\cite{Moraes:93, Berman:86, Caldwell:80}. }
\label{u_all-NNN}
\end{figure*}

\begin{figure*}[!tbh]
\subfigure{\includegraphics[trim={1.4cm 2.2cm 3.0cm 4.54cm},clip,width=0.5\textwidth]{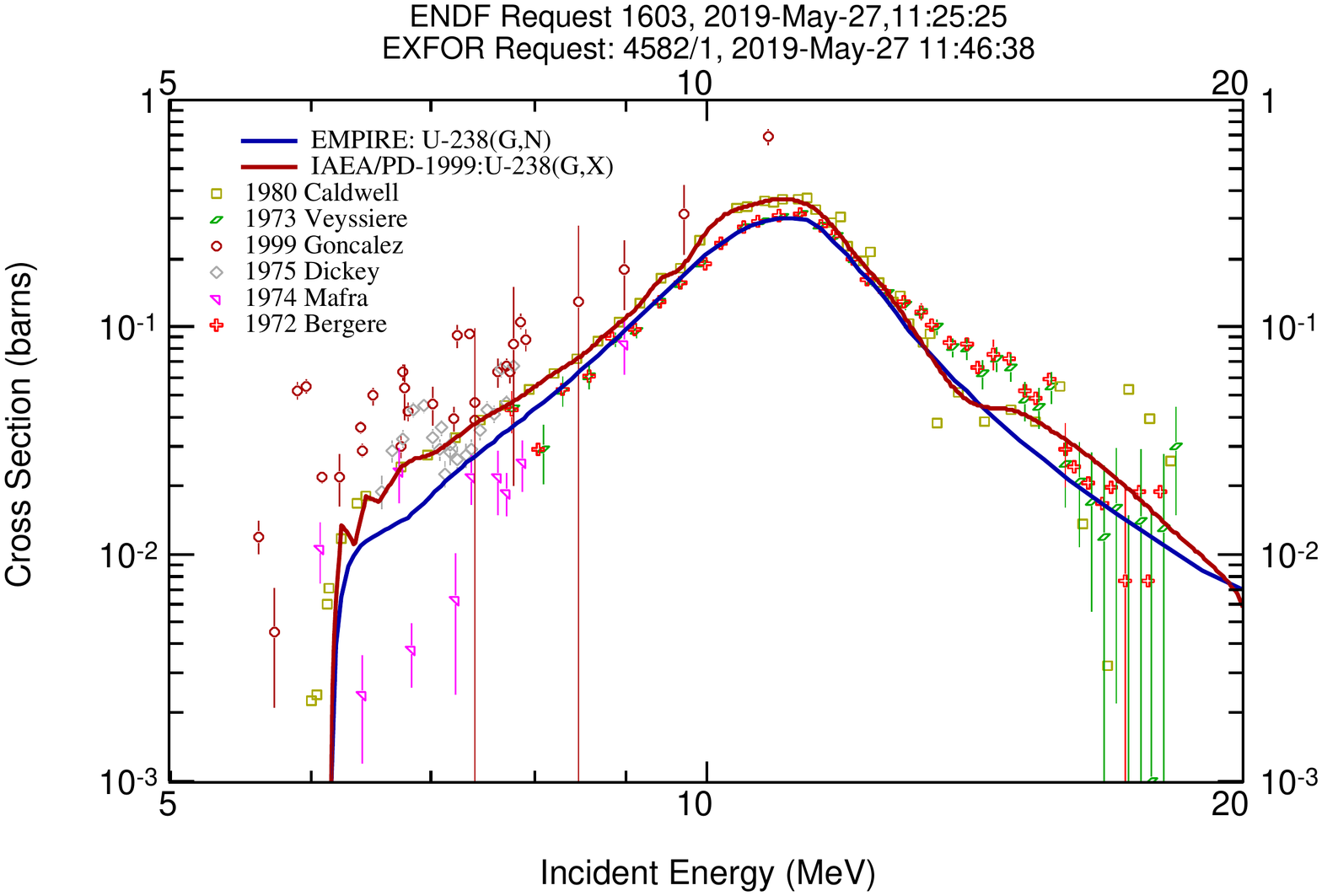}}
\put (-60, 140){\makebox{{\boldmath{$^{238}\rm{U(}\gamma,\rm{n)}$}}}}\hfill
\subfigure{\includegraphics[trim={1.4cm 2.2cm 3.1cm 4.54cm},clip,width=0.5\textwidth]{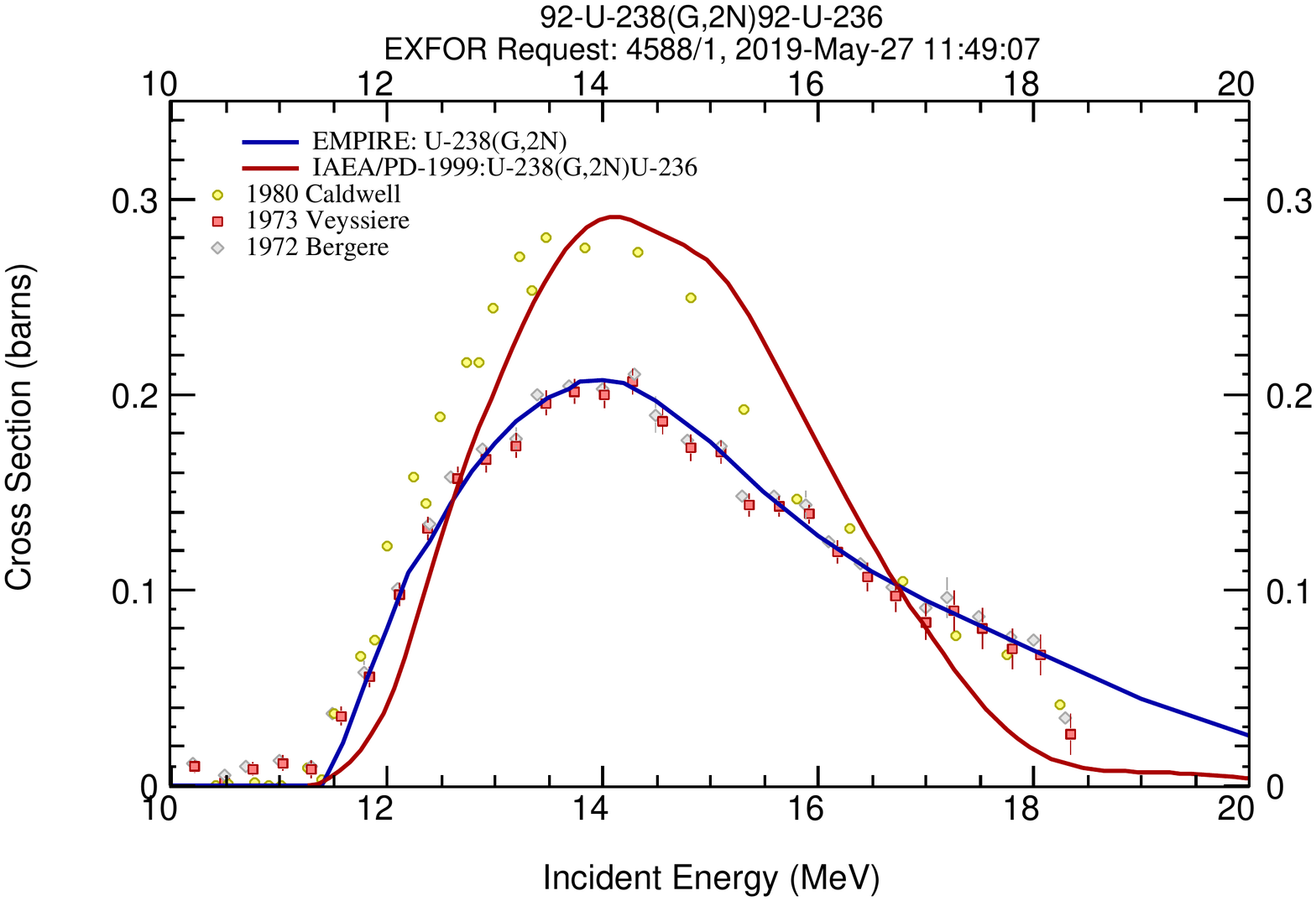}}
\put (-60, 140){\makebox{{\boldmath{$^{238}\rm{U(}\gamma,\rm{2n)}$}}}}\hfill
\vspace{-4mm}
\caption{IAEA-PD evaluation (red line) and EMPIRE calculation (blue line) for $^{238}$U($\gamma$,n) (left panel) and $^{238}$U($\gamma$,2n) (right panel) compared to experimental data from EXFOR~\cite{Caldwell:80, Veyssiere:73, Bergere:72, Goncalez:99, Dickey:75,Mafra:72}.}
\label{u238-N2N}
\end{figure*}

The decay probabilities are constrained by the consistency of the preequilibrium  and  compound nucleus models and by the input parameters (optical potentials, discrete level schemes, level densities, fission parameters). The absolute values of the cross sections can be scaled by adjusting the GDR parameters. As pointed out in Section \ref{models}, the decay of the photo-excited nuclei is treated with the same models and parameters used in \csin~for the decay of the compound systems formed in neutron induced reactions.

The EMPIRE neutron emission cross sections are compared in
Figs.~\ref{u_all-NNN},~\ref{u238-N2N} with JENDL-PD and IAEA-PD evaluations and with the experimental data from EXFOR.
The significance of the evaluated curves and of the experimental data from these figures is clarified in the next paragraphs.

Most of the EXFOR data are presented as the sum ($\gamma,n$)+($\gamma,np$) cross sections, but according to  calculations, the second contribution can be neglected. However, the high values above the ($\gamma,2n$) threshold would suggest that the experimental data include multiple neutron emission also. Considering these aspects, the EMPIRE neutron emission \xss~of $^{233-236}$U isotopes are in good agreement with the experimental data. An exception is the \ufive($\gamma,2n$) which overestimates
the experimental data of Caldwell \cite{Caldwell:80}.

JENDL-PD contains evaluations for the nonelastic (in reality total gamma) and fission cross sections, and a lumped \xs~for the other channels: ($\gamma,\gamma$), ($\gamma,n$), ($\gamma,2n$), and ($\gamma,3n$). This lumped cross section
symbolized as $(G,X)$
is represented in Figs.~\ref{u_all-NNN} and \ref{u238-N2N}.
One can identify in these cross sections the high tail at low energies as the ($\gamma,\gamma$) contribution  and the bumps at higher energies as produced by the ($\gamma,2n$), and ($\gamma,3n$) channels. For $^{233-236}$U isotopes the EMPIRE neutron emission \xs~is in good agreement  with JENDL-PD evaluation around the maximum.

IAEA-PD includes an explicit evaluation for the ($\gamma,2n$) \xs, but ignores the ($\gamma,\gamma$) and ($\gamma,3n$) contributions, so that the $(G,X)$ evaluation represents in fact the ($\gamma,n$) \xs.
The calculated ($\gamma,n$) \xs~ is in good agreement with IAEA-PD evaluation for \uthree, while for $^{234-236}$U isotopes, the agreement stops around 10 MeV incident energy. The differences between EMPIRE calculation and IAEA-PD evaluation above 10 MeV for the ($\gamma,n$) \xss~are reflected also  in the ($\gamma,2n$) \xs~where they become even larger, especially for \ufour.

In case of \useven, the significant discrepancy between the calculated photo-absorption \xs~and JENDL-PD evaluation is also reflected in the neutron-emission \xss~as depicted in Fig.~\ref{u_all-NNN}.

The same is true for \ueight~of which ($\gamma,n$) and ($\gamma,2n$) \xss~are plotted also separately in Fig.~\ref{u238-N2N} with the corresponding experimental data.
Using the parameters from \csin~and the GDR parameters from Table~\ref{t-gdr}, a simultaneous accurate  description of Veyssiere \cite{Veyssiere:73} and Bergere \cite{Bergere:72} data for the neutron emission (Fig.~\ref{u238-N2N}) and fission (Fig.~\ref{u_all-FISS}) \xss~has been obtained, but also of the Gurevich photo-absorption data \cite{Gurevich:86} (Fig.~\ref{u_all-ABS}). Impressive is the perfect fit of the ($\gamma,2n$) cross section. On the other hand, both evaluations follow the Caldwell data \cite{Caldwell:80} overestimating the experimental photo-absorption \xs.
Caldwell data \cite{Caldwell:80} are measured at Lawrence Livermore National Laboratory while Veyssiere \cite{Veyssiere:73} and Bergere \cite{Bergere:72} data are measured at CEA-Saclay Nuclear Research Centre. There is a controversy in literature regarding the experimental data measured by the Livermore and Saclay groups. There are not sufficient information to conclude that our model calculations support the data of one of these groups, but for \ueight~Saclay data seem to be more consistent.

\subsection{Fission cross sections}\label{fis-res}
The main purpose of the present calculations is to check the compatibility of the fission barriers deduced from the fit of the neutron induced fission \xss~of Uranium isotopes in \csin~with the input parameters specific to photo-reaction model calculations, and to test the accuracy of the predicted photo-fission \xss.
When judging how well the fission parameters from \csin~describe the
photon-induced fission cross sections one has to consider at least two specific features of the photo-excited compound nuclei: the access to lower excitation energies (due to the lack of photon separation energy), and the selectivity in spin and parity.

\begin{figure*}[!tbh]
\subfigure{\includegraphics[trim={2.0cm 2.6cm 3.0cm 4.54cm},clip,width=0.5\textwidth]{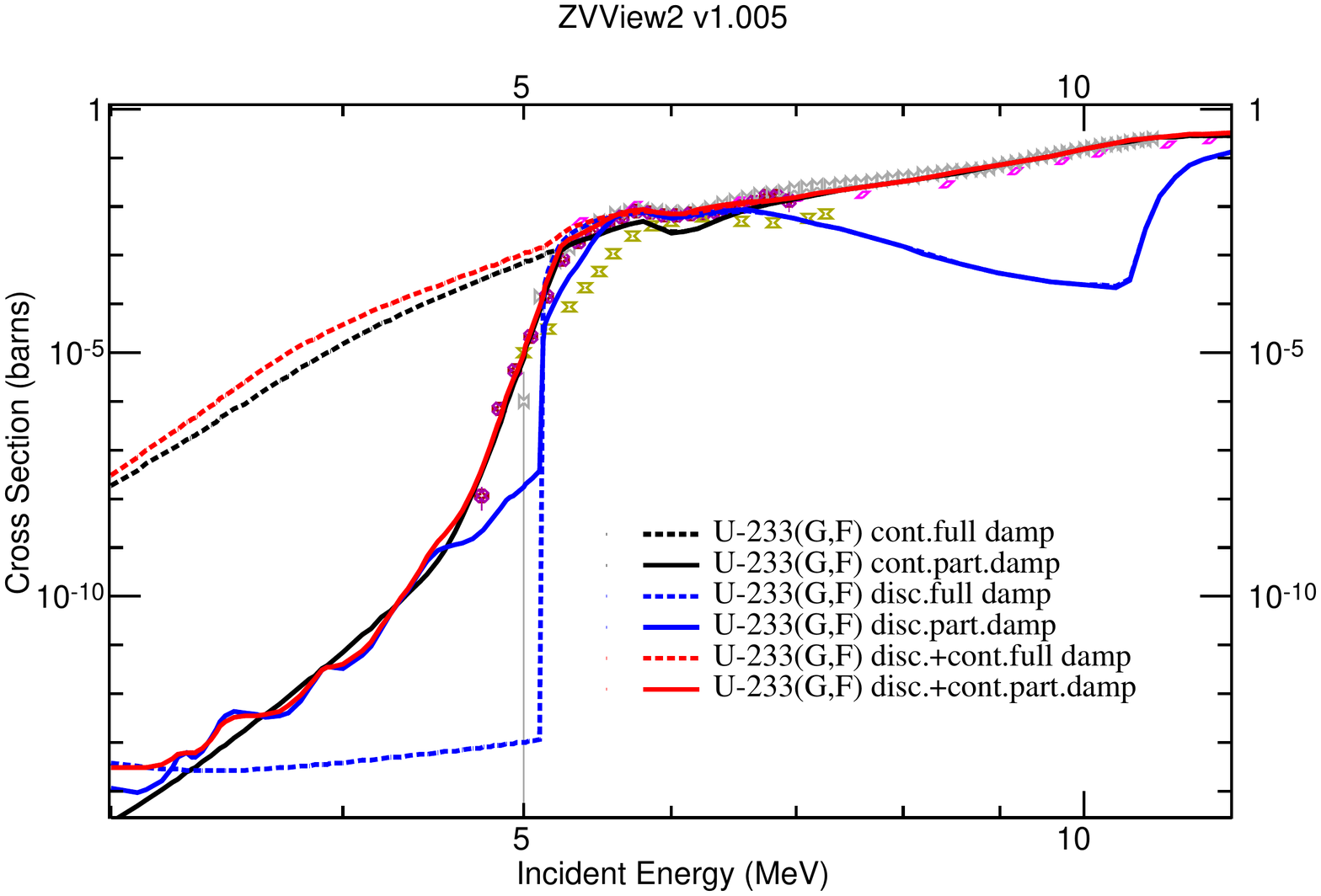}}
\put (-200, 140){\makebox{{\boldmath{$^{233}\rm{U(}\gamma,\rm{f)}$}}}}\hfill
\subfigure{\includegraphics[trim={2.0cm 2.6cm 3.0cm 4.54cm},clip,width=0.5\textwidth]{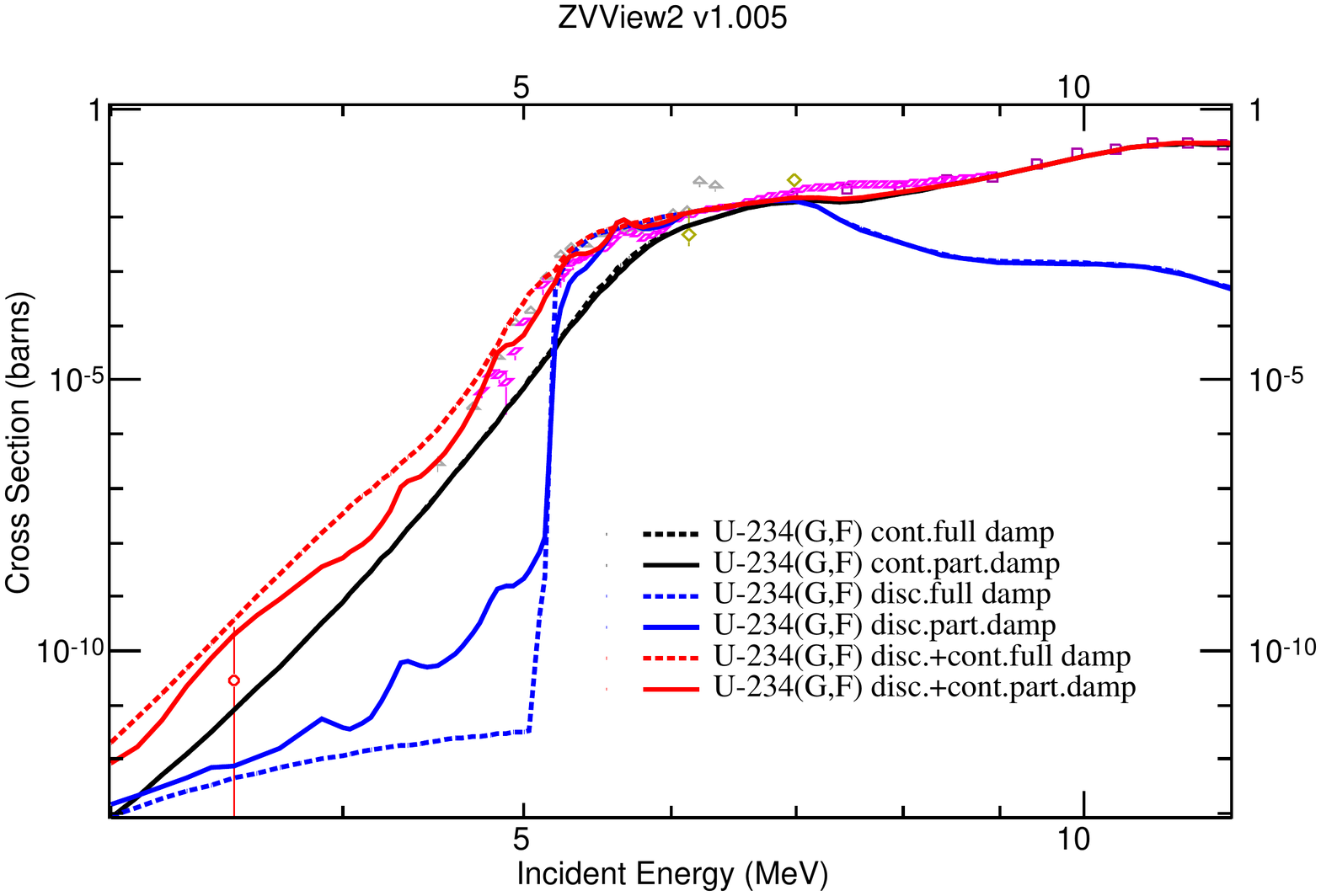}}
\put (-200, 140){\makebox{{\boldmath{$^{234}\rm{U(}\gamma,\rm{f)}$}}}}\hfill
\vspace{-4mm}
\caption{The effect of considering different degrees of damping of the class III vibrational states corresponding to discrete fission barriers and to barriers in the lower limit of continuum on the fission cross section of  $^{233,234}$U.
fission cross sections (red line), the contributions of the discrete fission channels (blue line) and of the channels in continuum (black line), considering partial (solid line) and full (dashed line)}
\label{u34-OMF}
\end{figure*}

In Fig.~\ref{u34-OMF} is presented the impact of the class II and III vibrational states' damping  on the fission cross sections  of \uthree~(representative for the odd-A isotopes), and \ufour~(representative for the even-even isotopes).
This figure reveals the striking behavior of the contributions of the discrete and continuum  fission channels corresponding to partial  and full  vibrational strength damping at excitation energies not reachable in the neutron-induced reactions. The first thing to notice is the different weight of the fission channels in continuum for the two types of nuclei at low energies.
The explanation is that for the odd-N nuclei the continuum starts at lower excitation energies and the level densities are higher than in the even-even nuclei. A similar behavior was observed to a lesser extent because of the higher excitation energies in the neutron-induced fission as shown in \r\cite{Sin:2016} for \ufive~and \usix~fissioning nuclei. For the odd-A nuclei it is impossible to describe the fission cross section below and above the ``threshold'' with the same parameters of the triple-humped barrier  without considering partial damping of the class III vibrational states associated to barriers with maxima in the lower part of the continuum spectrum.
Also of interest is the contribution of the transmission through the discrete barriers at excitation energies lower than the third well, especially the role of the resonant direct transmission at energies corresponding to the class II vibrational states partially damped.

\begin{table*}[!tbh]
\begin{center}
\caption{Fission barrier parameters.}
\label{t_fisparam}
\begin{tabular}{c|cc|cc|cc|cc|cc|l}
\hline\hline
\T        & $V_{A}$ & $\hbar\omega_{A}$ & $V_{I}$ & $\hbar\omega_{I}$
          & $V_{B}$ & $\hbar\omega_{B}$ & $V_{O}$ & $\hbar\omega_{O}$
          & $V_{C}$ & $\hbar\omega_{C}$ &  \\
 CN       & (MeV)   & (MeV) & (MeV) & (MeV) & (MeV) & (MeV) & (MeV) & (MeV) & (MeV) & (MeV) & Reaction \B \\
\hline
\T \uthree   & 4.70 & 0.70 & 1.70 & 1.00 & 5.85 & 1.30 & 5.00 & 1.00 & 5.80 & 1.30 &\uthree($\gamma,f)$ \\
          & 4.70 & 0.70 & 1.70 & 1.00 & 5.70 & 1.30 & 5.05 & 1.00 & 5.70 & 1.30 &\utwo($n,f$) \\
\hline
\T \ufour    & 4.70 & 0.60 & 1.70 & 1.00 & 5.83 & 1.40 & 5.00 & 1.00 & 5.83 & 1.40 &\ufour($\gamma,f)$ \\
          & 4.60 & 0.60 & 1.60 & 1.00 & 5.90 & 1.30 & 5.20 & 1.00 & 5.70 & 1.30 &\uthree($n,f$) \\
\hline
\T \ufive    & 4.90 & 0.60 & 1.75 & 1.00 & 6.20 & 1.45 & 5.18 & 1.00 & 5.90 & 1.45 &\ufive($\gamma,f)$ \\
          & 4.80 & 0.60 & 1.60 & 1.00 & 6.10 & 1.45 & 5.20 & 1.00 & 5.78 & 1.45 &\ufour($n,f$) \\
\hline
\T \usix     & 5.10 & 0.60 & 1.60 & 1.00 & 5.87 & 1.45 & 4.90 & 1.00 & 5.65 & 1.45 &\usix($\gamma,f)$ \\
          & 4.60 & 0.60 & 1.60 & 1.00 & 5.90 & 1.45 & 4.90 & 1.00 & 5.64 & 1.45 &\ufive($n,f$) \\
\hline
\T \useven   & 5.10 & 0.60 & 1.60 & 1.00 & 5.90 & 1.45 & 4.88 & 1.00 & 5.73 & 1.45 &\useven($\gamma,f)$ \\
          & 5.25 & 0.50 & 2.30 & 1.00 & 6.18 & 1.50 & 5.57 & 1.00 & 5.80 & 1.50 &\usix($n,f$) \\
\hline
\T \ueight   & 6.15 & 1.00 & 1.60 & 1.00 & 5.50 & 0.60 & - & - & - & - &\ueight($\gamma,f)$ \\
          & 6.30 & 1.00 & 1.60 & 1.00 & 5.50 & 0.60 & - & - & - & - &\useven($n,f$) \\
\hline\hline
\end{tabular}
\end{center}
\end{table*}
 In Table \ref{t_fisparam} the parameters of the fundamental fission barrier used in the present calculations are compared to those from Ref. \cite{Sin:2017}. The heights of the first hump are almost the same, excepting \usix, the new values showing an increase with increasing mass number. Even if no information on the  second well could be extracted from the neutron-induced reactions, the values adopted based on educated guess are confirmed by the present calculations (excepting the previous out of range value considered for \useven). The heights of the outer humps are also very close. Probably the most important confirmation is the value around 5 MeV for the bottom of the third well.
\\
\begin{figure*}[!thb]
\subfigure{\includegraphics[trim={1.4cm 2.2cm 3.0cm 4.54cm},clip,width=0.5\textwidth]{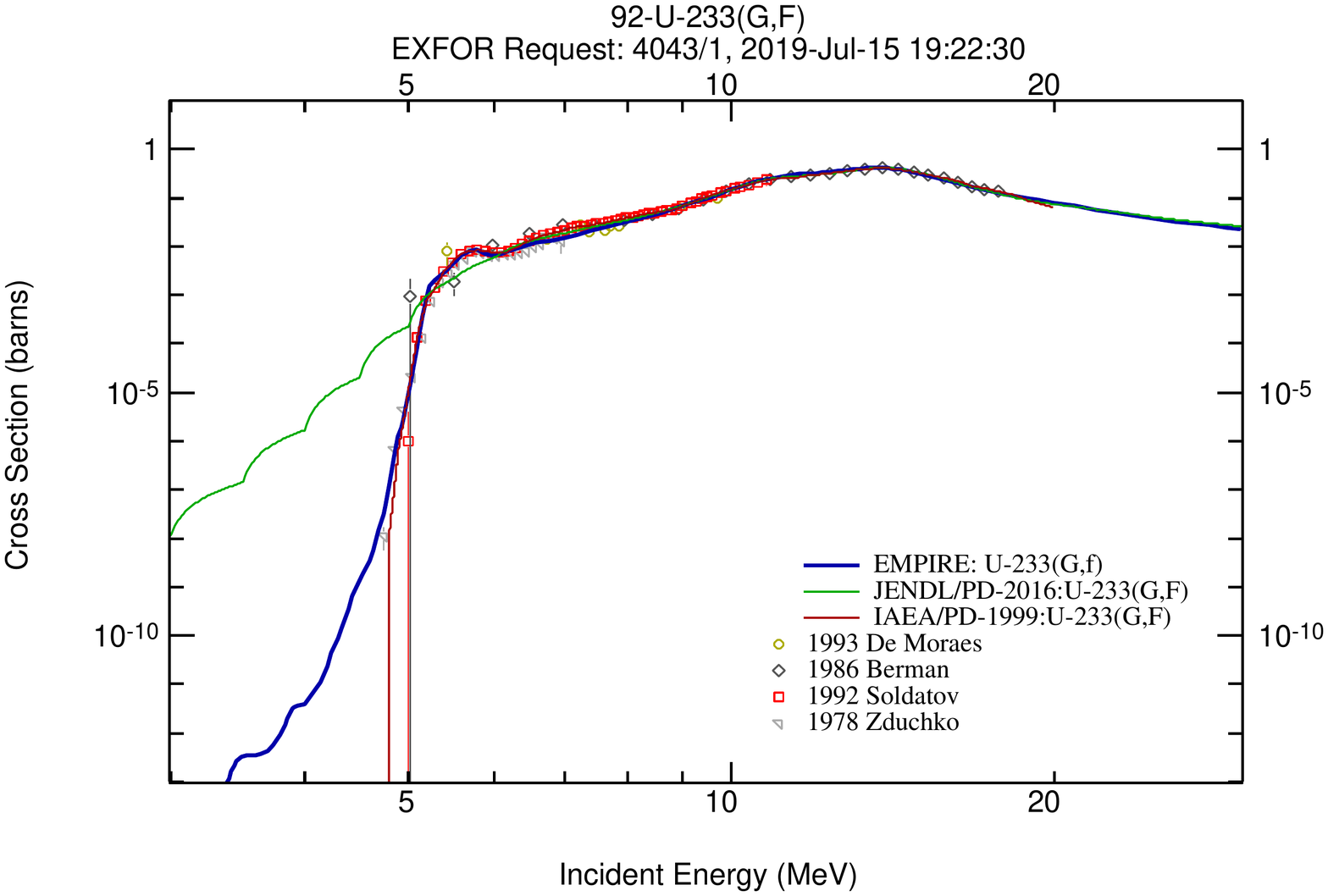}}
\put (-200, 140){\makebox{{\boldmath{$^{233}\rm{U(}\gamma,\rm{f)}$}}}}\hfill
\subfigure{\includegraphics[trim={1.4cm 2.2cm 3.0cm 4.54cm},clip,width=0.5\textwidth]{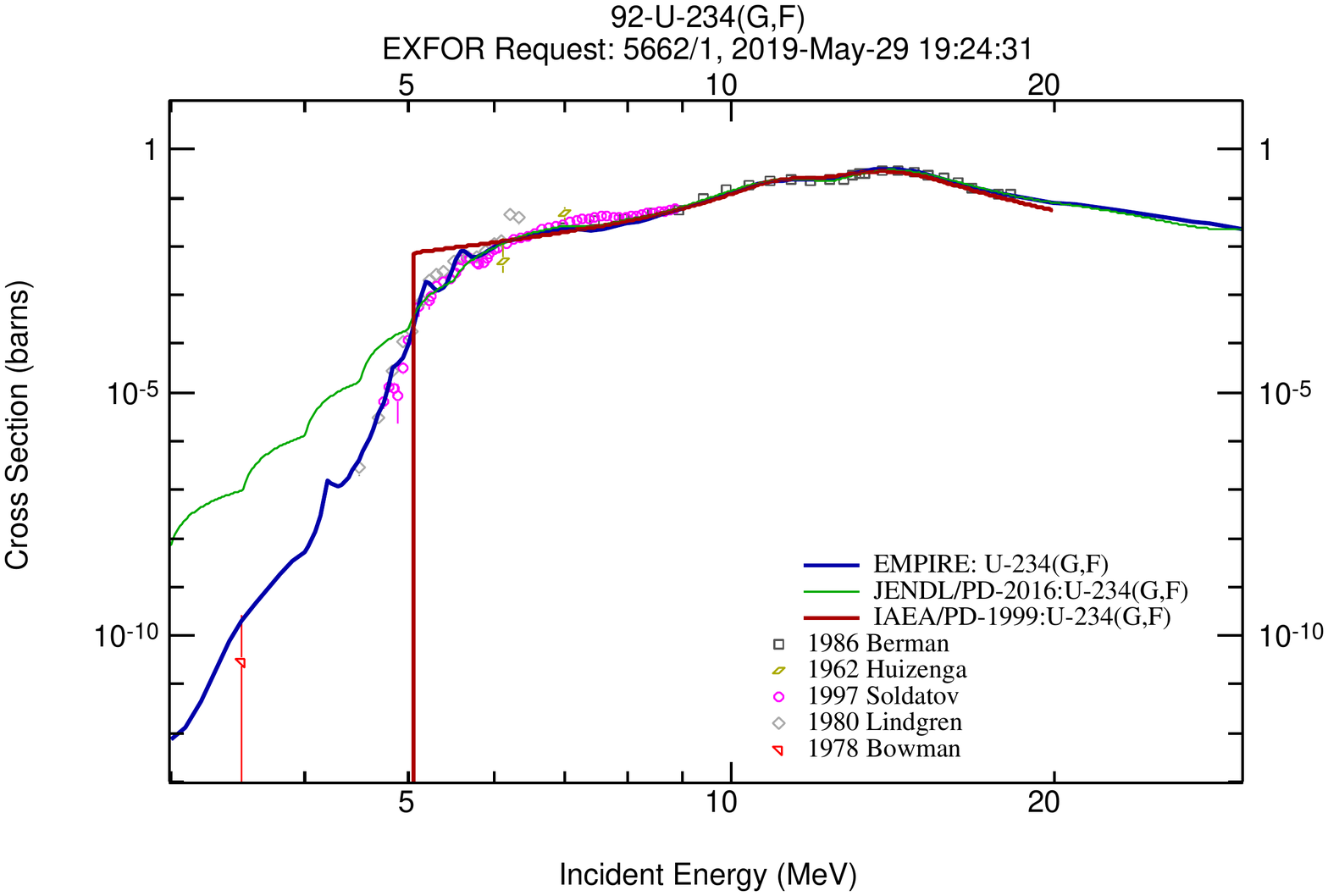}}
\put (-200, 140){\makebox{{\boldmath{$^{234}\rm{U(}\gamma,\rm{f)}$}}}}\hfill
\vspace{-5mm}
\\
\subfigure{\includegraphics[trim={1.4cm 2.2cm 3.0cm 4.54cm},clip,width=0.5\textwidth]{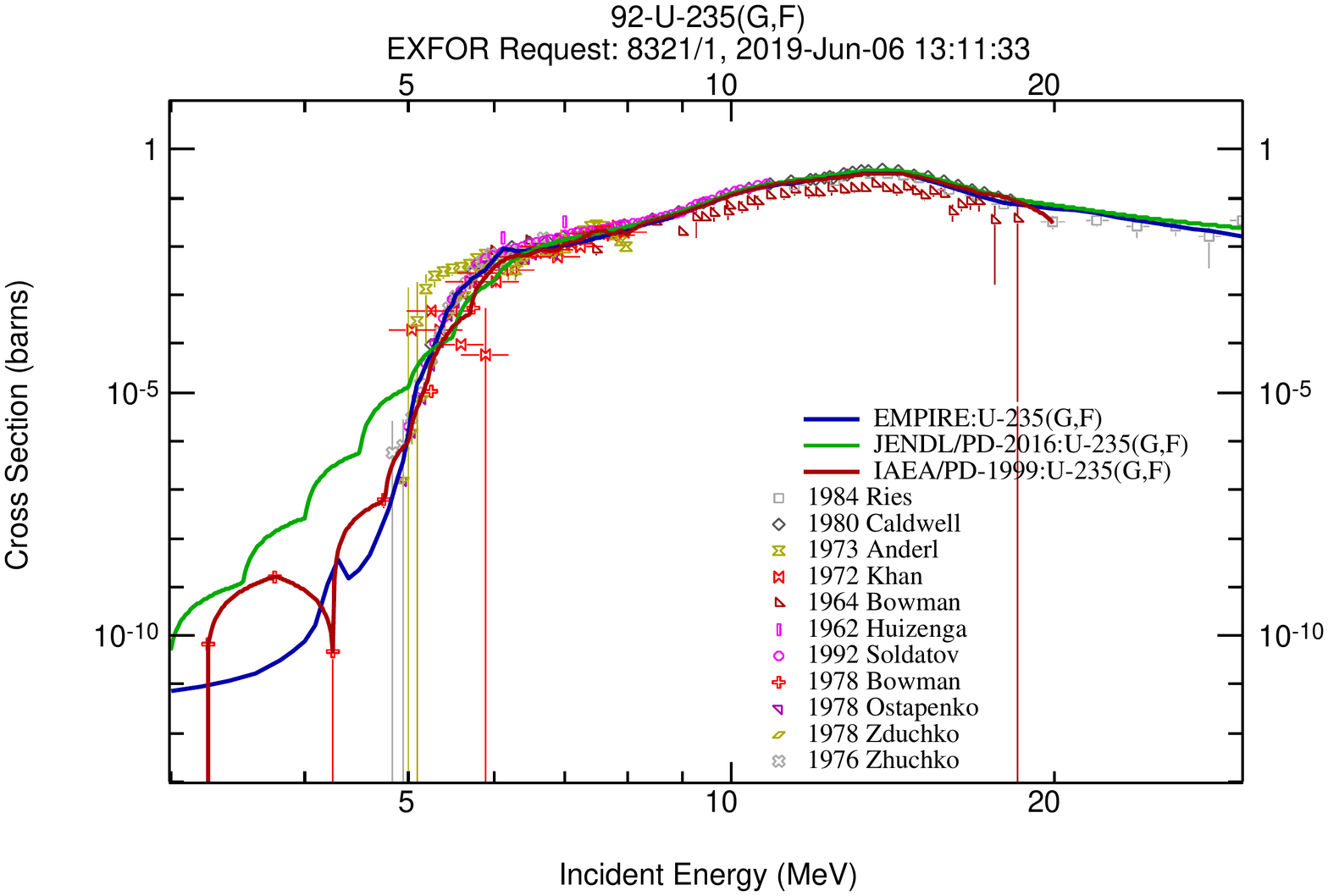}}
\put (-200, 140){\makebox{{\boldmath{$^{235}\rm{U(}\gamma,\rm{f)}$}}}}\hfill
\subfigure{\includegraphics[trim={1.4cm 2.2cm 3.0cm 4.54cm},clip,width=0.5\textwidth]{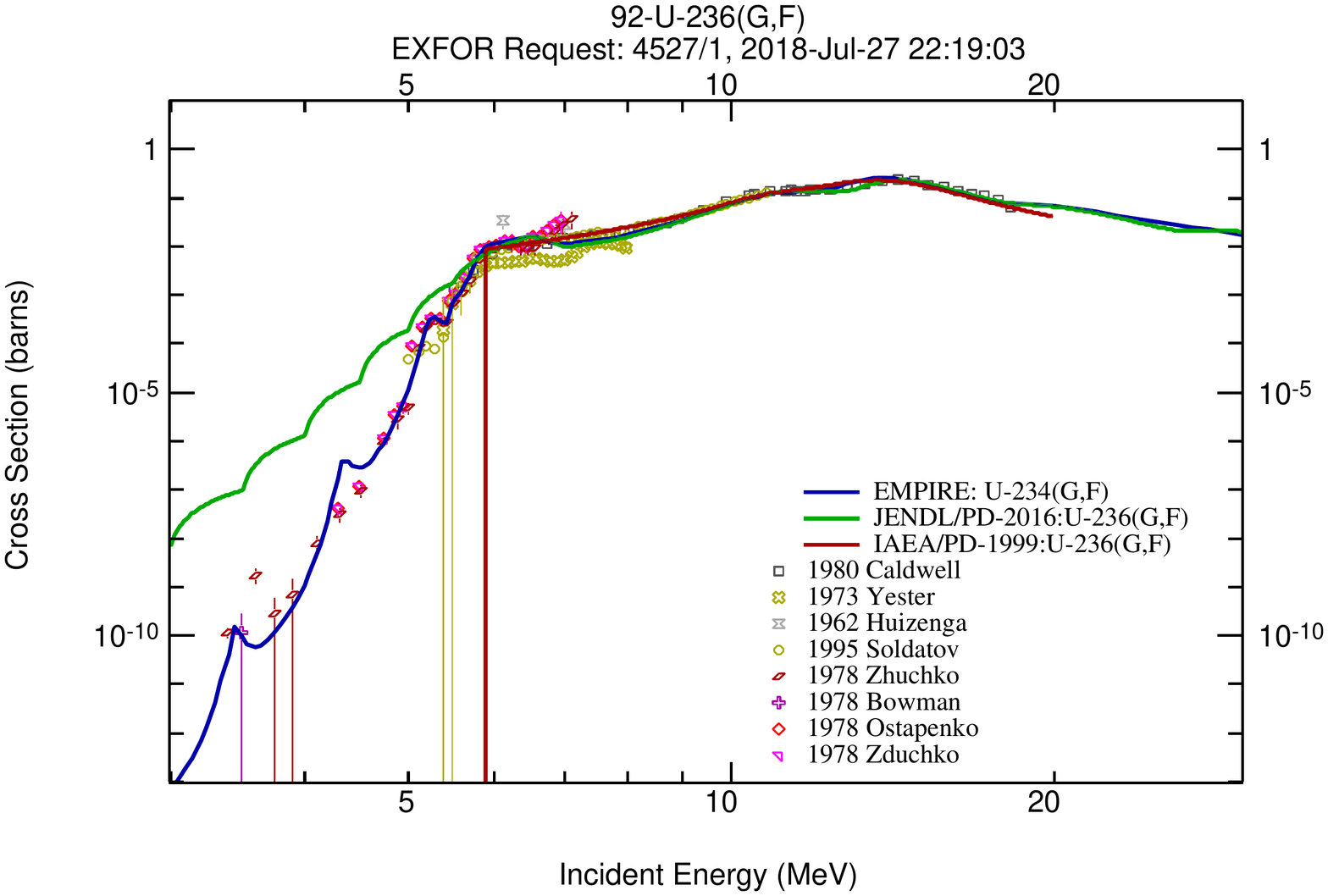}}
\put (-200, 140){\makebox{{\boldmath{$^{236}\rm{U(}\gamma,\rm{f)}$}}}}\hfill
\vspace{-5mm}
\\
\subfigure{\includegraphics[trim={1.4cm 2.2cm 3.0cm 4.54cm},clip,width=0.5\textwidth]{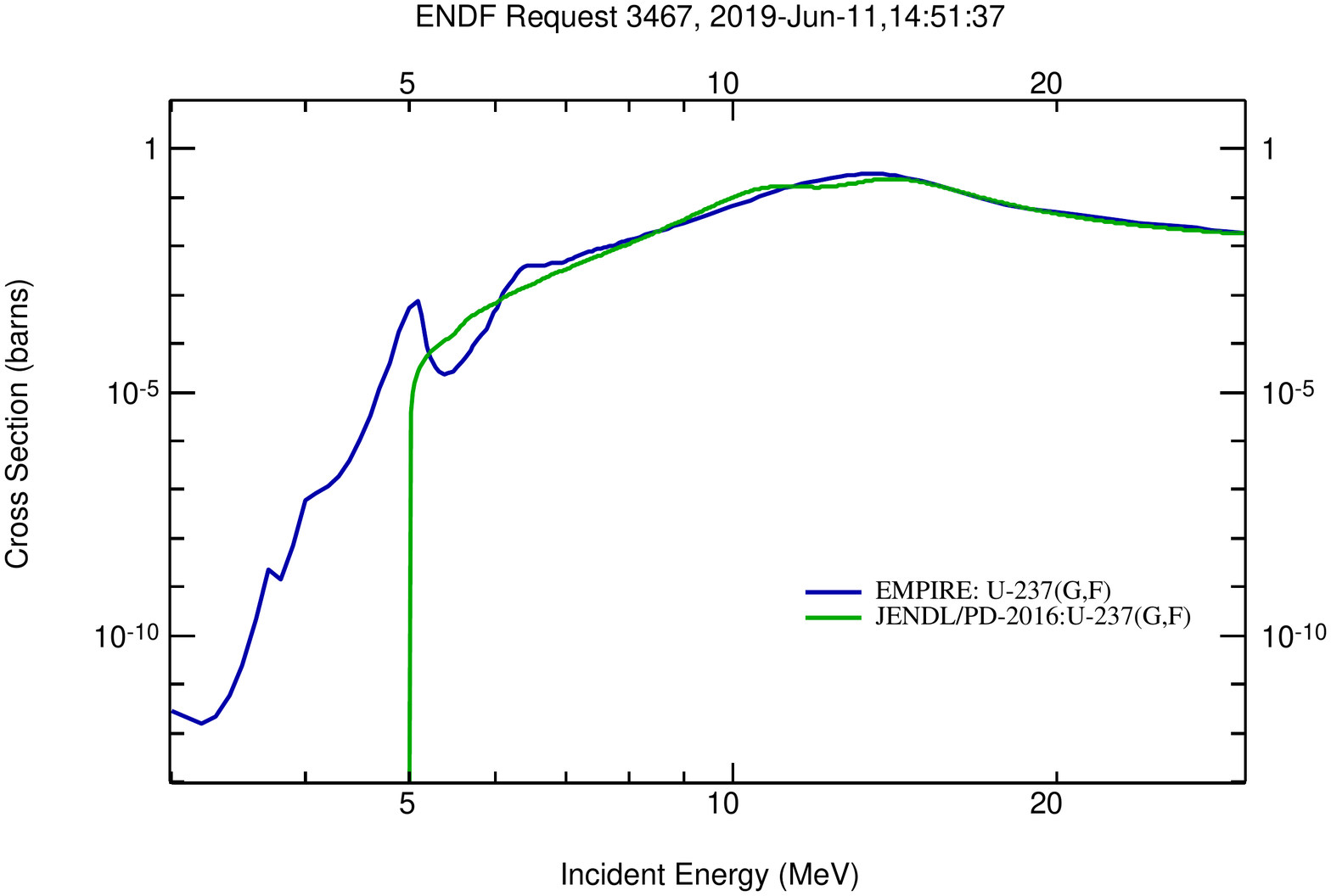}}
\put (-200, 140){\makebox{{\boldmath{$^{237}\rm{U(}\gamma,\rm{f)}$}}}}\hfill
\subfigure{\includegraphics[trim={1.4cm 2.2cm 3.0cm 4.54cm},clip,width=0.5\textwidth]{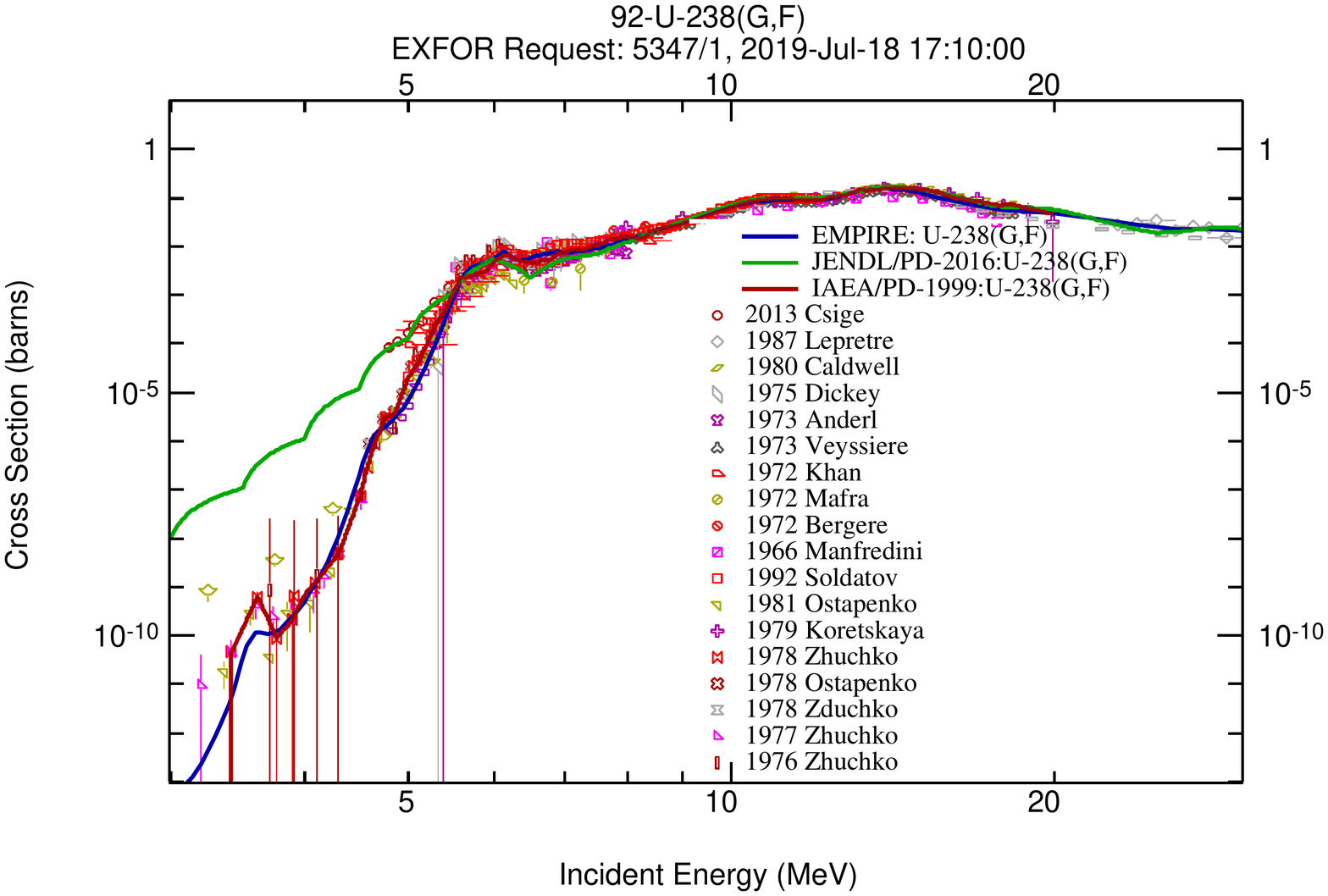}}
\put (-200, 140){\makebox{{\boldmath{$^{238}\rm{U(}\gamma,\rm{f)}$}}}}\hfill
\vspace{-4mm}
\caption{EMPIRE calculations (blue line) and JENDL-PD (green line), IAEA-PD (red line) evaluations for the photo-fission cross sections of $^{233-238}$U  compared to experimental data from EXFOR~\cite{Moraes:93,Caldwell:80, Berman:86,Veyssiere:73, Bergere:72, Dickey:75, Mafra:72, Soldatov:92, Zhuchko:78, Huizenga:62, Soldatov:93, Lindgren:80, Bowman:78, Ries:84, Anderl:73, Khan:72, Bowman:64, Ostapenko:78, Zhuchko:76, Yester:73,
Soldatov:95, Zhuchko:78a, Csige:2013, Lepretre:87, Manfredini:66, Ostapenko:81, Koretskaya:79, Arruda:76, Rabotnov:70, Katz:58}.}
\label{u_all-FISS}
\end{figure*}
\begin{figure*}[!thb]
\subfigure{\includegraphics[trim={1.4cm 2.2cm 3.0cm 4.54cm},clip,width=0.5\textwidth]{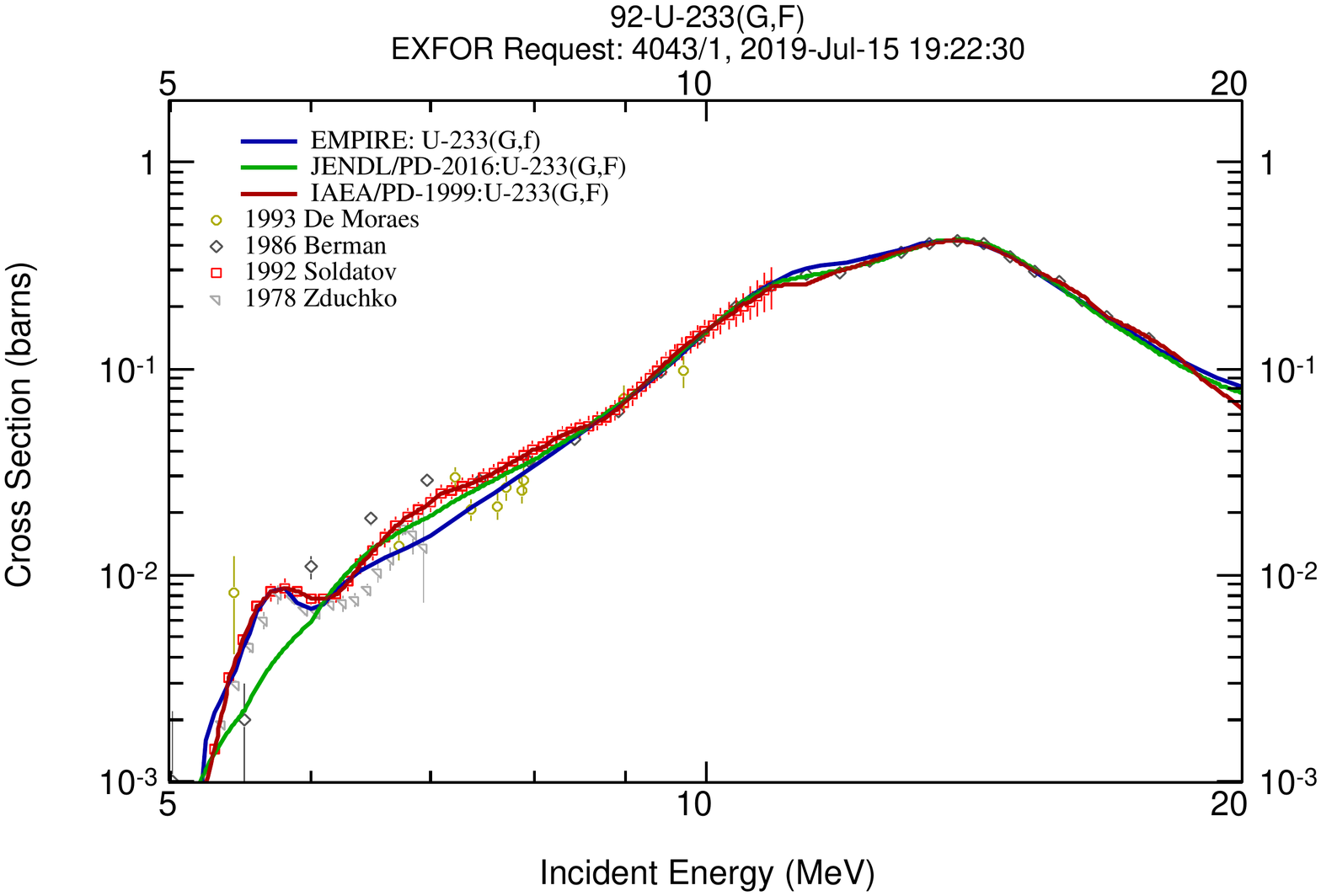}}
\put (-70, 140){\makebox{{\boldmath{$^{233}\rm{U(}\gamma,\rm{f)}$}}}}\hfill
\subfigure{\includegraphics[trim={1.4cm 2.2cm 3.0cm 4.54cm},clip,width=0.5\textwidth]{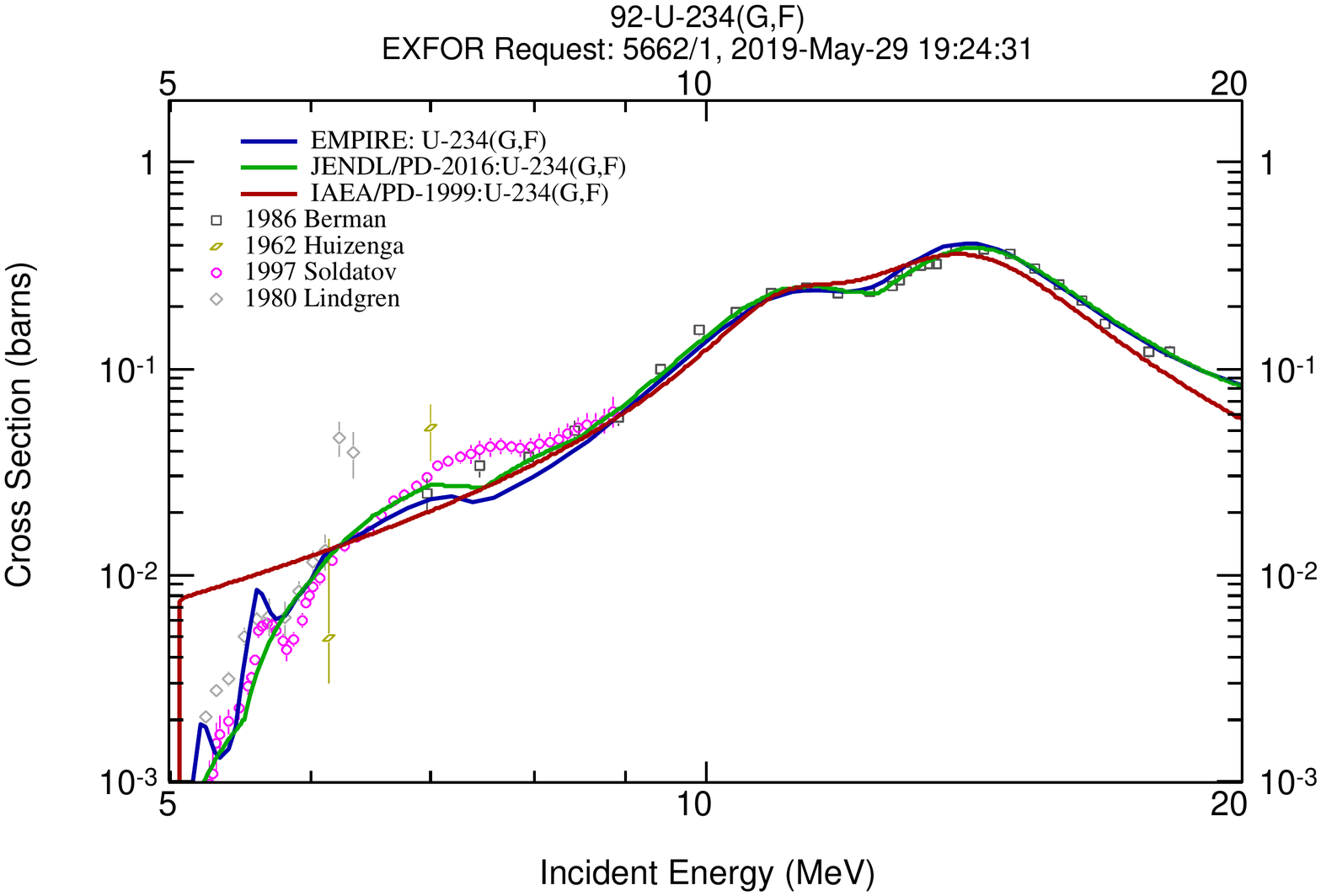}}
\put (-70, 140){\makebox{{\boldmath{$^{234}\rm{U(}\gamma,\rm{f)}$}}}}\hfill
\vspace{-4mm}
\\
\subfigure{\includegraphics[trim={1.4cm 2.2cm 3.0cm 4.54cm},clip,width=0.5\textwidth]{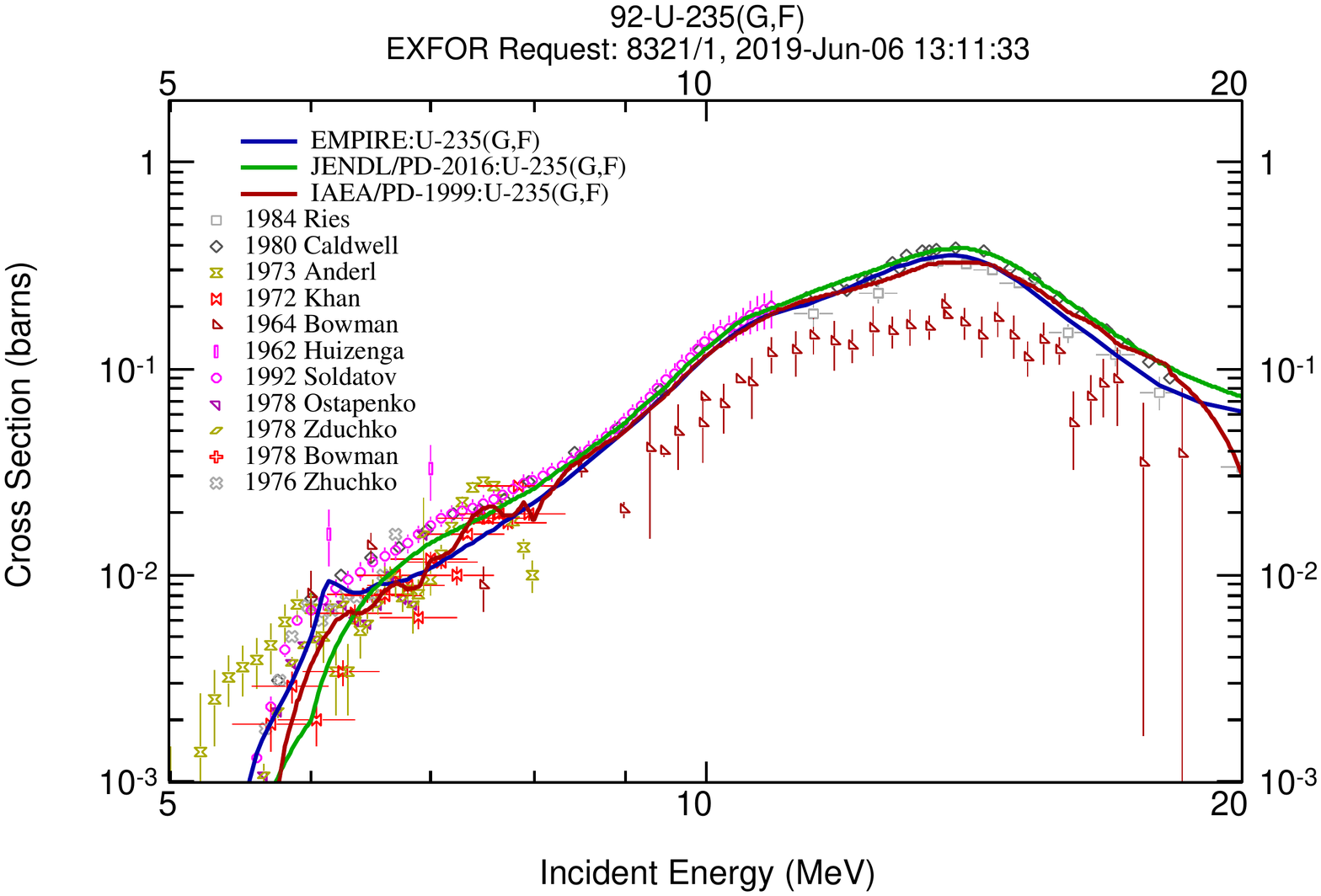}}
\put (-70, 140){\makebox{{\boldmath{$^{235}\rm{U(}\gamma,\rm{f)}$}}}}\hfill
\subfigure{\includegraphics[trim={1.4cm 2.2cm 3.0cm 4.54cm},clip,width=0.5\textwidth]{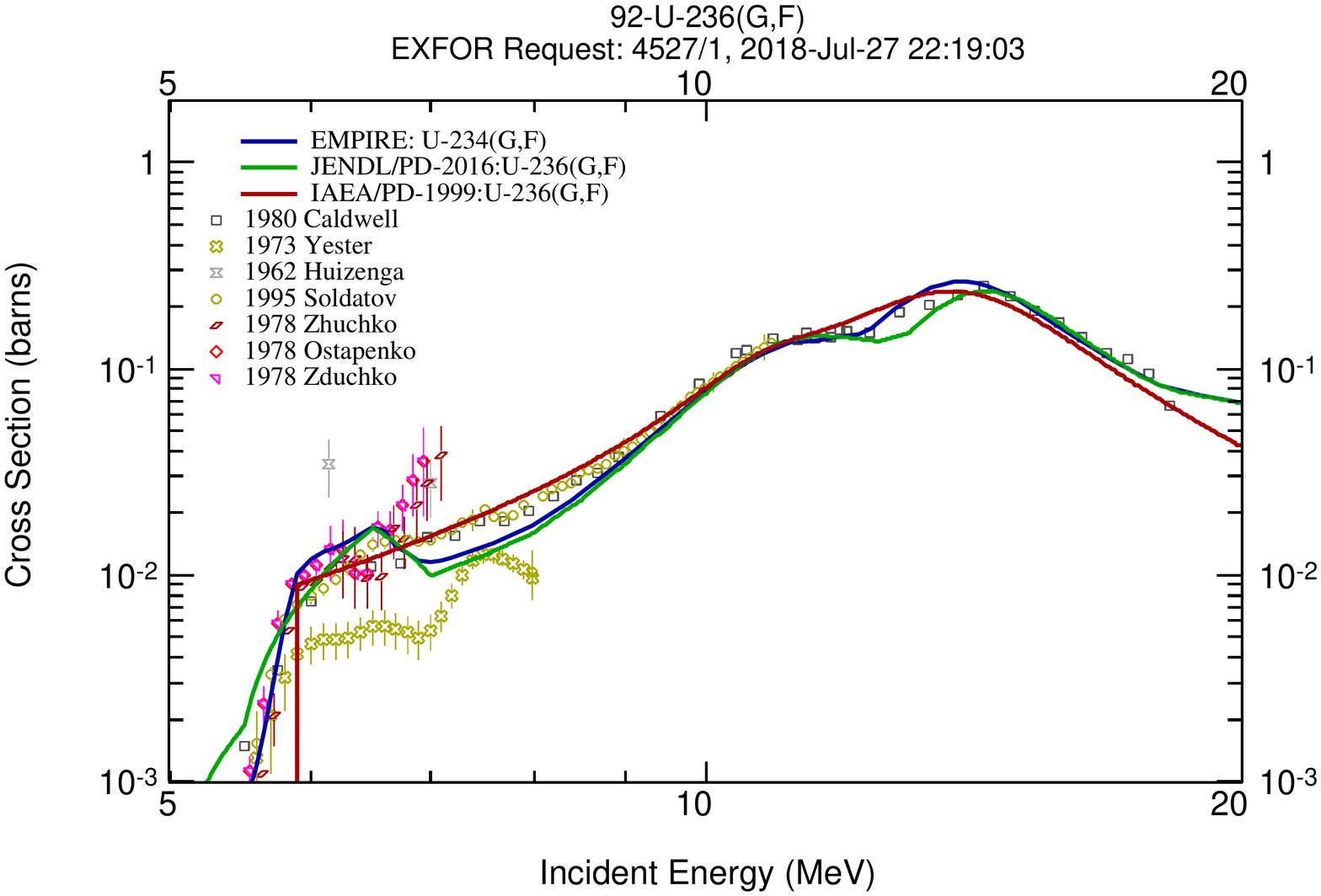}}
\put (-70, 140){\makebox{{\boldmath{$^{236}\rm{U(}\gamma,\rm{f)}$}}}}\hfill
\vspace{-4mm}
\\
\subfigure{\includegraphics[trim={1.4cm 2.2cm 3.0cm 4.54cm},clip,width=0.5\textwidth]{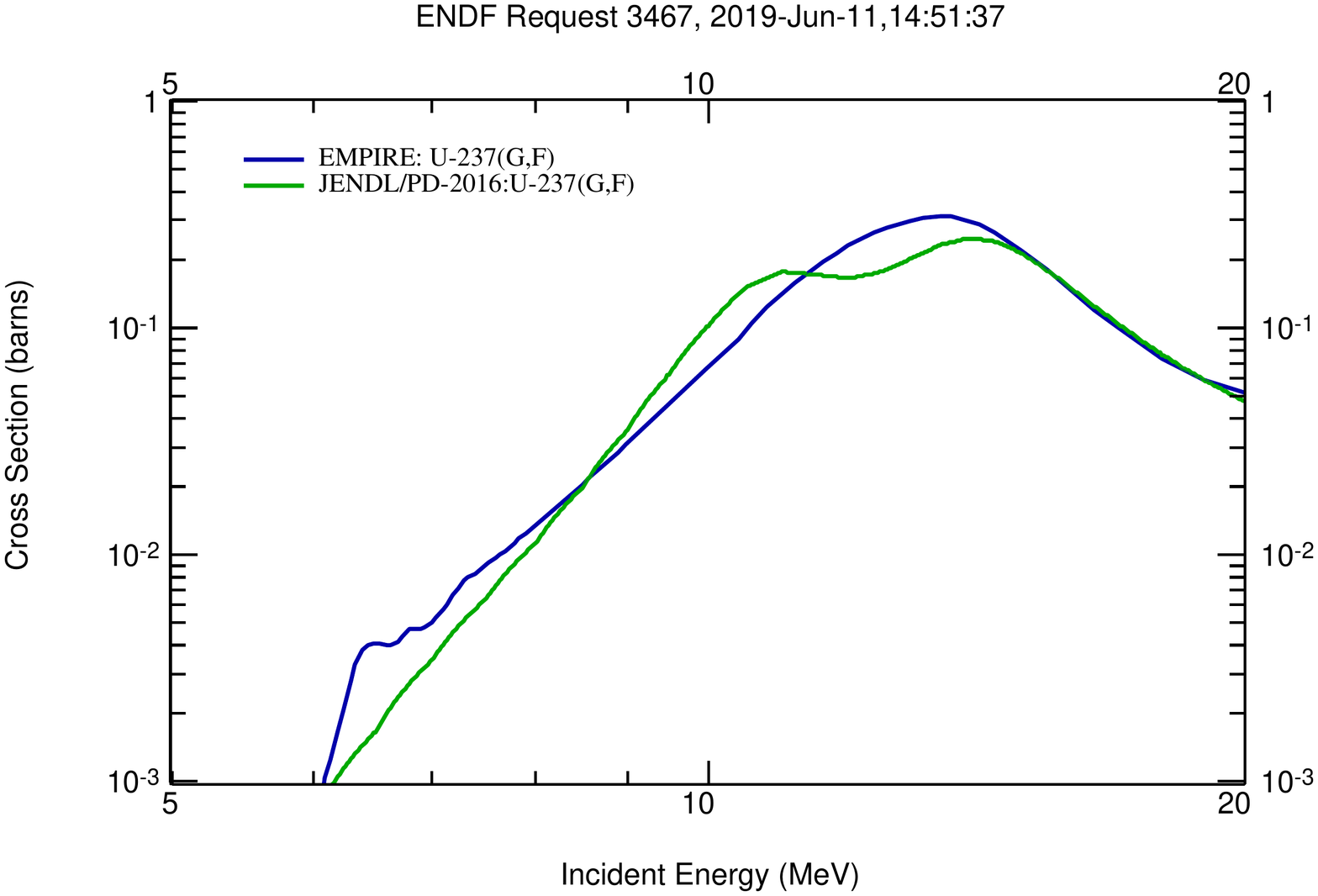}}
\put (-70, 140){\makebox{{\boldmath{$^{237}\rm{U(}\gamma,\rm{f)}$}}}}\hfill
\subfigure{\includegraphics[trim={1.4cm 2.2cm 3.0cm 4.54cm},clip,width=0.5\textwidth]{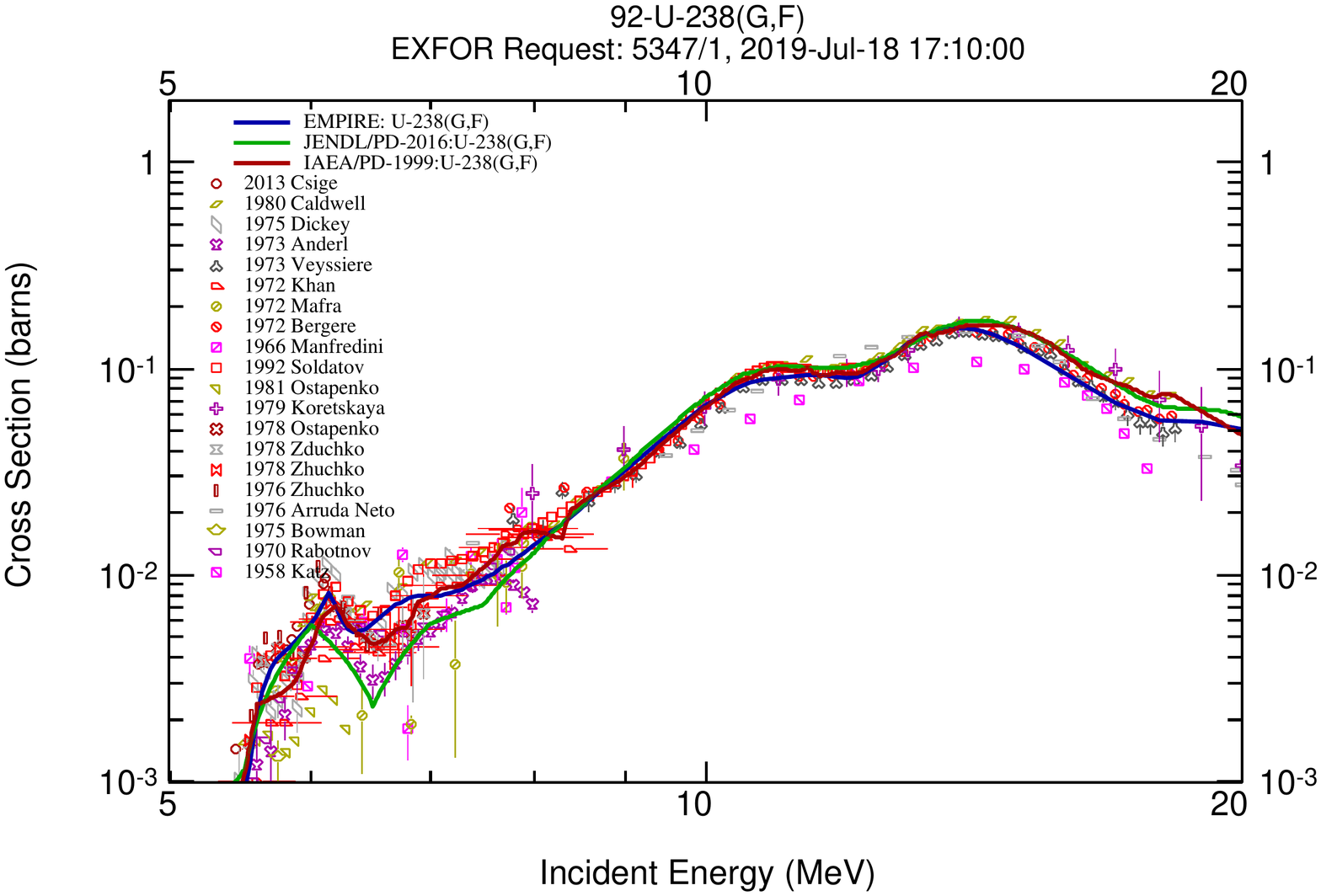}}
\put (-70, 140){\makebox{{\boldmath{$^{238}\rm{U(}\gamma,\rm{f)}$}}}}\hfill
\vspace{-4mm}
\caption{EMPIRE calculations (blue line) and JENDL-PD (green line), IAEA-PD (red line) evaluations for the photo-fission cross sections of $^{233-238}$U in the incident energy range 5-20 MeV compared to experimental data from EXFOR~\cite{Moraes:93,Caldwell:80, Berman:86,Veyssiere:73, Bergere:72, Dickey:75, Mafra:72, Soldatov:92, Zhuchko:78, Huizenga:62, Soldatov:93, Lindgren:80, Bowman:78, Ries:84, Anderl:73, Khan:72, Bowman:64, Ostapenko:78, Zhuchko:76, Yester:73,
Soldatov:95, Zhuchko:78a, Csige:2013, Lepretre:87, Manfredini:66, Ostapenko:81, Koretskaya:79, Arruda:76, Rabotnov:70, Katz:58}.}
\label{u_all-FISSp}
\end{figure*}
For a fair and realistic analysis one should remind the reader that the role of the fundamental barrier is affected by the selectivity in spin and parity.
If not known otherwise, the fundamental barrier is assigned spin projection and parity equal to the spin and parity of the ground state of the fissioning nucleus. Considering the spins and parities of the target nuclei, of the neutron and of the photon, it is obvious that by GDR photo-excitation the nuclei will never be populated in states of spin and parity which belong to the fundamental rotational band, while the compound nuclei formed by absorption of neutrons (which may carry higher orbital momenta) can. In other words, the transmission through the fundamental barrier and through the barriers associated to the rotational band built on it, which represent a significant contribution to the neutron-induced fission cross section at low energies, is practically forbidden in photo-fission.
 For example, for the even-even isotopes $^{234,236}$U which are populated in states with $J^{\pi}=1^-$, the transmission through the discrete barriers is determined by the absolute excitation energies of the rotational band-heads $K^{\pi}=0^-,1^-$ and less by the parameters of the fundamental barrier which has $K^{\pi}=0^+$.
\\
However, the heights and widths of the fundamental humps enter together with the level density functions in the calculation of the transmission coefficients through the barriers in the continuum spectrum (Eq~.\eqref{th-cont}). So, the role of the fundamental barrier remains very important for photo-fission also, especially at higher excitation energies, where the fission channels in the  continuum have the dominant contribution. Considering that the parameters of the level densities at saddles used in the present work are those from \csin~adjusted at most by 5\%, the agreement of the calculated photo-fission \xss~with the experimental data above 7 MeV represents a real test of the fundamental barrier parameters.

The most uncertain fission parameters are those of the discrete transition states, especially for the odd-A nuclei.
For even-even nuclei, there are collective states within the pairing gap with $K^{\pi}=0^+,2^+,0^-,1^-$ and with excitation energies at the saddle points correlated with the nuclear shape asymmetry at the corresponding deformations. But in general, if no other information is available, the sets ($\varepsilon, K\pi$) are chosen to fit the experimental fission cross section.

The results of EMPIRE calculations are presented in Figs.~\ref{u_all-FISS} and \ref{u_all-FISSp} together with JENDL-PD and IAEA-PD evaluations and the experimental data from EXFOR.
Once again, one can notice that JENDL-PD is based on model calculations. The shape of the JENDL-PD fission cross section at low energies indicates the use of a double-humped fission barrier in the full damping limit of the class II vibrational states. This is the reason why JENDL-PD fission cross sections overestimate the experimental data at those energies.  IAEA-PD on the other hand, is based mainly on a non-model fit of the experimental data, the most evident example being the \ufive~case.

The experimental data for $^{233-236}$U($\gamma$,f) cross sections are too scarce below 5 MeV to reveal a clear resonance structure. Still, for the even-even isotopes, for which the distance among the class II vibrational states is higher, hence their damping is lower, one can notice a resonance around 4.8 MeV for \ufour~and a sequence of three resonances around 3.4~MeV, 4.3~MeV and 5.3~MeV shown by the  $^{236}$U($\gamma$,f) \xs. These resonances are not sharp enough to be generated by the class III vibrational states, therefore one can confirm that the energy associated to the bottom of the third well should be around 5 MeV. The opening of the fission channel, the absolute value and the slope of the fission cross sections, the damping of the resonance and the distance between them have been used to extract information on the first hump and on the second well.
This information is not available in neutron induced reactions where the fission barrier cannot be explored at such low excitation energies.

In the energy range 5--8 MeV there are more experimental data, but the discrepancies among the different sets are significant. As shown in Fig.~\ref{u_all-ALL}, there is an abrupt behavior of three  cross sections in this range: gamma emission drops while fission and neutron emission rise.  Therefore, the interpretation  of the experimental fission cross section (Figs.~\ref{u_all-FISS},~\ref{u_all-FISSp}) must consider the behavior of all cross sections shown in Fig.~\ref{u_all-ALL} because what seems to be threshold or resonance might be a structure generated by other causes.

The neutron separation energy decreases as nuclei become neutron richer, but also has a strong odd-even effect, so that for \uthree, \ufour~and \usix~there is an energy interval between the opening of the fission and of the neutron emission channels. As gamma emission falls quickly once the fission channel opens, in this energy interval fission remains the dominant decay and the photo-absorption and fission cross sections become almost equal (see Fig.~\ref{u_all-ALL}). So, what looks at the first glance as a threshold is in fact a limitation imposed by the photo-absorption cross section and it is not directly related to the height of the fission barrier.

The neutron induced fission cross sections of even-N light actinide targets (\textit{e.g.}, $^{232}$Th, $^{231}$Pa, $^{234,236}$U) show in the same excitation energy range (5--8 MeV) a very clear resonance structure attributed to the low damped class III vibrational states \cite{Sin:2017,Sin:2006}. There are no similar resonances in the photo-fission cross sections, excepting \ufour~which has such a resonance around 5.6~MeV. More studies are needed to understand if this different behavior is related to experimental limitations or has other causes. The fission cross sections of \uthree~and \ueight~also have maxima  around 5.6~MeV and 6.1~MeV respectively, which can be mistaken as resonances. In fact the resonant-like shapes are the effect of the dips around 6~MeV and 6.6~MeV respectively, caused by the opening of the neutron emission channel.  The calculated fission cross sections of all isotopes have such a decrease, which does not appear to the same extent in the experimental data. One can notice the similar behavior in this region of the EMPIRE calculations and the JENDL-PD evaluations of $^{234,236}$U($\gamma$,f) cross sections.
\\
At these energies (5--8 MeV) the calculated fission \xss~are most sensitive to the excitation energies of the discrete transition states, especially at the second and third saddle points, but also to the density of the transition states in the continuum.

In the energy range 8--12 MeV the fission channels in the continuum are the dominant contribution. Those fission channels are described by the EGSM level densities for the first fission chance. Depending on the neutron separation energies, the second and third fission chances open around 6 and 12 MeV and become significant around 13 and 20 MeV respectively. For the first and second residual nuclei the same fission parameters from Table~\ref{t_fisparam} have been used.  Our calculations agree well with the experimental data, and with the JENDL-PD evaluation in this energy range, excepting the already discussed cases of \useven~and \ueight.

\section{Conclusions}
Photo-reaction cross section calculations for the Uranium isotopes have been performed with the EMPIRE 3.2 code in the energy range 3--30 MeV. The results give a comprehensive and systematic description of the experimental data better than current evaluations. A set of GDR parameters consistent with all available experimental data is provided.

Except for the incident channel,  the same reaction models and almost the same input parameters as for the neutron induced reaction calculations in \r\cite{Sin:2017} have been used. The extended optical model for fission proved again to describe accurately the experimental fission cross sections  at excitation energies below 7~MeV.
The parameters of the fundamental triple-humped fission barriers derived from the analysis of the neutron-induced reactions on the Uranium isotopes have been in general validated by the present photo-reaction calculations.
The access to low excitation energies allowed one to narrow the uncertainties of the first hump and second well fission parameters, and also confirmed the shallowness of the third well of which the energy of the bottom of the well is around 5 MeV.

This type of study which involve reactions induced by different projectiles leading to the same compound systems can identify data discrepancies, improve the models and reduce the uncertainties of the input parameters, and thus enhance the accuracy of the nuclear reaction data.

\subsection*{Acknowledgments}
The work of M.S. was partially supported by the IAEA under CRP F41033 Contract No.20364,  and UEFISCDI under PCE-2016-0014 Contract No.7/2017. BVC acknowledges support from grant 2017/05660-0 of the Sao Paulo Research Foundation (FAPESP), grant 306433/2017-6 of the CNPq and the INCT-FNA project 464898/2014-5. MWH acknowledges support from the National Nuclear Security Administration of the U.S. Department of Energy at Los Alamos National Laboratory under Contract No. 89233218CNA000001.


\begin{thebibliography}{999}

\bibitem{CRP-RIPL4} R.  Capote  Noy, S.  Goriely, S.  Hilaire, O.  Iwamoto, T. Kawano, A.  Koning, S.  Simakov, \ql Consultants'  Meeting  on  Recommended  Input  Parameters  for  Fission Cross Section Calculation, 17-18 December 2013,\qr Report {\bf IAEA(NDS)-0654}, International Atomic Energy Agency, Vienna (2014).

\bibitem{Sin:2017} M. Sin, R. Capote, M.W. Herman, A. Trkov, \ql Modelling Neutron-induced Reactions on $^{232-237}$U from 10 keV up to 30 MeV,\qr \textsc{ Nucl. Data Sheets }\textbf{139}, 138--170 (2017).


\bibitem{Carlson:2009} A. D. Carlson, V. G. Pronyaev, D. L. Smith \etals, \ql International Evaluation of Neutron Cross Section Standards,\qr \textsc{ Nucl. Data Sheets }\textbf{110}, 3215--3324~(2009).

\bibitem{Carlson:2018} A. D. Carlson, V. G. Pronyaev, R. Capote \etals, \ql Evaluation of Neutron Data Standards,\qr \textsc{ Nucl. Data Sheets }{\bf 148}, 142--187 (2018).

\bibitem{Capote:2014a} R. Capote, A. Trkov, M. Sin, M. Herman, A. Daskalakis, and Y. Danon, \ql Physics of Neutron Interactions with $^{238}$U: New Developments and Challenges,\qr \textsc{ Nucl. Data Sheets }\textbf{115}, 1--26 (2014).

\bibitem{Capote:2014b} R. Capote, A. Trkov, M. Sin, M. W. Herman, E. Sh. Soukhovitskii, \ql Elastic and inelastic scattering of neutrons on $^{238}$U nucleus,\qr \textsc{ EPJ Web of Conf. }\textbf{69}, 00008~(2014).

\bibitem{Capote:2014c} R. Capote, M. Sin, A. Trkov, M. W. Herman, D. Bernard, G. Noguere, A. Daskalakis, Y. Danon, \ql Evaluation of neutron induced reactions on U-238 nucleus,\qr {\sc ~Proc. NEMEA-7 Workshop} {\bf NEA/NSC/DOC(2014)13}, NEA, OECD~(2014).

\bibitem{CIELO-U} R. Capote, A. Trkov, M. Sin \etal, \ql IAEA CIELO Evaluation of Neutron-induced Reactions on $^{235}$U and $^{238}$U Targets,\qr  \textsc{ Nucl. Data Sheets }\textbf{148},254-292 (2018).

\bibitem{CIELO:2018} M.~B. Chadwick, R.~Capote, A.~Trkov,\textit{et~al.}, ``{CIELO} collaboration summary results: International evaluations of neutron reactions on uranium, plutonium, iron, oxygen and hydrogen,''  \textsc{Nucl. Data Sheets }\textbf{148}, 189--213 (2018).

\bibitem {ENDFVIII}D.A. Brown, M.B. Chadwick, R. Capote,..., M. Sin, \etal \ql ENDF/B-VIII.0: The 8th Major Release of the Nuclear Reaction Data Library with CIELO-project Cross Sections, New Standards and Thermal Scattering Data,\qr \textsc{Nucl. Data Sheets }{\bf 148}, 1 (2018).

\bibitem{EMPIRE} M.W. Herman, R. Capote, B. V. Carlson, P. Oblozinsk\'{y}, M. Sin, A. Trkov, H. Wienke and V. Zerkin, \ql  EMPIRE: Nuclear Reaction Model Code System for Data Evaluation,\qr {\sc Nucl. Data Sheets }{\bf 108}, 2655--2715 (2007).

\bibitem{EMPIRE-man} M. Herman \etals, \ql EMPIRE-3.2 Malta User's Manual\qrs, Report {\bf INDC(NDS)-0603}, IAEA, Vienna, Austria (2013).

\bibitem{Filipescu:2015} D. Filipescu, A. Anzalone, D. L. Balabanski, S. S. Belyshev, F. Camera, M. La Cognata, P. Constantin, L. Csige, P. V. Cuong, M. Cwiok \etal,,  Perspectives for photonuclear research at the Extreme Light Infrastructure - Nuclear Physics (ELI-NP) facility, \textsc{ Eur. Phys. J. }{\bf A51}, 185.1--30 (2015).

\bibitem{photo-CRP} IAEA Coordinated Research Project on Photonuclear Data and Photon Strength Functions. 

\bibitem{Kawano:2020} T. Kawano, Y.S. Cho, P. Dimitriou \etals, \ql  IAEA Photonuclear Data Library 2019\qrs, submitted to {\sc Nucl. Data Sheets}.

\bibitem{ripl3} \ R. Capote \textit{et~al.}, \ql RIPL--Reference Input Parameter
Library for Calculation of Nuclear Reactions and Nuclear Data Evaluations,\qr \textsc{ Nucl. Data Sheets }\textbf{110}, 3107--3214 (2009).

\bibitem{Plujko:2007}V. Plujko, I. Kadenko, S. Goriely, E. Kulich, O. Davidovskaya, and O. Gorbachenko,
\ql Models for photoabsorption cross section estimates,\qr \textit{Int. Conf. on Nuclear Data for Science and Technology 2007}, p.235, (2008).

\bibitem{Plujko:2011} V.A. Plujko, R. Capote, O.M. Gorbachenko, \ql Giant dipole resonance parameters with uncertainties from photonuclear cross sections,\qr \textsc{At. Data Nucl. Data Tables }\textbf{97}, 567--585 (2011).

\bibitem{Plujko:2018} V.A. Plujko, O.M. Gorbachenko, R. Capote, P. Dimitriou, \ql Giant dipole resonance parameters of ground-state photoabsorption: Experimental values with uncertainties,\qr \textsc{At. Data Nucl. Data Tables }\textbf{123--124}, 1--85 (2018).

\bibitem{Bohr:1936} N. Bohr, \ql Neutron Capture and Nuclear Constitution\qrs, \textsc{Nat. }\textbf{137}, 344 (1936).

\bibitem{PCROSS} R. Capote, V. Osorio, R. L\'opez, E. Herrera, and M. Piris, \ql Analysis of experimental data on neutron-induced reactions and development of code PCROSS for the calculation of differential pre-equilibrium emission spectra with modelling of the level density function\qrs, Final report on research contract 5472/RB, report \textbf{IAEA(CUB)-004} International Atomic Energy Agency, Vienna~(1991).

\bibitem{hf52} W. Hauser and H. Feshbach, \ql The inelastic scattering of neutrons\qrs, \textit{ Phys. Rev. }\textbf{87}, 366--373~(1952).

\bibitem{HRTW75} H. M. Hoffmann, J. Richert, J. W. Tepel, H. A. Weidenm\"uller, \ql Direct reactions and Hauser-Feshbach theory,\qr \textsc{ Ann. Phys. (N.Y.) }\textbf{90}, 403--437~(1975).

\bibitem{Kawano:2016} T. Kawano, R. Capote, S. Hilaire, P. Chau Huu-Tai, \ql Statistical Hauser-Feshbach theory with width-fluctuation correction including direct reaction channels for neutron-induced reactions at low energies,\qr  \textsc{ Phys. Rev. }\textbf{C94}, 014612~(2016).

\bibitem{Sin:2016} M. Sin, R. Capote, M. Herman, and A. Trkov, \ql Extended optical model for fission,\qr \textsc{ Phys. Rev. } \textbf{C93}, 034605 (2016).

%


\bibitem{Sin:2008} M. Sin and R. Capote, \ql Transmission through multi-humped fission barriers with absorption: A recursive approach,\qr \textsc{ Phys. Rev. }{\bf C77}, 054601 (2008).


\bibitem{Sin:2006} M. Sin, R. Capote, A. Ventura, M. Herman, and P. Oblozinsk\'{y}, \ql Fission of light actinides: $^{232}$Th(n,f) and $^{231}$Pa(n,f) reactions,\qr \textsc{ Phys. Rev. }\textbf{C74}, 014608~(2006).

\bibitem{Sin:2008a} M. Sin, R. Capote, S. Goriely, S. Hilaire and A.J. Koning, \ql Neutron-induced fission cross section on actinides using microscopic fission energy surfaces,\qr \textit{Int. Conf. on Nucl. Data for Sc. and Techn. 2007}, p.313 (2008).

\bibitem{Goriely:2009}S. Goriely, S. Hilaire, A. J. Koning, M. Sin, and R. Capote, \ql Towards a prediction of fission cross sections on the basis of microscopic nuclear inputs\qrs,
\textsc{ Phys. Rev. }{\bf C79}, 024612 (2009).

\bibitem{ENDF-6} M. Herman and A. Trkov, \ql ENDF-6 Formats Manual. Data Formats and Procedures for the Evaluated Nuclear Data File ENDF/B-VI and ENDF/B-VII,\qr CSEWG Document \textbf{ENDF-102}, Report \textbf{BNL-90365-2009}, Rev.2, SVN Commit 85 (2012).

\bibitem{EXFOR} N. Otuka, E. Dupont, V. Semkova, B. Pritychenko, \etals, \ql Towards a More Complete and Accurate Experimental Nuclear Reaction Data Library (EXFOR):
International Collaboration Between Nuclear Reaction Data Centres (NRDC),\qr \textsc{ Nucl. Data Sheets }, \textbf{120}, 272--276 (2014).

\bibitem{JENDL-PD} JENDL Photonuclear Data File 2016 (JENDL/PD-2016), N. Iwamoto, K. Kosako, T. Murata \ql Photonuclear Data File,\qr \textbf{JAEA-Conf 2016-004}, pp. 53-58 (2016). 

\bibitem{IAEA-PD} Technical report \textbf{IAEA-TECDOC-1178}, \textit{Handbook of photonuclear data for applications: Cross sections and spectra}, October 2000, IAEA Vienna, Austria.

\bibitem {CCONE} O.  Iwamoto,  \ql Development of a Comprehensive Code for Nuclear Data Evaluation, CCONE, and Validation Using Neutron-Induced Cross Sections for Uranium Isotopes,\qr  \textsc{J. Nucl.  Sc.  Tech. }{\bf 44}, 687 (2007).

\bibitem {Capote:2017} R. Capote, S. Hilaire, O. Iwamoto, T. Kawano, M. Sin, \ql Inter-comparison of Hauser-Feshbach model codes toward better actinide evaluations,\qr \textit{ND 2016: International Conference on Nuclear Data for Science and Technology}, \textsc{EPJ Web Conf. }{\bf 146}, 12034 (2017).

\bibitem{Bhandari:79} B. S. Bhandari, \ql Three-hump fission barrier in Th232,\qr \textsc{ Phys. Rev. }{\bf C19}, 1820--1826 (1979).

\bibitem{Froman:65} N. Fr\"{o}man and P.O. Fr\"{o}man, {\it JWKB Approximation, Contributions to the Theory}, North-Holland, Amsterdam (1965).

\bibitem{FD70} N. Fr\"{o}man and \"{O}. Dammert, \textsc{ Nucl. Phys. }{\bf A147}, 627 (1970).

\bibitem{Moraes:93} M.A.P.V. de Moraes, M.F. Cesar, \ql Photonuclear cross-sections of $^{233}$U using neutron capture gamma-rays, near threshold,\qr \textsc{Nuov. Cim. }\textbf{A106}, 1165--1173 (1993).


\bibitem{Gurevich:86} G.M. Gurevich, L.E. Lazareva, V.M. Mazur, G.V. Solodukhov, B.A. Tulupov, \ql Giant resonance in the total photoabsorption cross section near of Z=90 nuclei,\qr \textsc{ Nucl. Phys. }\textbf{A273}, 326 (1976).

\bibitem{Berman:86} B.L. Berman, J.T. Caldwell, E.J. Dowdy, S.S. Dietrich, P. Meyer, and R. A. Alvarez, \ql Photofission and photoneutron cross sections and photofission neutron multiplicities for 233U, 234U, 237Np, and 239Pu\qrs, \textsc{ Phys. Rev. }\textbf{C34}, 2201~(1986).


\bibitem{Caldwell:80} J.T. Caldwell, E.J. Dowdy, B.L. Berman, R.A. Alvarez, P. Meyer, \ql Giant resonance for the actinide nuclei: photoneutron and photofission cross sections for 235U, 236U, 238U, and 232Th,\qr \textsc{ Phys. Rev. }\textbf{C21}, 1215~(1980).


\bibitem{Moraes:89} M.A.P.V. de Moraes and M.F. Cesar, \ql Photofission cross sections of 233U and 239Pu near threshold induced by gamma rays from thermal neutron capture,\qr \textsc{Nucl. Instr. Meth. }\textbf{A277}, 467 (1989)

\bibitem{Veyssiere:73}A. Veyssiere, H. Beil, R. Bergere, P. Carlos, A. Lepretre, \ql A study of the photofission and photoneutron processes in the giant dipole resonance of 232Th, 238U and 237Np,\qr \textsc{Nucl. Phys. }\textbf{A199}, 45 (1973).

\bibitem{Bergere:72}R. Bergere, H. Beil, B. Carlos, A. Veyssiere, A. Lepretre, \ql Study of the giant resonance of fissile nuclei,\qr \textit{Conf. Nucl. Structure Studies}, Sendai, Japan, 273 (1972).

\bibitem{Goncalez:99}O.L. Goncalez, L.P. Geraldo, R. Semmler, \ql Measurement of neutron photoproduction cross sections for 232Th and 238U using capture gamma rays,\qr \textsc{Nucl. Sc. Eng. }\textbf{132}, 135 (1999).

\bibitem{Dickey:75}P.A. Dickey and P. Axel, \ql 238U and 232Th Photofission and Photoneutron Emission near Threshold,\qr \textsc{Phys. Rev. Lett. }\textbf{35}, 501 (1975).

\bibitem{Mafra:72} O.Y. Mafra, S. Kuniyoshi, and J. Goldemberg, \ql Intermediate structure in the photoneutron and photofission cross sections in U-238 and Th-232,\qr \textsc{Nucl. Phys. }\textbf{A186}, 110 (1972).

\bibitem{Soldatov:92} A.S. Soldatov and G.N. Smirenkin,  \ql Yield and cross section for fission of odd nuclei by gamma-rays with energy up to 11 MeV. Results of relative measuring of photofission yields and cross sections for nuclei 233,235U, 237Np, 239,241Pu and 241Am in the energy region 5--11 MeV,\qr \textsc{Yad. Fiz. }\textbf{55}, 3153 (1992).

\bibitem{Zhuchko:78} V.E. Zhuchko, Ya.B. Ostapenko, G.N. Smirenkin, A.S.Soldatov, Ya.M. Tsipenyuk, \ql Investigation of probability of the near-threshold fission of Th, U, Np, Pu, Am isotopes by bremsstrahlung gamma-quanta,\qr \textsc{Yad. Fiz. }\textbf{28}, 1170 (1978).

\bibitem{Huizenga:62} J.R. Huizenga, K.M. Clarke, J.E. Gindler, R. Vandenbosch, \ql Photofission cross sections of several nuclei with mono-energetic gamma rays,\qr \textsc{Nucl. Phys. }\textbf{34}, 439 (1962).

\bibitem{Soldatov:93} A.S. Soldatov, \ql Photofission cross section of Uranium-234 in the energy region from 5 to 9 MeV and its comparison with the data for Thorium-232 and Neptunium-237 in subbarrier region,\qr \textsc{Yad. Fiz. }\textbf{56}, 16 (1993).

\bibitem{Lindgren:80} L.J. Lindgren, A.S. Soldatov, and Yu.M. Tsipenyuk, \ql Under-barrier photofission of U-234,\qr \textsc{Yad. Fiz. }\textbf{32}, 335 (1980).

\bibitem{Bowman:78} C.D. Bowman, I.G. Schroder, K.C. Duvall, C.E. Dick, \ql Subthreshold photofission of U-235 and Th-232,\qr \textsc{Phys. Rev. }\textbf{C17}, 1086 (1978).

\bibitem{Ries:84} H. Ries, U. Kneissl, G. Mank, H. Stroher, W. Wilke, R. Bergere, P. Bourgeois, P. Carlos, J.L. Fallou, P. Garganne, A. Veyssiere, L.S. Cardman,
\ql Absolute photofission cross sections of U-235,238 measured with tagged photons between 40 and 105 MeV,\qr \textsc{Phys. Lett. }\textbf{B139}, 254 (1984).

\bibitem{Anderl:73} R.A. Anderl, M.V. Yester, R.C. Morrison, \ql Photofission cross sections of U-238 and U-235 from 5.0 to 8.0 MeV,\qr \textsc{Nucl. Phys. }\textbf{A212}, 221 (1973).

\bibitem{Khan:72} A.M. Khan and J.W. Knowles, \ql Photofission of Th-232, U-238 and U-235 near threshold using a variable energy beam of g-rays,\qr \textsc{Nucl. Phys. }\textbf{A179}, 333 (1972).

\bibitem{Bowman:64}C.D. Bowman, C.F. Auchampaugh, S.C. Fultz, \ql Photodisintegration of 235U,\qr \textsc{Phys. Rev. }\textbf{133}, B676 (1964).

\bibitem{Ostapenko:78} Yu.B. Ostapenko, G.N. Smirenkin, A.S. Soldatov, V.E. Zhuchko, Yu.M.Tsipenyuk, \ql Yields and cross sections of photofission for isotopes Th, U, Np, and Am in energy range from 4.5 MeV to 7.0 MeV,\qr \textsc{Vop. At.Nauki i Tekhn.,Ser.Yad. Konst. }\textbf{30}, 3 (1978).


\bibitem{Zhuchko:76} V.E. Zhuchko, Yu.B. Ostapenko, A.S. Soldatov, Yu.M. Tsipenyuk, \ql Restoration of photofission cross-sections from bremsstrahlung experiments,\qr \textsc{Nucl. Instr. Meth. Phys. Res. }\textbf{136}, 373 (1976).

\bibitem{Yester:73}M.V. Yester, R.A. Anderl, R.C. Morrison, \ql Photofission cross sections of Th-232 and U-236 from threshold to 8 MeV\qrs, \textsc{Nucl. Phys. }\textbf{A206}, 593 (1973).

\bibitem{Soldatov:95} A.S. Soldatov and G.N. Smirenkin, \ql Yield and cross section of 232Th and 236U fission induced by gamma-quanta with energy below 11 MeV,\qr \textsc{Yad. Fiz. }\textbf{58}, 224 (1995).

\bibitem{Zhuchko:78a} V.E. Zhuchko, Yu.B. Ostapenko, G.N. Smirenkin, A.S. Soldatov, Yu.M. Tsipenyuk, \ql Experimental investigations of the effect of the`isomer shelf` in photofission cross sections of heavy nuclei,\qr \textsc{Yad. Fiz. }\textbf{28}, 1185 (1978).

\bibitem{Csige:2013} L. Csige, D.M. Filipescu, T. Glodariu, J. Gulyas, M.M. Gunther, D. Habs, H.J. Karwowski, A. Krasznahorkay, G.C. Rich, M. Sin, L. Stroe, O. Tesileanu, P.G. Thirolf,
\ql Exploring the multihumped fission barrier of 238U via sub-barrier photofission,\qr \textsc{Phys. Rev. }\textbf{C87}, 044321 (2013).

\bibitem{Lepretre:87} A. Lepretre, R. Bergere, P. Bourgeois, P. Carlos, J. Fagot,J.L.Fallou, P. Garganne, A. Veyssiere, H. Ries, R. Gobel, U. Kneissl, G. Mank, H. Stroher, W. Wilke, D. Ryckbosch, J. Jury, \ql Absolute photofission cross sections for Th-232 and U-235,238 measured with monochromatic tagged photons (20 MeV $<$ EG $<$ 110 MeV),\qr \textsc{Nucl. Phys.} \textbf{A472}, 593 (1987).

\bibitem{Manfredini:66} A. Manfredini, M. Muchnik, L. Fiore, C. Ramorino, H.G. De Carvalho, R. Bosch, W.Wolfli,
\ql Results on the cross section of 238U fission induced by low energy monoenergetic gamma rays,\qr \textsc{Nuovo Cim. }\textbf{B44}, 218 (1966).

\bibitem{Ostapenko:81} Yu.B. Ostapenko, G.N. Smirenkin, A.S. Soldatov, Yu.M. Tsipenyuk, \ql Isomer shelf in the photofission of Th-232 and U-238\qrs, \textsc{Phys. Rev. }\textbf{C24}, 529 (1981).

\bibitem{Koretskaya:79} I.S. Koretskaya, V.L. Kuznetsov, L.E.Lazareva, V.G. Nedoresov, N.V. Nikitina, \ql Photofission cross sections for the nuclei Am-241 and Am-243 in the region of the E1 giant resonance,\qr \textit{Yad. Fiz. }\textbf{30}, 910 (1979).

\bibitem{Arruda:76}J.D.T. Arruda Neto, S.B. Herdade, B.S. Bhandari, I.C. Nascimento, \ql Electrofission and photofission of U-238 in theenergy range 6--60 MeV,\qr \textsc{Phys. Rev. }\textbf{C14}, 1499 (1976).

\bibitem{Rabotnov:70} N.S. Rabotnov, G.N. Smirenkin, A.S. Soldatov, L.N. Usachev, S.P. Kapitza, Yu.M. Tsipenyuk, \ql Photofission of Th-232, U-238, Pu-238, Pu-240, and Pu-242 nuclei and structure of fission barrier,\qr \textsc{Yad. Fiz. }\textbf{11}, 508 (1970).

\bibitem{Katz:58} L. Katz, A.P. Baerg, F. Brown, \ql Photofission in heavy elements\qrs, \textit{Sec. Int. At. En. Conf. }, Geneva, Vol.15,  p.188 (1958).

\end{thebibliography}
\end{document}